\documentclass[12pt,preprint]{aastex}
\usepackage{soul}
\usepackage{ulem}
\newcommand{\eg}{e.g.}

\newcommand{\Msun}{\hbox{M$_{\odot}$}}

\newcommand{\ha}{H$\alpha$}
\newcommand{\HST}{{\sl HST}}

\slugcomment{Accepted to ApJ}

\shorttitle{Young M dwarfs within 25 pc}
\shortauthors{Shkolnik et al.}

\begin{document}


\title{Identifying the young low-mass stars within 25 pc. I. Spectroscopic Observations\altaffilmark{1}\\}


\author{Evgenya~Shkolnik\altaffilmark{2}}
\affil{Department of Terrestrial Magnetism, Carnegie Institution of Washington, 5241 Broad Branch Road, NW, Washington, DC 20015}
\email{shkolnik@dtm.ciw.edu}

\author{Michael C. Liu\altaffilmark{3}}
\affil{Institute for Astronomy, University of Hawaii at Manoa\\ 2680 Woodlawn Drive, Honolulu, HI 96822}
\email{mliu@ifa.hawaii.edu}

\and

\author{I. Neill Reid}
\affil{Space Telescope Science Institute, Baltimore, MD 21218}
\email{inr@stsci.edu}

\altaffiltext{1}{Based on
observations collected at the W. M. Keck Observatory and the Canada-France-Hawaii Telescope.  The Keck Observatory is operated as a scientific partnership between the California Institute of
Technology, the University of California, and NASA, and was made possible by the generous
financial support of the W. M. Keck Foundation. The CFHT is operated by the National Research Council of Canada,
the Centre National de la Recherche Scientifique of France, and the University of Hawaii.}
\altaffiltext{2}{Carnegie Fellow}
\altaffiltext{3}{Alfred P. Sloan Research Fellow}

\begin{abstract}

We have completed a high-resolution (R$\approx$60,000) optical spectroscopic survey of 185 nearby M dwarfs identified using ROSAT data to
select active, young objects with fractional X-ray luminosities comparable to or greater than Pleiades members. Our targets are drawn
from the {\it NStars} 20-pc census and the {\it Moving-M} sample with distances determined from parallaxes or spectrophotometric
relations. We limited our sample to 25 pc from the Sun, prior to correcting for pre-main sequence over-luminosity or binarity. Nearly
half of the resulting M dwarfs are not present in the Gliese catalog and have no previously published spectral types. We identified 30
spectroscopic binaries (SBs) from the sample, which have strong X-ray emission due
to tidal spin-up rather than youth. This is equivalent to a 16\% spectroscopic binary fraction, with at most a handful of undiscovered SBs.  We estimate upper limits on the age of the remaining M dwarfs using spectroscopic youth indicators
such as surface gravity-sensitive indices (CaH and K I). We find that for a sample of field stars with no metallicity measurements, a
single CaH gravity index may not be sufficient, as higher metallicities mimic lower gravity. This is demonstrated in a sub-sample of
metal-rich RV standards, which appear to have low surface gravity as measured by the CaH index, yet show no other evidence of youth. We
also use additional youth diagnostics such as lithium absorption and strong H$\alpha$ emission to set more stringent age limits. Eleven M
dwarfs with no \ha\/ emission or absorption are likely old ($>$400 Myr) and were caught during an X-ray flare. We estimate that our final
sample of the 144 youngest and nearest low-mass objects in the field is less than 300 Myr old, with 30\% of them being younger than 150
Myr and 4 very young ($\lessapprox$10 Myr), representing a generally untapped and well-characterized resource of M dwarfs for intensive
planet and disk searches.

\end{abstract}


\section{Introduction}\label{intro}

Observational studies of planet formation have been energized
by the discovery of young ($<$100 Myr) solar-type stars close to
Earth (e.g.~\citealt{jeff95,webb99,mont01}), identified from multiple indicators of youth
including chromospheric activity and strong X-ray.  The
combination of distances, proper motions, and radial velocities has
allowed many of these stars to be kinematically linked to coeval
moving groups (e.g., \citealt{zuck04,torr08}).  These young moving groups (YMGs)
are several times closer to Earth than the traditional
well-studied star-forming regions such as Taurus and Orion
($\sim$150--500 pc).  More importantly, these groups have ages of
$\sim$10--100 Myr, a time period in stellar evolution that has largely been underrepresented
in previous studies.  This is expected to be a key epoch for
understanding planet formation, coinciding with the end of giant planet
formation and the active phase of terrestrial planet formation (e.g.~\citealt{mand07,ida08}).

While these YMGs have been a boon for observers, the current census has severe limitations --- in
particular, it is mostly restricted to the higher-mass (AFGK-type) stars and contains very few low-mass M~dwarfs.
This paucity is striking given that M~dwarfs dominate the stellar mass function by number$\colon$ roughly 3 out of
4 stars in a volume-limited sample of the solar neighborhood are M~dwarfs (\citealt{reid95,boch08}). For instance, in the
12-Myr old $\beta$ Pictoris moving group, 17 of the 44 identified members (\citealt{torr06}) are M dwarfs, whereas we would expect 75\% based on the stellar mass function: $\sim$60 M dwarfs are missing from this group's known members.
%i.e.~$(17+x) / (44+x) = 0.75 \rightarrow x = 64$: 
%80\% of the M~dwarfs are missing from this group's known members.

%2/5 of the $\approx$40 identified members are M~dwarfs

\begin{figure}
\epsscale{1.0}
\plottwo{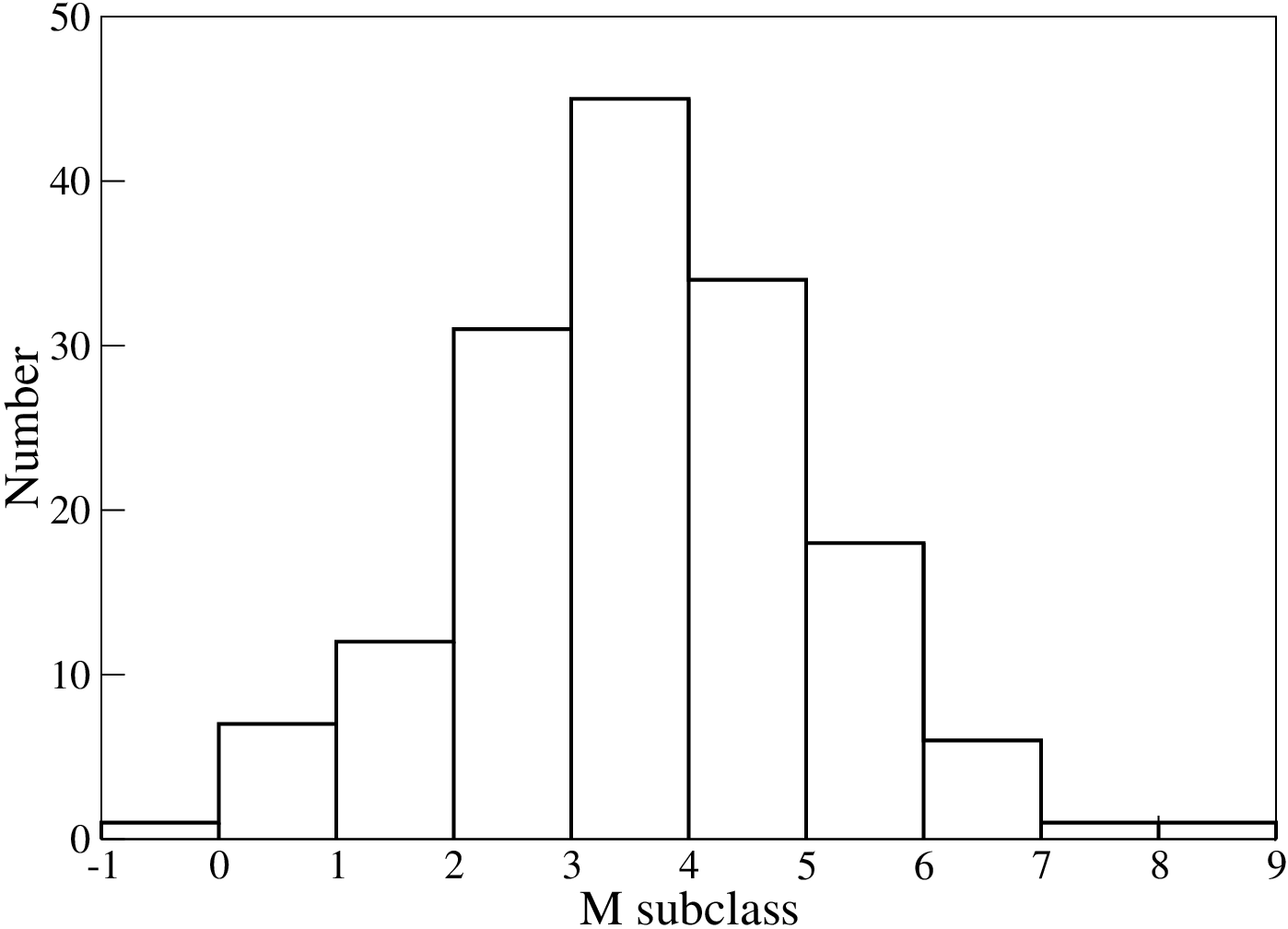}{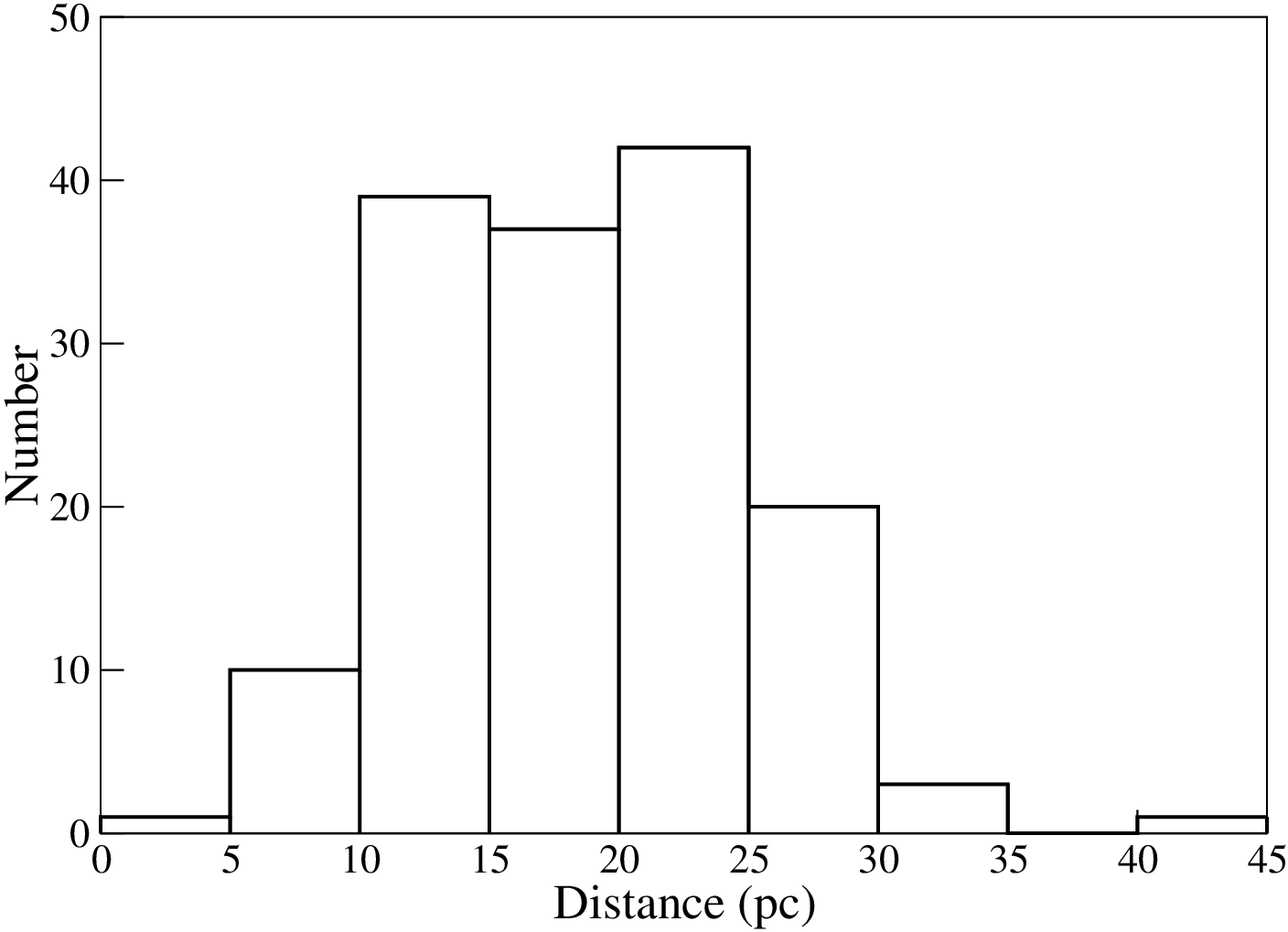}
\caption{Spectral type and distance histograms of the ROSAT-selected sample of M dwarfs. Distance greater than 25 pc (within error bars) are due to subsequent discoveries of visual binaries.
\label{hist}}
\end{figure}

The incompleteness in the current young low-mass census
arises from a combination of two factors:  (1) Young star searches to
date have relied on optical catalogs for distances and/or proper
motions, e.g., the Hipparcos and Tycho catalogs \citep{perr97}, the former limited to V $<$ 7 with parallaxes, while the latter limited to V $<$ 7 but without parallaxes.  However, M dwarfs are
optically faint and thus make a disproportionately small contribution to these catalogs.  (2) Until very recently, the
only all-sky surveys suited for finding X-ray active stars have been the ROSAT\footnote{The R\"ontgensatellit (ROSAT) was a joint German, US and British X-ray observatory operational from 1990 to 1999.}
catalogs (\citealt{voge99,voge00}),  
and since the X-ray luminosities of M dwarfs are
$\sim$10--300$\times$ lower than solar-type stars, ROSAT detections are mostly
limited to the nearest, earliest-M stars (Figure~\ref{hist}).

Previous searches for young, X-ray-active, late-type stars focused on a satellite's detection limits rather than a volume-limited sample. E.g.~\cite{riaz06} collected low-resolution spectra of $\approx$800 M dwarfs (M0 -- M5) detected with ROSAT with photometric distances ranging from 3 to 750 pc. They determine their sample to be generally young since the measured coronal activity of most of the stars is higher than that of Hyades members ($\sim$600 Myrs). \cite{zick05} collected high-resolution spectra of 118 ROSAT-selected G, K and M stars extending out to 600 pc and from lithium measurements, estimate that 25\% of their sample is near the age of the Pleiades ($\sim$120 Myr). And more recently, \cite{cove08} published a catalog of 348 stars identified from correlating the Chandra X-ray Observatory's archival data with the Sloan Digital Sky Survey (SDSS). Of these, 36 are newly identified M dwarfs with distances out to 1000 pc and a wide range of X-ray and \ha\/ luminosities.
%($L_X/L_{bol} \sim 10^{-6} - 10^{-3}$). 

The goal of this program is to identify the best possible M dwarf targets for direct imaging searches of extrasolar planets and
circumstellar disks. This requires a young, well-characterized, and nearby sample. The proximity is an essential benefit for
studies requiring high sensitivity 
as large distances limit both the linear resolution and available flux with which we can
image planets and detect disks.
Also, M~dwarfs in
principle could represent the most common and nearest to Earth hosts of planetary systems, and provide a
potentially much larger population of targets than has been studied to date.

It is imperative that we add youth as a criterion to our M dwarf sample as planets cool and fade significantly between 10 Myr and 1 Gyr. For example, the luminosity of a 5 M$_{Jup}$ planet drops by 2 -- 3 orders of magnitude in this time span, while its effective temperature decreases from 1300 to 400 K (\citealt{bara03}). Also, the fraction of debris disks around AFGK, and possibly M stars, decreases significantly by $\sim$150 Myr \citep{riek05,hill08}.

 The value of young low-mass stars is demonstrated by the
 $\approx$12~Myr-old star AU~Mic, the first robustly identified M~dwarf
 debris disk system (\citealt{song02,liu04b}).  The AU~Mic disk
 has been the subject of intense scrutiny since its discovery.  At a
 distance of only 10~pc, its disk is seen in scattered light as far as
 20\arcsec\ in radius \citep{kala04}.  Adaptive optics
 and \HST\ imaging of the disk achieves a spatial resolution of 0.4~AU
 (\eg, \citealt{liu04b,kris05b,fitz07}) and
 reveals a rich variety of substructure.  The proximity and edge-on
 geometry of the disk allow for very sensitive studies of the
 circumstellar gas content (\citealt{robe06,fran07}) and
 searches for transiting planets (\citealt{hebb07}).  However, thus far
 this system remains the singular example of a resolved debris disk
 around a low-mass star.  Thus, the scientific potential of young M dwarfs remains largely untapped due to the very limited current
 census.

\section{Sample Selection}\label{sample}

The Two Micron All Sky Survey (2MASS) is optimal for finding low-mass stars, since the SEDs of cool stars peak in the near-IR (e.g.~\citealt{hawl02}). However, the $JHK$ infrared passbands provide less distinctive spectral classification of early- and mid-M dwarfs (i.e.~M2--M7 dwarfs have ($J-K$) colors which span only 0.2 mag; \citealt{reid07b}) impeding the photometric distance determination. Thus, in order to fully characterize a volume-limited sample of young M dwarfs, a proper motion requirement of $\mu >$0.18\arcsec\ yr$^{-1}$ was implemented, equivalent to a tangential velocity of 21 km~s$^{-1}$ at 25 pc.

We drew $\approx$800 targets from the NStars 20-pc census \citep{reid03,reid04} constructed from the 2MASS catalogs (\citealt{skru06}) along with the \cite{lepi02} and \cite{lepi05} proper motion catalogs. In addition to these, we included $\sim$300 newly-catalogued M dwarfs that exhibit significant proper motion between the POSSI and 2MASS surveys (i.e.~the Moving-M sample; \citealt{reid07a}). \cite{reid03,reid04,reid07a} used a combination of photometric (i.e.~$(V-J)\geq3$ and $(H-K_s)\geq0.23$ to largely eliminate stars earlier than M0) and proper-motion methods to identify candidate cool objects, which were subsequently confirmed with follow-up moderate-resolution (R$\sim$1800) optical spectroscopy. 
Distances are available either from parallaxes or spectrophotometric relations and are limited to 25 pc from the Sun, good to $\lesssim$15\% assuming the stars are single and on the main-sequence (e.g., \citealt{reid02c,cruz03}). Nearly half of the resulting M dwarfs are not present in the Gliese catalog and have no previously-published spectral types.   

As stellar activity is a powerful indicator of youth (open clusters show a consistent and relatively rapid decline in activity with age at a given spectral type; \citealt{prei05}), we cross-referenced our total sample of 1103 M dwarfs in the immediate solar neighborhood against the ROSAT All-Sky Survey Bright Source Catalog and Faint Source Catalog \citep{voge99,voge00}. Our query was limited to a search radius of 25\arcsec\ around the 2MASS coordinates, the 2$\sigma$ positional error determined by \cite{voge99}. This returned 364 sources.

\begin{figure}
\epsscale{.60}
\plotone{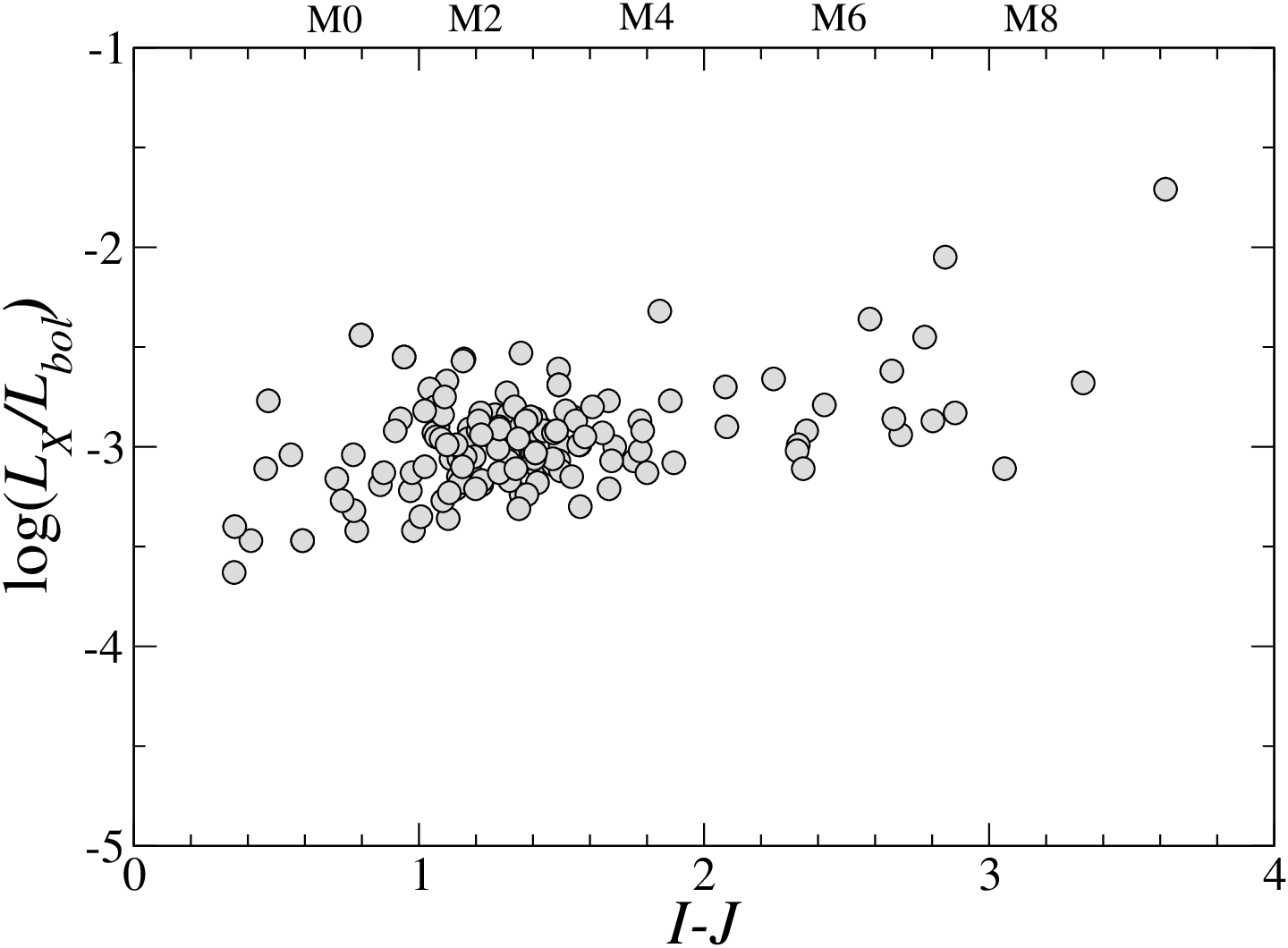}
\caption{The fractional X-ray luminosity ($L_X/L_{bol}$) as a function of $I-J$ for our sample of single low-mass star.
\label{ij_lxlbol}}
\end{figure}

\begin{figure}
\epsscale{.60}
\plotone{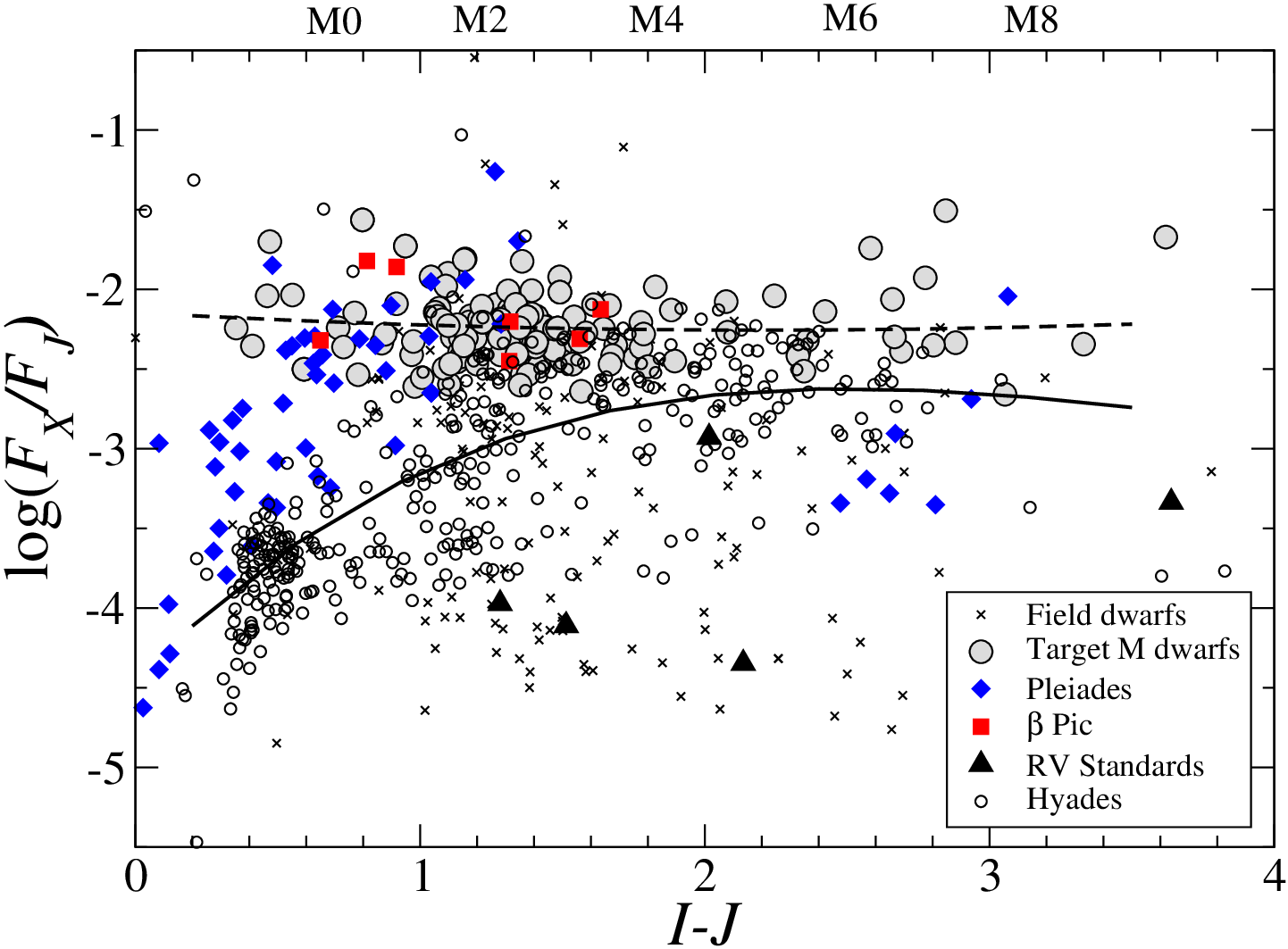}
\caption{The fractional X-ray luminosity as a function of $I-J$ for our sample of single low-mass stars compared with seven M-dwarf members of the $\beta$ Pic young moving group at 12 Myr \citep{torr06}, Pleiades members at 120 Myr (\citealt{mice98}), Hyades M dwarfs at 650 Myr (\citealt{ster95}), field stars (\citealt{huns99}), and the RV standard stars we observed that have been detected by ROSAT. The lines are quadratic fits to the M dwarf (dashed) and Hyades (solid) samples.
\label{ij_fxfj}}
\end{figure}

Of these, 196 (18\% of the full sample and with a median offset between the ROSAT and 2MASS coordinates of 9\arcsec) are strong coronal X-ray emitters with fractional luminosities hovering near the ``saturation'' level $log(L_X/L_{bol}) \sim -3$ (e.g.~\citealt{riaz06}), ranging across all M dwarf subclasses (Figure~\ref{hist} \& \ref{ij_lxlbol}). Mismatches with other sources would have been easily identified as the fractional X-ray luminosity of normal stars relative to optical magnitudes is always on the order of 0.1 or less, where as other X-ray sources, such as galaxies and quasars have fractional luminosities of 1 or greater (\citealt{stoc91,zick03}). In Figure~\ref{ij_fxfj} we plot the fractional X-ray flux, $F_X/F_J$ as a function of ($I-J$), where $F_X$ is the empirically calibrated X-ray flux using the count-rate conversion equation of Schmitt et al.~(1995), and $F_J$ is the 2MASS $J$-band flux. Target stars were chosen to have high X-ray emission ($log(F_X/F_J) > -2.5$) comparable to or greater than the fractional luminosities of Pleiades members (120 Myr, \citealt{mice98}) and $\beta$~Pic members (12 Myr, \citealt{torr06}). The 650-Myr-old Hyades stars emit two orders of magnitude less X-ray radiation for early-Ms than the younger samples \citep{ster95}.
Though data at later spectral types are sparse and the X-ray fluxes of the samples plotted in Figure~\ref{ij_fxfj} appear to converge, our targets on average are still significantly stronger X-ray emitters than the late-Ms of the Hyades.

\section{The Spectra}\label{spectra}

We acquired high-resolution \'echelle spectra of 185 low-mass stars\footnote{We were unable secure spectra of 11 of the 196 X-ray sources because they were either too far south to be observed from Mauna Kea and/or too faint, even with the Keck I telescope.} over 4 nights with the High Resolution \'Echelle Spectrometer (HIRES; \citealt{vogt94}) on the Keck I 10-m telescope and over 6 nights with the \'Echelle SpectroPolarimetric Device for
the Observation of Stars (ESPaDOnS; \citealt{dona06}) on the Canada-France-Hawaii 3.6-m telescope, both located on the summit of Mauna Kea.

The relevant properties of the targets confirmed to not be spectroscopic binaries (SB) are listed in Table~\ref{targets}. Three of these 155 M dwarfs do not have ROSAT detections but were nonetheless included in the sample due to previously identified youth indicators.  (See Table~\ref{ages} for details.) 

We used the 0.861$\arcsec$ slit with HIRES to give a spectral resolution of $\lambda$/$\Delta\lambda$$\approx$58,000. The upgraded detector consists of a mosaic of three 2048 $\times$ 4096 15-$\micron$ pixel CCDs, corresponding to a blue, green and red chip spanning 4900 -- 9300 \AA. To maximize the throughput near the peak of a M dwarf spectral energy distribution, we used the GG475 filter with the red cross-disperser. The data product of each exposure is a multi-extension FITS file from which we reduce and extract the data from each chip separately.  

ESPaDOnS is fiber fed from the Cassegrain to Coud\'e focus where the fiber image is projected onto a Bowen-Walraven
slicer at the spectrograph entrance. With a 2048$\times$4608-pixel CCD detector, ESPaDOnS'
`star+sky' mode records the full spectrum over 40 grating orders covering 3700 to 10400 \AA\/ at a spectral
resolution of $\lambda$/$\Delta\lambda$$\approx$68,000. The data were reduced using {\it Libre Esprit}, a fully automated reduction package
provided for the instrument and described in detail by \cite{dona97,dona07}.

Each stellar exposure was bias-subtracted and flat-fielded for pixel-to-pixel sensitivity variations. After optimal
extraction, the 1-D spectra were wavelength calibrated with a Th/Ar arc. Finally
the spectra were divided by a flat-field response and corrected for the heliocentric velocity.
The final spectra were of moderate S/N reaching  20 -- 50 per pixel at 7000 \AA. Each night, spectra were
also taken of an A0V standard star for telluric line correction and an early-, mid-, and/or late-M radial velocity (RV) standard, which are listed in Table~\ref{standards}.

The high resolution of the data provides RV measurements to better than 1 km~s$^{-1}$ in almost all cases, which can be used in conjunction with a star's distance and proper motion to measure its three-dimensional space velocity (UVW).  This provides a promising way to determine stellar ages by linking stars kinematically to one of the several known YMGs or associations, which span ages between 8 and 300 Myr (e.g.~\citealt{zuck04}). However, only 11\% of our sample has the required distance precision ($<$10\%) from trigonometric parallaxes to unambiguously associate a star with a single YMG. The bulk of our targets have photometric distances (\citealt{reid02b}), which require an iterative analysis similar to the convergence method developed by \cite{torr06} to make meaningful conclusions about group membership. This UVW analysis of our ROSAT sample will be presented in a follow-up paper. Though it should be noted that kinematics alone should not be used to identify young M stars, as demonstrated by the lack of ROSAT detections in several of the proposed YMG members by \cite{lope06}, e.g.~HD 233153, HIP 53020, GJ 466, and  HIP 51317.

\subsection{Culling of the Spectroscopic Binaries}

High-resolution spectra are necessary to identify and remove spectroscopic binaries (SBs) from the sample, since tight
binaries have enhanced activity which would erroneously suggests youth. To search for single-lined SBs (SB1), we observed two
epochs of 65 targets, none of which showed a significant RV variation between visits. This implies that the
single-lined binary fraction in our sample is very low, less than 1.5\%, and that M dwarf binaries with low mass ratios ($M_2/M_1 \ll 1$) are rare. We had originally planned to observe all targets twice, but when it became clear that the SB1
fraction was so low, it was no longer good use of telescope time to continue multi-epoch observations of the entire
list. These 65 targets are identified in the last column of Table~\ref{targets}.

To search for multi-lined binaries, we cross-correlated each order between 7000 and 9000 \AA\/ of each stellar spectrum
with a RV standard of similar spectral type using IRAF's\footnote{IRAF (Image Reduction and Analysis Facility) is
distributed by the National Optical Astronomy Observatories, which is operated by the Association of Universities for
Research in Astronomy, Inc.~(AURA) under cooperative agreement with the National Science Foundation.} {\it fxcor}
routine \citep{fitz93}. We excluded the Ca II infrared triplet (IRT)\footnote{The target stars exhibit Ca II emission
that is not present in the RV standards.} and regions of strong telluric absorption in the cross-correlation.

We find a low-mass spectroscopic binary fraction, and therefore contamination rate, of 16\%. These 30 SBs
are composed of 28 SB2s, 1 SB3 and 1 SB4 \citep{shko08}, effectively doubling the number of known low-mass SBs and
proving that strong X- ray emission is an extremely efficient way to find multi-lined SBs.\footnote{These details pertaining to these SBs, i.e., orbital velocities, $v$sin$i$'s, mass ratios, etc., will be published in an upcoming paper.} It is possible that up to 4\%\footnote{This 4\% limit is based on the time a close-in binary would spend near conjunction such that the RVs of the two components would not produce a resolved cross-correlation function.} of the 90 stars with a single observation are indeed double-lined SBs if the systems were in conjunction at the times of the observation. Combining this with the $\leq$1.5\% chance of observing an SB1, there are at most a handful of undiscovered SBs in Table 1.

\section{Spectral Types}\label{spt}

\begin{figure}
\epsscale{.80}
\plotone{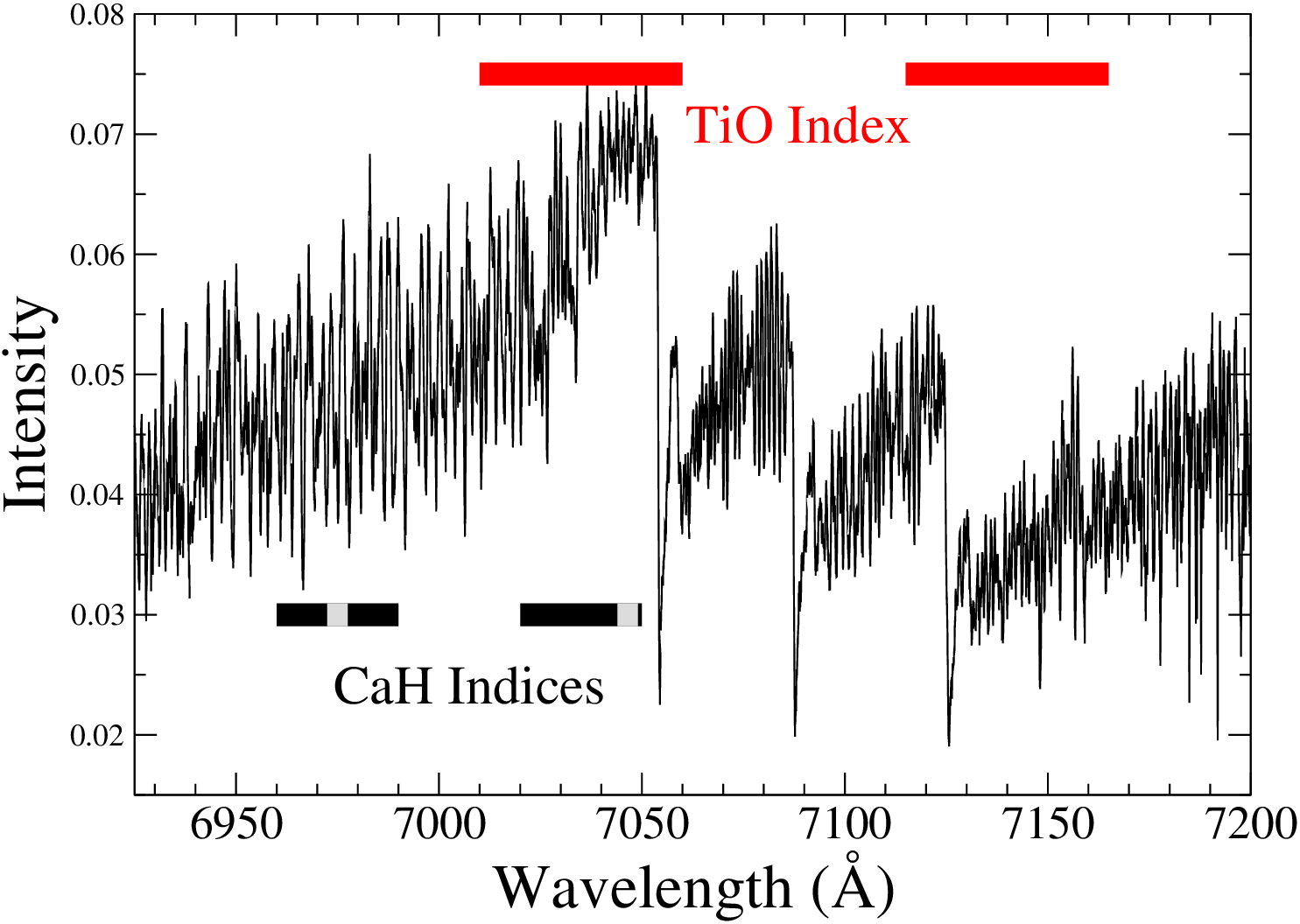}
\caption{A single order of a CFHT/ESPaDoNS spectrum of GJ 875.1 (SpT=M3). The three indices discussed in the text are marked by the bars: TiO-7140: [7010--7060]/[7115--7165] (red), CaH-wide: [7020--7050]/[6960-6990] (blue), and CaH-narrow: [7044--7049]/[6972.5--6977.5] (green).
\label{spec_indices}}
\end{figure}

The spectra of M dwarfs are dominated by the strong TiO molecular bands (Figure~\ref{spec_indices}), particularly diagnostic of the star's temperature.\footnote{See Section~\ref{metallicity} for a discussion on metallicity effects.}  Several TiO band indices have been used throughout the literature to determine the spectral types of M dwarfs, most commonly the TiO5 index of \citet{reid95}. 
The more recent TiO-7140 index, defined by \citet{wilk05}, is the ratio of the mean flux in two 50-\AA\/ bands: the `continuum' band centered on 7035 \AA\/ and the TiO band on 7140 \AA. We chose to use this latter index as our primary diagnostic of SpT because both of the TiO-7140 flux bands appear in a single order in all of our CFHT and Keck spectra.

\begin{figure}
\epsscale{.60}
\plotone{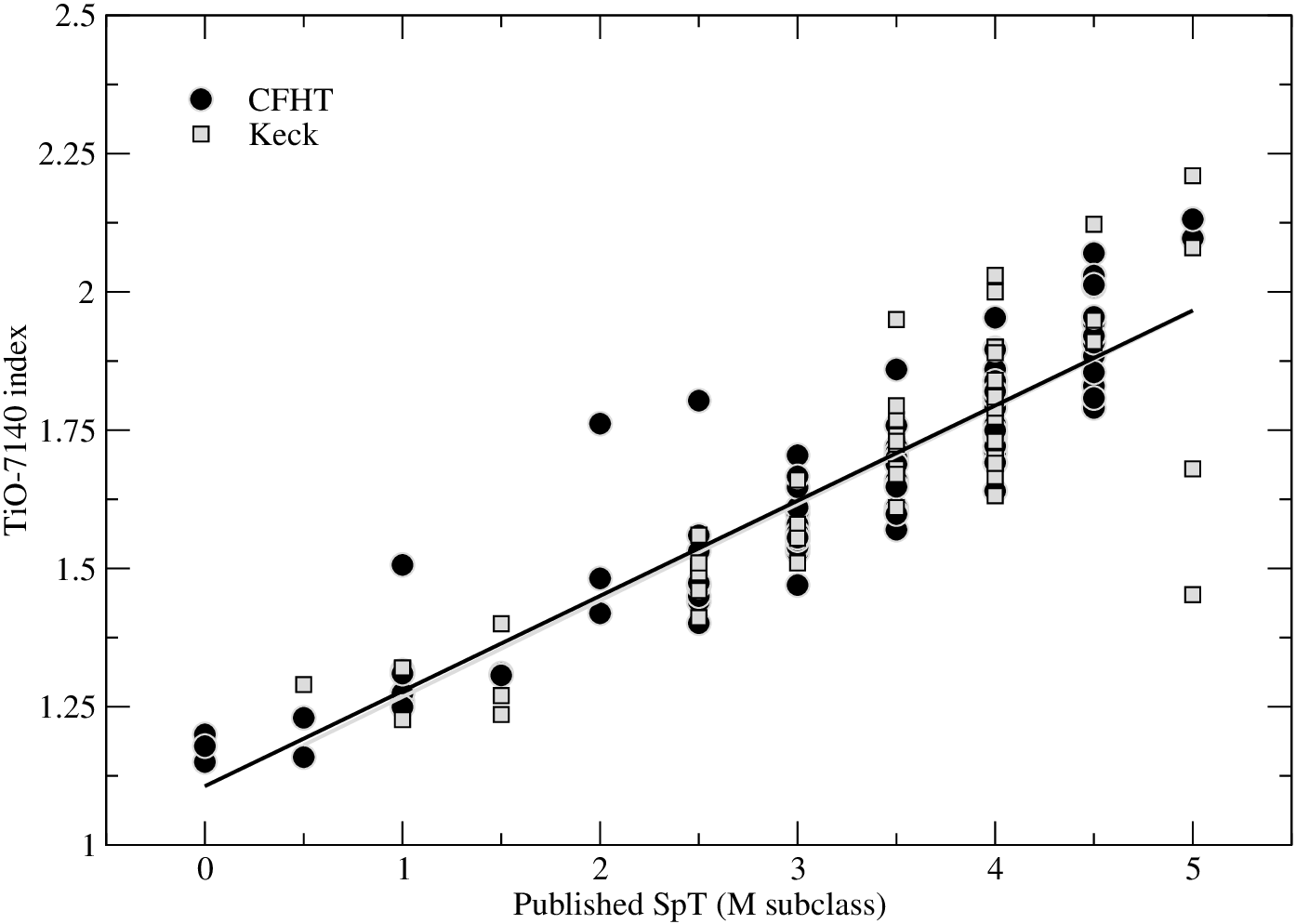}
\caption{Calibration of the TiO-7140 index using published SpTs for those stars in our sample, plus a few RV standards and known $\beta$ Pic members. The best-fit line to the CFHT data is virtually indistinguishable from that to the Keck data.  The best fit to full data set results in a calibration of SpT=(TiO$_{7140}-1.0911)/0.1755$.
\label{TiO7140_SpT}}
\end{figure}

We calibrated the TiO-7140 index for our data sets against previously measured (\citealt{reid95}) and published ($SIMBAD$ and references therein; \citealt{weng07}) spectral types of 136 M dwarfs we had observed, including several RV standards and known members of the $\beta$ Pic moving group. Both the linear fits to the CFHT and Keck data sets for stars with SpT earlier than M6 derive spectral types which agree to better than 0.2 subclasses (Figure~\ref{TiO7140_SpT}). We therefore combined the sets to get an average fit for the entire sample. The linear relationship used to convert the TiO-7140 index to SpT for M0 -- M5 stars is: 

\begin{center}
SpT=(TiO$_{7140}-1.0911)/0.1755$, {\it rms}= 0.6~~~~~~~~(1)
\end{center}

Here, M0 corresponds to SpT=0, M1 $\rightarrow$ 1, M2 $\rightarrow$ 2, etc. We used this calibration to determine the SpT of the 78 stars in our ROSAT sample which had no previously published spectral types, as well as refine the published values of the others.

We determine the errors of our measurements for the TiO index (and subsequent indices discussed below) by taking the average difference in values determined for the same stars observed on different nights. We have 25 Keck targets\footnote{Though we have an additional 19 Keck targets with repeat observations, the seeing on one of the two nights was too poor to measure the TiO and CaH indices.} and 21 CFHT targets with repeat observations. The measurement errors of the TiO index are larger for the Keck data than for the CFHT data: 0.061 and 0.015, corresponding to 0.33 and 0.11 M subclasses using Eq.~1, respectively. The bulk of the discrepancy in these two errors is attributed to the fiber feed of ESPaDOnS, which illuminates the spectrograph slit uniformly and is thus less affected by variable seeing. Most of the data collected at Keck on 11 May 2006 are particularly plagued by poor seeing, degrading the spectrophotometry used in measuring the indices. If no repeat observation was made, we defer to the published SpT for this run. 
Although the errors of the TiO index are relatively small, the calibration is based on a sample with SpTs binned to half a subclass, imposing a 0.5 subclass uncertainty in the calculated SpTs listed in Table~\ref{targets}.

For stars of spectral type M6 or later, the TiO band begins to weaken due to saturation and condensation onto grains (\citealt{jone98}) and additional absorption features which depress the ``continuum'' reference bandpass of the index. Similarly, the VO band at 7300 \AA\/ becomes stronger with spectral type and then weakens, but not until after M7, allowing us to unambiguously classify late-type M dwarfs at least until M7.
Therefore, for the 17 targets in the sample that are M6 or later, we derived spectral types from visual classification by comparison of the TiO and VO bands with standard stars of known spectral types (\citealt{reid02a,reid02b,reid03}). Again, due to the half-a-subclass binning of published spectral types, the error in the derived SpTs for these late Ms is also $\sim$0.5 subclasses.

\section{Age-Dating the Sample} \label{age_dating}

X-ray emission is ubiquitous amongst low-mass stars and is indicative of active stellar coronae throughout their lifetimes (e.g.~94\% of all K and M dwarfs within 6 pc exhibited detectable X-ray emission as observed by ROSAT; \citealt{schm95}). Fractional X-ray luminosities have also been shown to be ``saturated'' across a wide range of spectral types, H$\alpha$ equivalent widths, and ages at the value of log($L_X/L_{\rm bol}$) $\sim -3$, with the bulk of the dispersion in both field and cluster samples between  log($L_X/L_{\rm bol}$) of --2 and --4 due to variations in stellar rotation (\citealt{stau97,delf98}).\footnote{It has been well-established that the chromospheric activity and coronal emission of FGKM stars steadily decreases with age due to the reduced dynamo production of magnetic fields as the star spins down. Unlike the spin-down time-scale for higher-mass stars ($<$1 Gyr; e.g.~\citealt{skum72}), the spin-down time-scales for field M dwarfs range from 1 to 10 Gyr, taking longer with decreasing stellar mass (\citealt{delf98}).} As shown in Figures~\ref{ij_lxlbol} and \ref{ij_fxfj}, we have selected our sample of M dwarfs with high fractional X-ray luminosities as compared with the Pleiades, all near the X-ray saturation level. And though data at later spectral types are sparse, our targets do not have X-ray fluxes as low as Hyades members. 

X-ray emission of M dwarfs declines almost linearly in log-log space from $\sim$1 Myr to the about 650 Myr, the age of the Hyades, with a more rapid drop off after that (\citealt{prei05}). Using the \cite{prei05} relation of $L_X \sim t^{-0.75}$ for X-ray luminosity decline, we determine that the bulk of our objects are less than 280 Myr old.  We thus estimate an age of $\lesssim$300 Myr for our sample of 144 targets, with the early Ms ($I-J <$ 1.2, SpT $<$ M2.5), which have $F_X/F_J$ well above Hyades members, likely less than 150 Myr. This proportion of young stars (144 of 1103) is roughly consistent with expected number of young stars found in the Galactic disk assuming a uniform star-formation history.
 
We cannot however use X-ray activity to refine the stellar ages beyond this point, as the 150 Myr and 300 Myr limits discussed above are statistical in nature rather than applicable to individual stars.". Additional age diagnostics are necessary to better characterize and date individual stars. The spectroscopic age indicators available to us such as surface gravity, lithium absorption, and \ha\/ emission are discussed below, each with its implications and limitations.

\subsection{Surface Gravity}\label{grav}

A pre-main-sequence (PMS) star exhibits lower surface gravity as it has not yet fully contracted onto the main-sequence (MS). Even without measuring accurate values for surface gravities, the relative metric of a gravity index provides upper limits on the age of a low-g star using PMS stellar evolution models. Such models show that lower-mass stars take longer to contract to the MS, e.g.~a 0.5 \Msun~star (SpT $\approx$ M1) will reach the main sequence within 100 Myr whereas a 0.1 \Msun~star (SpT $\approx$ M8.5) will do so in 1 Gyr \citep{dant94}.

\begin{figure}
\epsscale{.60}
\plotone{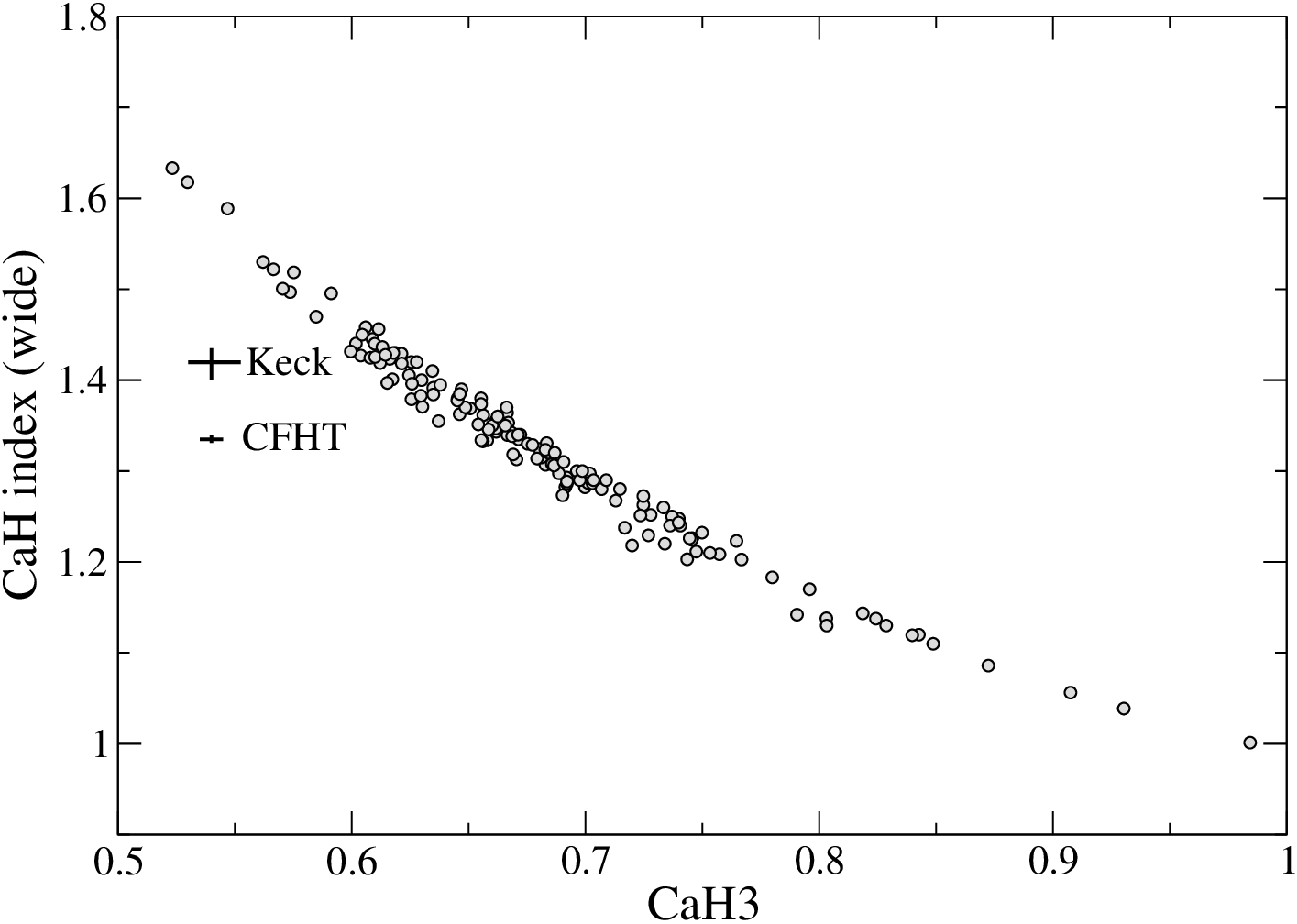}
\caption{Correlation of the CaH3 index for \cite{reid95} and CaH (wide) of \cite{kirk91}. See text for their definitions.
\label{CaH3_CaHw}}
\end{figure}

The prominent CaH molecular absorption bands (Figure~\ref{spec_indices}) in M dwarf optical spectra are often used as gravity indicators. Typically, M dwarf spectra are collected with lower resolution due to their intrinsic faintness and as such require broad 30--50\AA-wide indices.
We measured two indices defined in the literature: (1) ``Ratio A" from \cite{kirk91}, which is defined as the ratio of the mean intensity in two passbands, a ``continuum'' band and a molecular absorption band of CaH $\lambda$6975: [7020-7050\AA]/[6960-6990\AA], and (2) the ``CaH3'' ratio from \cite{reid95}, [6960-6990\AA]/[7042-7046\AA]. Figure~\ref{CaH3_CaHw} shows how the two indices are strongly correlated. We focus the rest of our discussion using the first index for which the relative errors in our sample are slightly smaller. 
Since we have 15--20 times the resolution of previous M dwarf surveys, we also defined a narrower 5-\AA\/ CaH index [7044--7049]/[6972.5--6977.5] providing a more discriminating scale with which to identify low-gravity stars. Both indices are plotted as function of SpT in Figure~\ref{SpT_CaH}.

\subsubsection{The effects of higher metallicity}\label{metallicity}

\begin{figure}
\epsscale{1.0}
\plottwo{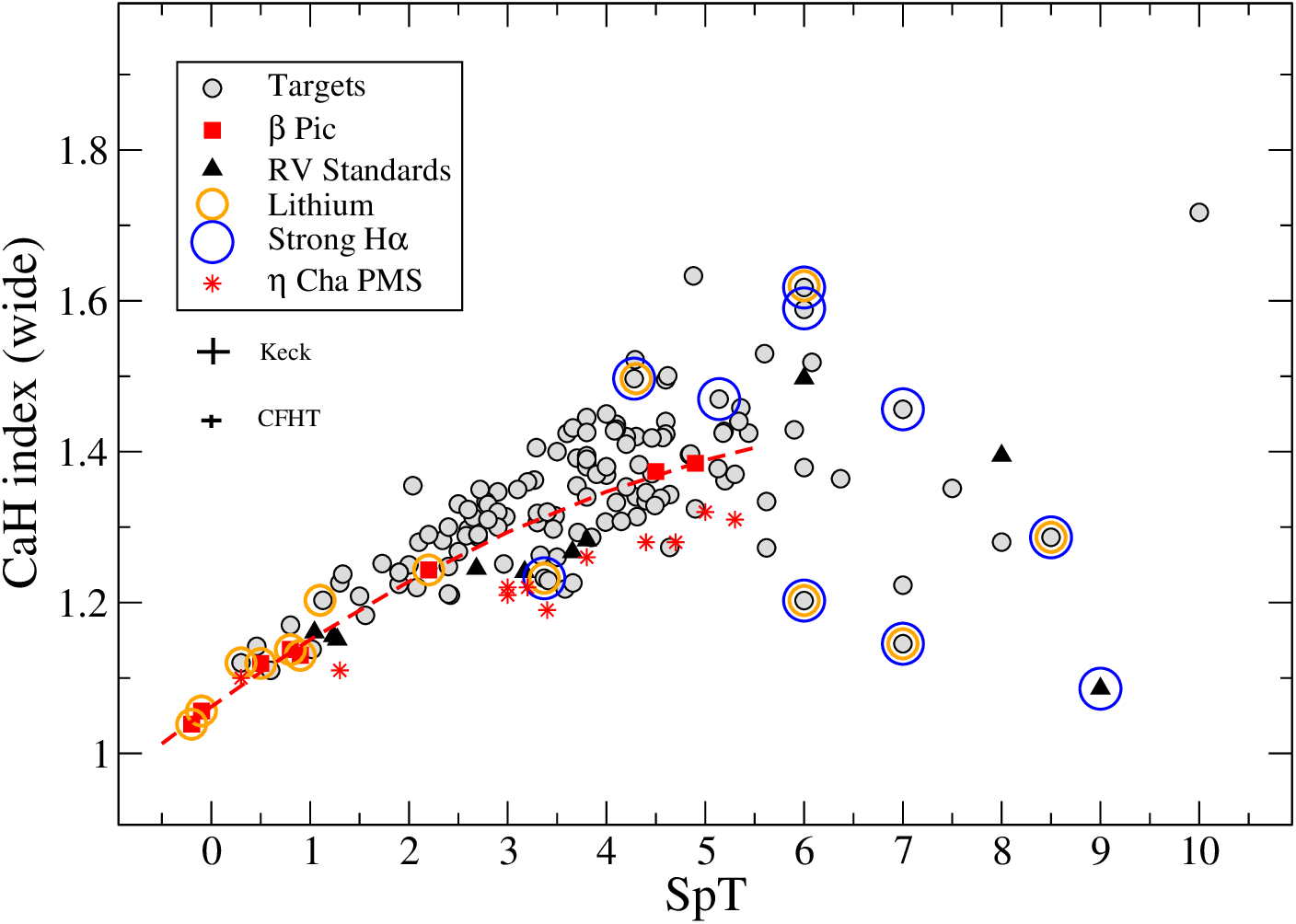}{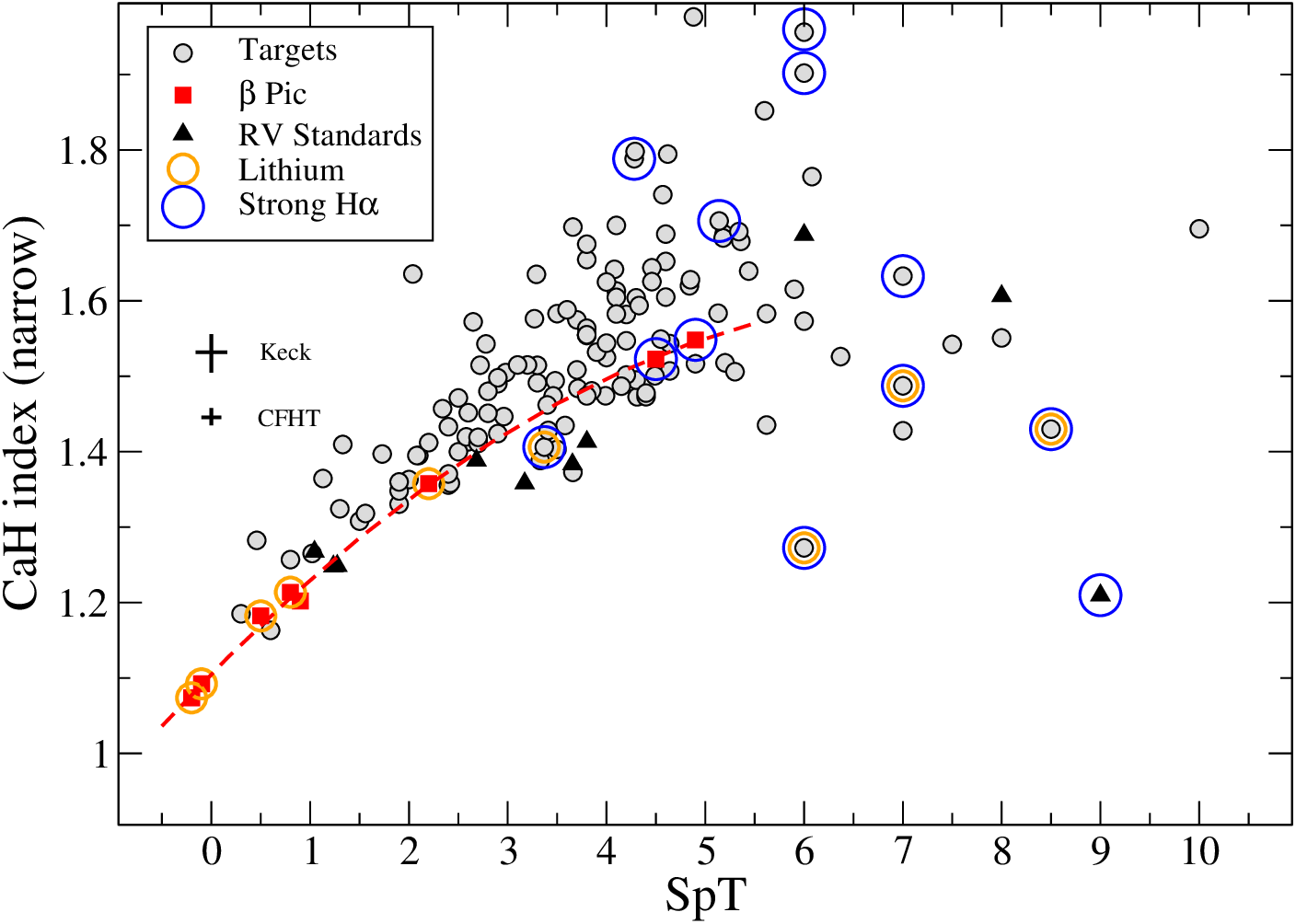}
\caption{Surface gravity sensitive indices, CaH (wide; \citealt{kirk91}) and CaH (narrow; this work) as a function M-type subclasses. The red dashed curves are polynomial fits to our observations of $\beta$ Pic M dwarfs: CaH$_{wide}$ = --0.0067 SpT$^2$ + 0.0986 SpT + 1.0633 and CaH$_{narr}$ = --0.00891 SpT$^2$ + 0.13364 SpT + 1.1053, respectively. The values in the left figure for the $\eta$ Cha PMS cluster members (9 Myr) were measured from low-resolution ($R\approx$900) spectra provided by \cite{lyo04}.
\label{SpT_CaH}}
\end{figure}

An important caveat to using the TiO and CaH molecules as temperature and gravity diagnostics is their
dependence on metallicity. Higher metallicity will mimic later spectral types and lower surface gravities.
\cite{wool06} demonstrate this metallicity dependence for M dwarfs using the TiO5 and CaH2 indices defined in
\cite{reid95}. Though the data are quite limited, we can estimate from Figure 4 of \cite{wool06} that in order to
increase the calculated SpT by 0.5 subclasses or decrease the CaH-wide index by 0.1, an M dwarf would need to have a
[Fe/H] enhancement of $\approx$0.5 dex relative to the population's mean metallicity. Unfortunately, measuring
metallicities directly from atomic lines in M dwarf spectra remains difficult and calibration errors usually range
from 0.2--0.3 dex and are likely higher (e.g.~\citealt{bonf05, wool06, bean06}). However, using the age-metallicity
relation for FGK stars in the local Galactic disk as measured by \cite{reid07b} and assuming the same metallicity distribution for nearby Ms as for higher mass stars, our young targets should have a mean [Fe/H] of 0.11 with a dispersion of 0.18 dex, implying that very few, if any, of our targets will have high enough metallicities to falsely appear as low-g stars in Figure~\ref{SpT_CaH}.

\begin{figure}
\epsscale{.60}
\plotone{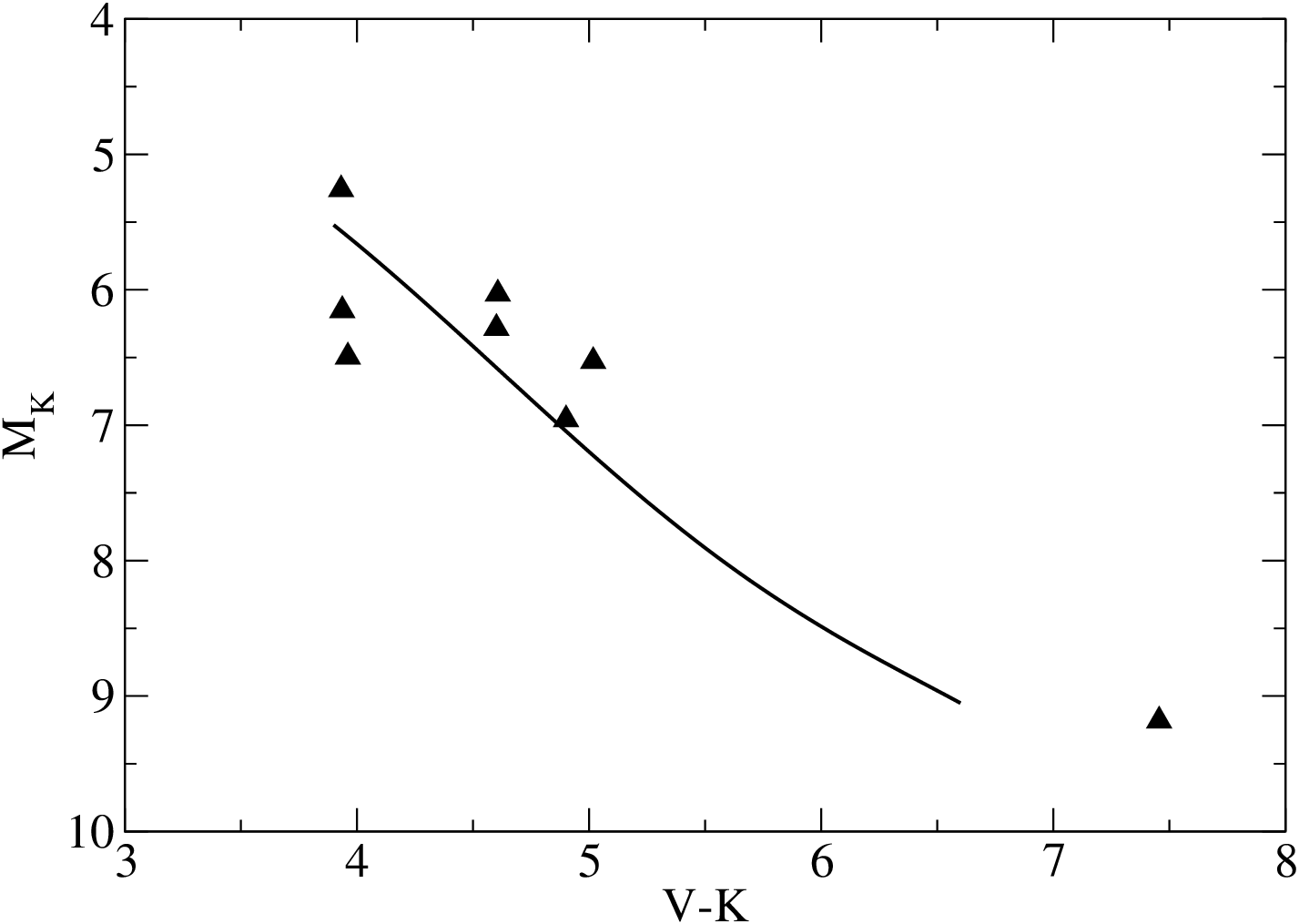}
\caption{Color-magnitude plot for the RV standards.  The solid line is the \cite{john09} mean main-sequence.  Those stars above the line are metal rich, while those below are metal poor. 
\label{rvstd_colors}}
\end{figure}

This metallicity effect is, however, apparent in the placement of the RV standards in the CaH-SpT distribution. They are near or below the locus of the $\beta$ Pic members and would therefore be flagged as low-g, when they should, in general, be old field stars. None of the standards appears to have low-g from its EW$_{\rm KI}$ and only 5 of 10 have ROSAT detections, all lying well below the target sample in Figure~\ref{ij_fxfj}.  The RV standards exhibit no other evidence of youth, less LHS 2065 and GJ 406, which are both known to be flare stars (\citealt{schm02, schm08}). We attribute their apparent low surface gravity to our choice of the most stable RV standards from \cite{nide02} and \cite{marc89}, likely due to their stronger absorption lines from higher metallicity. 
Published [Fe/H] values based on the \cite{bonf05} calibration range from --0.16 to +0.15 (see Table~\ref{standards}), and do not agree with this conclusion. However, recently \cite{john09} pointed out errors in the Bonfils calibration that underestimates M dwarf metallicities by 0.32 dex on average.  Using the revised photometric calibration of Johnson \& Apps, it is clear from Figure~\ref{rvstd_colors} that the bulk of our RV standards are indeed metal rich.

It is clear now that for a sample of field stars with no metallicity measurements, a single gravity index may not be sufficient.  The atomic alkali lines of K I ($\lambda$7665 and 7699\AA) and Na I ($\lambda$ 8183 and 8195\AA) may also be used as gravity indicators (e.g.~\citealt{sles06}). Yet, care is required with these lines as well as they are affected by stellar activity such that higher levels of chromospheric emission fill in the absorption cores and reduce the measured EWs. We settled on the K I line as this filling-in is stronger in the Na I (\citealt{reid99}). 

Combining the effects of the chromosphere on K I  with the uncertainties in metallicity on the TiO and CaH indices,
we consider a target as having low-g only if both the CaH {\it and} K I measurements indicate that it is so.  We flag a target as such in Table~\ref{ages} if it falls on or below (within error bars) the best-fit curves to the observed $\beta$ Pic members in Figures~\ref{SpT_CaH} and \ref{SpT_KI}. The equations for these curves are listed in the figure captions.

Though we do not have spectra of known young stars at SpTs later than M6.5 with which to calibrate the CaH indices, extrapolating the $\beta$ Pic curve might imply that all the late Ms in our sample lie in the low-g regime of Figure~\ref{SpT_CaH}. This is reassuring since PMS evolutionary models (e.g.~\citealt{dant94}) predict a M6.5 (0.15 \Msun) star will reach the main-sequence in 500 Myr. Therefore, we would expect that for a sample of X-ray-selected targets with ages  $\lesssim$300 Myr, all late Ms will have low surface gravities.\footnote{Alternative models by \cite{burr93} predict that the radius, and thus the gravity, of a 0.15\Msun star will reach its final size in 250 Myrs.}

\subsection{Lithium detection}

PMS stars across all M subclasses are not hot enough to destroy their primordial Li abundance (T $>$2.5$\times$10$^6$ K) and are easily identified by their strong lithium absorption at 6708 \AA\/ (EW$_{\rm Li}>$0.6\AA).\footnote{\cite{kirk08} recently pointed out that both theory and observations (albeit weakly) suggest a weakened Li absorption at very young ages of very-late-M and L dwarfs due to their extremely low gravity. As the pre-MS star contracts, it reaches a maximum observed Li strength at about 100 Myr before it begins to weaken again due to Li burning.}  Very young late-K and early-M dwarfs deplete their lithium by a factor of 2 in less than 10 Myr, providing a robust way to discriminate between 8-Myr old stars, such as TWA or $\eta$ Cha members, and 12-Myr old $\beta$ Pic members. This depletion is substantially slowed in lower mass stars, such that by M6 a lithium detection sets a 90-Myr upper limit on the star's age (\citealt{chab96,stau98}).

We have measured lithium EWs\footnote{The lithium abundances have not been corrected for possible contamination with the Fe I line at 6707.44 \AA. Uncertainties in the setting of continuum levels prior to measurement induce EW errors of about 10-20 m\AA\/ with a dependence on the S/N in the region. We therefore consider our 2$\sigma$ detection limit to be 0.05 \AA.} in 8 of our targets, listed in Table~\ref{lithium_stars} and plotted in Figure~\ref{lithium_EW}. The lithium in the two earliest of these, GJ 9809 (M0.3) and GJ 3305 (M1.1) sets a limit on their ages of 20--30 Myr, consistent with their known membership of the AB Dor \citep{lope06} and $\beta$ Pic \citep{torr06} YMGs, respectively.  
The spectra of the four strongest lithium absorbers 2MASS~J1553 (M3.5), 2MASS~J2234 (M6), 2MASS~J0557 (M7), and 2MASS~J0335 (M8.5), are shown in Figures~\ref{lithium_spectra}.\footnote{2MASS~J1553 appeared as a visual binary with the acquisition camera on Keck I and the observations presented in this paper are of the southern component. Allers et al.~(in press) have observed 2MASS J2234 as a 0.16'' binary with Keck's laser guide star adaptive optics system, but as we were unable to resolve the system, our spectra are a composite of the two.} 
According to stellar evolution models by \cite{chab96}, a M3.5 star such as 2MASS~J1553 with lithium absorption must be less than 40 Myr old. For the ages of other stars, the models can only place upper limits between 90 and 150 Myr. However, 3 of these 4 stars (2MASS~J1553, 2MASS~J2234, and 2MASS~J0335) have EW$_{\rm Li}$ $>$ 0.6 \AA\/, implying that they are PMS stars and must be younger than 10 Myr \citep{zuck04}, consistent with their accretion-level \ha\/ emission discussed in the following section.

\subsection{H$\alpha$ from accretion}

Accretion by PMS stars produces a strong emission-line spectrum, which in the visible includes the Balmer series, caused by accreting gas falling along magnetic field lines from the circumstellar disk onto the star. The near free-fall velocities result in accretion shocks producing both strong and extremely velocity-broadened ($\sim$ hundreds of km~s$^{-1}$) \ha\/ emission profiles.  The equivalent width and/or velocity width of \ha\/ is often used as an indicator of accretion and can set a very young upper limit of 10 Myr on the age of a given star (e.g.~\citealt{barr03}). 

We measured the H$\alpha$ EWs of all targets and plot them in Figure~\ref{SpT_Ha} as a function of SpT.  Eleven of the targets have no significant emission (i.e.~$>-1$ \AA) and none has H$\alpha$ in absorption. The bright X-ray emission detected by ROSAT from these chromospherically {\it in}active stars was likely then the result of a stellar flare, rather than enhanced activity due to youth. The remaining 93\% of the ROSAT-selected objects show \ha\/ in emission with a clear trend towards higher emission with later spectral type.  This is in general attributed to the ``contrast effect'' as the photospheric luminosity decreases in the $R$-band with lower effective temperatures (\citealt{basr95}). However, comparisons within a narrow range of spectral types can be made to identify more chromospherically active and potentially accreting young objects.

Eight stars in Figure~\ref{SpT_Ha} have \ha\/ emission near the empirical accretion/non-accretion boundary (\citealt{barr03}) depicted by
the dashed curve in the figure. Though the accretion curve is not thought to be very robust for objects of SpT
later than M5.5, due to the few late-M cluster members used to calibrate the sequence, it does serve as an outer envelope of the chromospheric emission. Unfortunately, this ``saturation'' criterion is not robust enough to be used exclusively. This is demonstrated by the flaring stars 1RXS J0414, GJ 316.1 and LHS 2065 (\citealt{schm94,schm02}). Though they are likely much younger than 300 Myr, none is accreting, as indicated by the lack of lithium in their spectra.

The empirical ``10\%-width'' metric, which measures the full width at the 10\% level of the \ha\/ velocity profile peak, is a more promising diagnostic of accretion. \cite{whit03} use a combination of EW$_{\rm H\alpha}$ $<-10$~\AA\/ and a velocity width of $>$200 km/s to identify accreting T Tauri stars. 
\cite{moha05} use He I $\lambda$6678 emission as yet another accretion indicator, though several of the objects they determined to be accreting do not exhibit He I emission, while others with He I emission show no signs of accretion.  This appears to be the case in our sample as well. Seventeen of our targets have detectable He I emission though the bulk of them are certainly not accretors.

2MASS~J2234, 2MASS~J1553, 2MASS~J0557, and 2MASS~J0335 all have lithium absorption, indications of low-g from both the
CaH {\it and} K~I diagnostics, and broad \ha\/ profiles, plotted in Figure~\ref{Halpha_vel} with average 10\%-widths of 306,
446, 207, and 273 km~s$^{-1}$, respectively. Only two however exhibit He I emission. (See Table~\ref{lithium_stars}.) Though 2MASS~J0335 does not have a strong enough EW$_{\rm H\alpha}$ to exceed the \cite{barr03} accretion limit, its EW$_{\rm H\alpha}$ of --10.7 \AA\/ does qualify it as an accretor using the \cite{whit03} criteria. We conclude that 2MASS~J2234, 2MASS~J1553 and 2MASS~J0335 are accreting T Tauri stars while 2MASS~J0557 is on the borderline, with upper limits on the ages of 5, 3, 10, and 10 Myr, respectively, based on the criteria devised by \cite{whit03}, \cite{moha05}, and \cite{barr03}. 
Using gravity-sensitive spectral features in the near-IR, Allers et al.~(in press) further refined the age of 2MASS~J2234 to a very young $\sim$1 Myr.

\section{Summary}\label{summary}

We present our ground-based spectroscopic survey of 185 X-ray-bright low-mass stars, most of which are within 25 pc. 
Our high-resolution optical spectra allowed us to identify 30 SBs which are strong X-ray emitters, but not necessarily young. For the remainder of the sample, we measured youth indicators such as wide- and narrow-band gravity indices of the CaH molecular band as well as K I equivalent widths. We find that for a sample of field stars with no metallicity measurements, a single gravity index may not be sufficient, as high metallicities mimic low-g on a CaH-SpT plot. 
This is apparent in a metal-rich sub-sample of our RV standards, which appear to have low surface gravity as measured by the CaH index, yet show no other evidence of youth. 
Combining the effects of chromospheric activity on EW$_{\rm KI}$  with the uncertainties in metallicity on the TiO and CaH indices, we require that both the CaH {\it and} K I measurements indicate low-g before classifying a star as such.  We have also detected lithium absorption, a sure sign of youth, in 8 targets, 4 of which show evidence of accretion from their velocity-broadened \ha\/ profiles.

We estimate that our final sample of the 144 youngest and nearest low-mass objects in the field is less than $\sim$300 Myr old, with 30\% of them being younger than 150 Myr old and 4 very young
($\lesssim$10 Myr old). This sample provides a rich set of well-characterized targets with the primary requirements of youth and proximity for intensive disk evolution and planet formation studies with ground-based (e.g.~adaptive optics) and space-based observations. Two such programs are currently underway: (1) We are using the Multiband Imaging Photometer for the Spitzer Space Telescope (MIPS) to extend debris
disk studies to low-mass stars with sufficient sensitivity for
meaningful comparison with higher mass stars. Also, the newly discovered disks will be among the closest
to Earth, making them prime targets for multiwavelength and high
angular resolution follow−up.  (2) The Gemini Planet-Finding Campaign (\citealt{liu09}), an extensive (500 hours) search for massive extrasolar planets using the high-contrast AO Near-Infrared Coronagraphic Imager (NICI), has observed many of our targets and will continue to do so over the next two years. This program will provide key constraints for the separation distributions of extrasolar planets at 5 -- 10 AU, the dependence on planet frequency on stellar host mass, and the spectral properties of extrasolar giant planets.

\acknowledgements

E.S thanks the CFHT and Keck staff for their care in setting up the instruments and support in the control rooms, and to
Jean-Francois Donati for making Libre-ESpRIT available to CFHT. E.S. also thanks John Johnson for useful discussions and
stepping in to observe one night, and the anonymous referee for a detailed and helpful review of the manuscript.
Research funding from the NASA Postdoctoral Program (formerly the NRC Research Associateship) and the Carnegie
Institution of Washington for E.S.~is gratefully acknowledged. This material is also based upon work supported by the
National Aeronautics and Space Administration through the NASA Astrobiology Institute and the NASA/GALEX grant program
under Cooperative Agreement Nos. NNA04CC08A and NNX07AJ43G issued through the Office of Space Science. M.C.L.
acknowledges support from the Alfred P. Sloan Research Fellowship. This publication makes use of data products from the
Two Micron All Sky Survey, which is a joint project of the University of Massachusetts and the Infrared Processing and
Analysis Center/California Institute of Technology, funded by the National Aeronautics and Space Administration and the
National Science Foundation.

\clearpage
\bibliography{refs}{}

\begin{thebibliography}{99}
\expandafter\ifx\csname natexlab\endcsname\relax\def\natexlab#1{#1}\fi

\bibitem[{{Allen} \& {Reid}(2008)}]{alle08}
{Allen}, P.~R., \& {Reid}, I.~N. 2008, \aj, 135, 2024

\bibitem[{{Allers} {et~al.}(2009){Allers}, {Liu}, {Shkolnik}, {Cushing},
  {Dupuy}, {Mathews}, {Reid}, {Cruz}, \& {Vacca}}]{alle09}
{Allers}, K.~N., {et~al.} 2009, ArXiv e-prints, 0902.4742

\bibitem[{{Baraffe} {et~al.}(1998){Baraffe}, {Chabrier}, {Allard}, \&
  {Hauschildt}}]{bara98}
{Baraffe}, I., {Chabrier}, G., {Allard}, F., \& {Hauschildt}, P.~H. 1998, \aap,
  337, 403

\bibitem[{{Baraffe} {et~al.}(2003){Baraffe}, {Chabrier}, {Barman}, {Allard}, \&
  {Hauschildt}}]{bara03}
{Baraffe}, I., {Chabrier}, G., {Barman}, T.~S., {Allard}, F., \& {Hauschildt},
  P.~H. 2003, \aap, 402, 701

\bibitem[{{Barrado y Navascu{\'e}s} \& {Mart{\'{\i}}n}(2003)}]{barr03}
{Barrado y Navascu{\'e}s}, D., \& {Mart{\'{\i}}n}, E.~L. 2003, \aj, 126, 2997

\bibitem[{{Basri} \& {Marcy}(1995)}]{basr95}
{Basri}, G., \& {Marcy}, G.~W. 1995, \aj, 109, 762

\bibitem[{{Bean} {et~al.}(2006){Bean}, {Sneden}, {Hauschildt}, {Johns-Krull},
  \& {Benedict}}]{bean06}
{Bean}, J.~L., {Sneden}, C., {Hauschildt}, P.~H., {Johns-Krull}, C.~M., \&
  {Benedict}, G.~F. 2006, \apj, 652, 1604

\bibitem[{{Beuzit} {et~al.}(2004){Beuzit}, {S{\'e}gransan}, {Forveille},
  {Udry}, {Delfosse}, {Mayor}, {Perrier}, {Hainaut}, {Roddier}, {Roddier}, \&
  {Mart{\'{\i}}n}}]{beuz04}
{Beuzit}, J.-L., {et~al.} 2004, \aap, 425, 997

\bibitem[{{Bochanski} {et~al.}(2008){Bochanski}, {Hawley}, {Reid}, {Covey},
  {West}, {Golimowski}, \& {Ivezic}}]{boch08}
{Bochanski}, J.~J., {Hawley}, S.~L., {Reid}, I.~N., {Covey}, K.~R., {West},
  A.~A., {Golimowski}, D.~A., \& {Ivezic}, Z. 2008, ArXiv e-prints, 0810.2343

\bibitem[{{Bonfils} {et~al.}(2005){Bonfils}, {Delfosse}, {Udry}, {Santos},
  {Forveille}, \& {S{\'e}gransan}}]{bonf05}
{Bonfils}, X., {Delfosse}, X., {Udry}, S., {Santos}, N.~C., {Forveille}, T., \&
  {S{\'e}gransan}, D. 2005, \aap, 442, 635

\bibitem[{{Burrows} {et~al.}(1993){Burrows}, {Hubbard}, {Saumon}, \&
  {Lunine}}]{burr93}
{Burrows}, A., {Hubbard}, W.~B., {Saumon}, D., \& {Lunine}, J.~I. 1993, \apj,
  406, 158

\bibitem[{{Casagrande} {et~al.}(2008){Casagrande}, {Flynn}, \&
  {Bessell}}]{casa08}
{Casagrande}, L., {Flynn}, C., \& {Bessell}, M. 2008, \mnras, 389, 585

\bibitem[{{Chabrier} {et~al.}(1996){Chabrier}, {Baraffe}, \& {Plez}}]{chab96}
{Chabrier}, G., {Baraffe}, I., \& {Plez}, B. 1996, \apjl, 459, L91+

\bibitem[{{Covey} {et~al.}(2008){Covey}, {Ag{\"u}eros}, {Green}, {Haggard},
  {Barkhouse}, {Drake}, {Evans}, {Kashyap}, {Kim}, {Mossman}, {Pease}, \&
  {Silverman}}]{cove08}
{Covey}, K.~R., {et~al.} 2008, \apjs, 178, 339

\bibitem[{{Cruz} {et~al.}(2007){Cruz}, {Reid}, {Kirkpatrick}, {Burgasser},
  {Liebert}, {Solomon}, {Schmidt}, {Allen}, {Hawley}, \& {Covey}}]{cruz07}
{Cruz}, K.~L., {et~al.} 2007, \aj, 133, 439

\bibitem[{{Cruz} {et~al.}(2003){Cruz}, {Reid}, {Liebert}, {Kirkpatrick}, \&
  {Lowrance}}]{cruz03}
{Cruz}, K.~L., {Reid}, I.~N., {Liebert}, J., {Kirkpatrick}, J.~D., \&
  {Lowrance}, P.~J. 2003, \aj, 126, 2421

\bibitem[{{Daemgen} {et~al.}(2007){Daemgen}, {Siegler}, {Reid}, \&
  {Close}}]{daem07}
{Daemgen}, S., {Siegler}, N., {Reid}, I.~N., \& {Close}, L.~M. 2007, \apj, 654,
  558

\bibitem[{{D'Antona} \& {Mazzitelli}(1994)}]{dant94}
{D'Antona}, F., \& {Mazzitelli}, I. 1994, \apjs, 90, 467

\bibitem[{{Delfosse} {et~al.}(1998){Delfosse}, {Forveille}, {Perrier}, \&
  {Mayor}}]{delf98}
{Delfosse}, X., {Forveille}, T., {Perrier}, C., \& {Mayor}, M. 1998, \aap, 331,
  581

\bibitem[{{Donati} {et~al.}(2006){Donati}, {Catala}, {Landstreet}, \&
  {Petit}}]{dona06}
{Donati}, J.-F., {Catala}, C., {Landstreet}, J.~D., \& {Petit}, P. 2006, in
  Astronomical Society of the Pacific Conference Series, Vol. 358, Astronomical
  Society of the Pacific Conference Series, ed. R.~{Casini} \& B.~W. {Lites},
  362--+

\bibitem[{{Donati} {et~al.}(2007){Donati}, {Jardine}, {Gregory}, {Petit},
  {Bouvier}, {Dougados}, {M{\'e}nard}, {Cameron}, {Harries}, {Jeffers}, \&
  {Paletou}}]{dona07}
{Donati}, J.-F., {et~al.} 2007, \mnras, 380, 1297

\bibitem[{{Donati} {et~al.}(1997){Donati}, {Semel}, {Carter}, {Rees}, \&
  {Collier Cameron}}]{dona97}
{Donati}, J.-F., {Semel}, M., {Carter}, B.~D., {Rees}, D.~E., \& {Collier
  Cameron}, A. 1997, \mnras, 291, 658

\bibitem[{{Fitzgerald} {et~al.}(2007){Fitzgerald}, {Kalas}, {Duch{\^e}ne},
  {Pinte}, \& {Graham}}]{fitz07}
{Fitzgerald}, M.~P., {Kalas}, P.~G., {Duch{\^e}ne}, G., {Pinte}, C., \&
  {Graham}, J.~R. 2007, \apj, 670, 536

\bibitem[{{Fitzpatrick}(1993)}]{fitz93}
{Fitzpatrick}, M.~J. 1993, in Astronomical Society of the Pacific Conference
  Series, Vol.~52, Astronomical Data Analysis Software and Systems II, ed.
  R.~J. {Hanisch}, R.~J.~V. {Brissenden}, \& J.~{Barnes}, 472--+

\bibitem[{{France} {et~al.}(2007){France}, {Roberge}, {Lupu}, {Redfield}, \&
  {Feldman}}]{fran07}
{France}, K., {Roberge}, A., {Lupu}, R.~E., {Redfield}, S., \& {Feldman}, P.~D.
  2007, \apj, 668, 1174

\bibitem[{{Gershberg} {et~al.}(1999){Gershberg}, {Katsova}, {Lovkaya},
  {Terebizh}, \& {Shakhovskaya}}]{gers99}
{Gershberg}, R.~E., {Katsova}, M.~M., {Lovkaya}, M.~N., {Terebizh}, A.~V., \&
  {Shakhovskaya}, N.~I. 1999, \aaps, 139, 555

\bibitem[{{Hawley} {et~al.}(2002){Hawley}, {Covey}, {Knapp}, {Golimowski},
  {Fan}, {Anderson}, {Gunn}, {Harris}, {Ivezi{\'c}}, {Long}, {Lupton},
  {McGehee}, {Narayanan}, {Peng}, {Schlegel}, {Schneider}, {Spahn}, {Strauss},
  {Szkody}, {Tsvetanov}, {Walkowicz}, {Brinkmann}, {Harvanek}, {Hennessy},
  {Kleinman}, {Krzesinski}, {Long}, {Neilsen}, {Newman}, {Nitta}, {Snedden}, \&
  {York}}]{hawl02}
{Hawley}, S.~L., {et~al.} 2002, \aj, 123, 3409

\bibitem[{{Hebb} {et~al.}(2007){Hebb}, {Petro}, {Ford}, {Ardila}, {Toledo},
  {Minniti}, {Golimowski}, \& {Clampin}}]{hebb07}
{Hebb}, L., {Petro}, L., {Ford}, H.~C., {Ardila}, D.~R., {Toledo}, I.,
  {Minniti}, D., {Golimowski}, D.~A., \& {Clampin}, M. 2007, \mnras, 379, 63

\bibitem[{{Hillenbrand} {et~al.}(2008){Hillenbrand}, {Carpenter}, {Kim},
  {Meyer}, {Backman}, {Moro-Mart{\'{\i}}n}, {Hollenbach}, {Hines}, {Pascucci},
  \& {Bouwman}}]{hill08}
{Hillenbrand}, L.~A., {et~al.} 2008, \apj, 677, 630

\bibitem[{{H{\"u}nsch} {et~al.}(1999){H{\"u}nsch}, {Schmitt}, {Sterzik}, \&
  {Voges}}]{huns99}
{H{\"u}nsch}, M., {Schmitt}, J.~H.~M.~M., {Sterzik}, M.~F., \& {Voges}, W.
  1999, \aaps, 135, 319

\bibitem[{{Ida} \& {Lin}(2008)}]{ida08}
{Ida}, S., \& {Lin}, D.~N.~C. 2008, \apj, 673, 487

\bibitem[{{Jeffries}(1995)}]{jeff95}
{Jeffries}, R.~D. 1995, \mnras, 273, 559

\bibitem[{{Johnson} \& {Apps}(2009)}]{john09}
{Johnson}, J.~A., \& {Apps}, K. 2009, ArXiv e-prints, 0904.3092

\bibitem[{{Jones} \& {Tsuji}(1998)}]{jone98}
{Jones}, H.~R.~A., \& {Tsuji}, T. 1998, in Astronomical Society of the Pacific
  Conference Series, Vol. 134, Brown Dwarfs and Extrasolar Planets, ed.
  R.~{Rebolo}, E.~L. {Martin}, \& M.~R. {Zapatero Osorio}, 423--+

\bibitem[{{Kalas} {et~al.}(2004){Kalas}, {Liu}, \& {Matthews}}]{kala04}
{Kalas}, P., {Liu}, M.~C., \& {Matthews}, B.~C. 2004, Science, 303, 1990

\bibitem[{{Kirkpatrick} {et~al.}(2008){Kirkpatrick}, {Cruz}, {Barman},
  {Burgasser}, {Looper}, {Tinney}, {Gelino}, {Lowrance}, {Liebert},
  {Carpenter}, {Hillenbrand}, \& {Stauffer}}]{kirk08}
{Kirkpatrick}, J.~D., {et~al.} 2008, \apj, 689, 1295

\bibitem[{{Kirkpatrick} {et~al.}(1991){Kirkpatrick}, {Henry}, \&
  {McCarthy}}]{kirk91}
{Kirkpatrick}, J.~D., {Henry}, T.~J., \& {McCarthy}, Jr., D.~W. 1991, \apjs,
  77, 417

\bibitem[{{Krist} {et~al.}(2005){Krist}, {Ardila}, {Golimowski}, {Clampin},
  {Ford}, {Illingworth}, {Hartig}, {Bartko}, {Ben{\'{\i}}tez}, {Blakeslee},
  {Bouwens}, {Bradley}, {Broadhurst}, {Brown}, {Burrows}, {Cheng}, {Cross},
  {Demarco}, {Feldman}, {Franx}, {Goto}, {Gronwall}, {Holden}, {Homeier},
  {Infante}, {Kimble}, {Lesser}, {Martel}, {Mei}, {Menanteau}, {Meurer},
  {Miley}, {Motta}, {Postman}, {Rosati}, {Sirianni}, {Sparks}, {Tran},
  {Tsvetanov}, {White}, \& {Zheng}}]{kris05b}
{Krist}, J.~E., {et~al.} 2005, \aj, 129, 1008

\bibitem[{{L{\'e}pine} \& {Shara}(2005)}]{lepi05}
{L{\'e}pine}, S., \& {Shara}, M.~M. 2005, \aj, 129, 1483

\bibitem[{{L{\'e}pine} {et~al.}(2002){L{\'e}pine}, {Shara}, \& {Rich}}]{lepi02}
{L{\'e}pine}, S., {Shara}, M.~M., \& {Rich}, R.~M. 2002, \aj, 124, 1190

\bibitem[{{Li} \& {Hu}(1998)}]{li98}
{Li}, J.~Z., \& {Hu}, J.~Y. 1998, \aaps, 132, 173

\bibitem[{{Liu} {et~al.}(2004){Liu}, {Matthews}, {Williams}, \&
  {Kalas}}]{liu04b}
{Liu}, M.~C., {Matthews}, B.~C., {Williams}, J.~P., \& {Kalas}, P.~G. 2004,
  \apj, 608, 526

\bibitem[{{Liu} {et~al.}(2009){Liu}, {Wahhaj}, {Biller}, {Shkolnik}, {Chun},
  {Ftaclas}, {Toomey}, {Close}, {Nielsen}, {Hayward}, {Hartung}, \&
  {Artigau}}]{liu09}
{Liu}, M.~C., {et~al.} 2009, in American Institute of Physics Conference
  Series, Vol. 1094, American Institute of Physics Conference Series, ed.
  E.~{Stempels}, 461--464

\bibitem[{{L{\'o}pez-Santiago} {et~al.}(2006){L{\'o}pez-Santiago}, {Montes},
  {Crespo-Chac{\'o}n}, \& {Fern{\'a}ndez-Figueroa}}]{lope06}
{L{\'o}pez-Santiago}, J., {Montes}, D., {Crespo-Chac{\'o}n}, I., \&
  {Fern{\'a}ndez-Figueroa}, M.~J. 2006, \apj, 643, 1160

\bibitem[{{Lyo} {et~al.}(2004){Lyo}, {Lawson}, \& {Bessell}}]{lyo04}
{Lyo}, A.-R., {Lawson}, W.~A., \& {Bessell}, M.~S. 2004, \mnras, 355, 363

\bibitem[{{Mandell} {et~al.}(2007){Mandell}, {Raymond}, \&
  {Sigurdsson}}]{mand07}
{Mandell}, A.~M., {Raymond}, S.~N., \& {Sigurdsson}, S. 2007, \apj, 660, 823

\bibitem[{{Marcy} \& {Benitz}(1989)}]{marc89}
{Marcy}, G.~W., \& {Benitz}, K.~J. 1989, \apj, 344, 441

\bibitem[{{McCarthy} {et~al.}(2001){McCarthy}, {Zuckerman}, \&
  {Becklin}}]{mcca01}
{McCarthy}, C., {Zuckerman}, B., \& {Becklin}, E.~E. 2001, \aj, 121, 3259

\bibitem[{{Micela} {et~al.}(1998){Micela}, {Sciortino}, {Harnden}, \&
  {Rosner}}]{mice98}
{Micela}, G., {Sciortino}, S., {Harnden}, Jr., F.~R., \& {Rosner}, R. 1998,
  \apss, 261, 105

\bibitem[{{Mohanty} {et~al.}(2005){Mohanty}, {Jayawardhana}, \&
  {Basri}}]{moha05}
{Mohanty}, S., {Jayawardhana}, R., \& {Basri}, G. 2005, \apj, 626, 498

\bibitem[{{Montes} {et~al.}(2001){Montes}, {L{\'o}pez-Santiago},
  {Fern{\'a}ndez-Figueroa}, \& {G{\'a}lvez}}]{mont01}
{Montes}, D., {L{\'o}pez-Santiago}, J., {Fern{\'a}ndez-Figueroa}, M.~J., \&
  {G{\'a}lvez}, M.~C. 2001, \aap, 379, 976

\bibitem[{{Nidever} {et~al.}(2002){Nidever}, {Marcy}, {Butler}, {Fischer}, \&
  {Vogt}}]{nide02}
{Nidever}, D.~L., {Marcy}, G.~W., {Butler}, R.~P., {Fischer}, D.~A., \& {Vogt},
  S.~S. 2002, \apjs, 141, 503

\bibitem[{{Norton} {et~al.}(2007){Norton}, {Wheatley}, {West}, {Haswell},
  {Street}, {Collier Cameron}, {Christian}, {Clarkson}, {Enoch}, {Gallaway},
  {Hellier}, {Horne}, {Irwin}, {Kane}, {Lister}, {Nicholas}, {Parley},
  {Pollacco}, {Ryans}, {Skillen}, \& {Wilson}}]{nort07}
{Norton}, A.~J., {et~al.} 2007, \aap, 467, 785

\bibitem[{{Perryman} \& {ESA}(1997)}]{perr97}
{Perryman}, M.~A.~C., \& {ESA}, eds. 1997, ESA Special Publication, Vol. 1200,
  {The HIPPARCOS and TYCHO catalogues. Astrometric and photometric star
  catalogues derived from the ESA HIPPARCOS Space Astrometry Mission}

\bibitem[{{Pettersen}(1980)}]{pett80}
{Pettersen}, B.~R. 1980, \pasp, 92, 188

\bibitem[{{Preibisch} \& {Feigelson}(2005)}]{prei05}
{Preibisch}, T., \& {Feigelson}, E.~D. 2005, \apjs, 160, 390

\bibitem[{{Reid} \& {Cruz}(2002)}]{reid02c}
{Reid}, I.~N., \& {Cruz}, K.~L. 2002, \aj, 123, 2806

\bibitem[{{Reid} {et~al.}(2003){Reid}, {Cruz}, {Allen}, {Mungall}, {Kilkenny},
  {Liebert}, {Hawley}, {Fraser}, {Covey}, \& {Lowrance}}]{reid03}
{Reid}, I.~N., {et~al.} 2003, \aj, 126, 3007

\bibitem[{{Reid} {et~al.}(2004){Reid}, {Cruz}, {Allen}, {Mungall}, {Kilkenny},
  {Liebert}, {Hawley}, {Fraser}, {Covey}, {Lowrance}, {Kirkpatrick}, \&
  {Burgasser}}]{reid04}
---. 2004, \aj, 128, 463

\bibitem[{{Reid} {et~al.}(2007{\natexlab{a}}){Reid}, {Cruz}, \&
  {Allen}}]{reid07a}
{Reid}, I.~N., {Cruz}, K.~L., \& {Allen}, P.~R. 2007{\natexlab{a}}, \aj, 133,
  2825

\bibitem[{{Reid} {et~al.}(2002{\natexlab{a}}){Reid}, {Gizis}, \&
  {Hawley}}]{reid02a}
{Reid}, I.~N., {Gizis}, J.~E., \& {Hawley}, S.~L. 2002{\natexlab{a}}, \aj, 124,
  2721

\bibitem[{{Reid} \& {Hawley}(1999)}]{reid99}
{Reid}, I.~N., \& {Hawley}, S.~L. 1999, \aj, 117, 343

\bibitem[{{Reid} {et~al.}(1995){Reid}, {Hawley}, \& {Gizis}}]{reid95}
{Reid}, I.~N., {Hawley}, S.~L., \& {Gizis}, J.~E. 1995, \aj, 110, 1838

\bibitem[{{Reid} {et~al.}(2002{\natexlab{b}}){Reid}, {Kirkpatrick}, {Liebert},
  {Gizis}, {Dahn}, \& {Monet}}]{reid02b}
{Reid}, I.~N., {Kirkpatrick}, J.~D., {Liebert}, J., {Gizis}, J.~E., {Dahn},
  C.~C., \& {Monet}, D.~G. 2002{\natexlab{b}}, \aj, 124, 519

\bibitem[{{Reid} {et~al.}(2007{\natexlab{b}}){Reid}, {Turner}, {Turnbull},
  {Mountain}, \& {Valenti}}]{reid07b}
{Reid}, I.~N., {Turner}, E.~L., {Turnbull}, M.~C., {Mountain}, M., \&
  {Valenti}, J.~A. 2007{\natexlab{b}}, \apj, 665, 767

\bibitem[{{Reid} \& {Walkowicz}(2006)}]{reid06}
{Reid}, I.~N., \& {Walkowicz}, L.~M. 2006, \pasp, 118, 671

\bibitem[{{Riaz} {et~al.}(2006){Riaz}, {Gizis}, \& {Harvin}}]{riaz06}
{Riaz}, B., {Gizis}, J.~E., \& {Harvin}, J. 2006, \aj, 132, 866

\bibitem[{{Rieke} {et~al.}(2005){Rieke}, {Su}, {Stansberry}, {Trilling},
  {Bryden}, {Muzerolle}, {White}, {Gorlova}, {Young}, {Beichman},
  {Stapelfeldt}, \& {Hines}}]{riek05}
{Rieke}, G.~H., {et~al.} 2005, \apj, 620, 1010

\bibitem[{{Roberge} {et~al.}(2006){Roberge}, {Feldman}, {Weinberger},
  {Deleuil}, \& {Bouret}}]{robe06}
{Roberge}, A., {Feldman}, P.~D., {Weinberger}, A.~J., {Deleuil}, M., \&
  {Bouret}, J.-C. 2006, \nat, 441, 724

\bibitem[{{Schmitt}(1994)}]{schm94}
{Schmitt}, J.~H.~M.~M. 1994, \apjs, 90, 735

\bibitem[{{Schmitt} {et~al.}(1995){Schmitt}, {Fleming}, \& {Giampapa}}]{schm95}
{Schmitt}, J.~H.~M.~M., {Fleming}, T.~A., \& {Giampapa}, M.~S. 1995, \apj, 450,
  392

\bibitem[{{Schmitt} \& {Liefke}(2002)}]{schm02}
{Schmitt}, J.~H.~M.~M., \& {Liefke}, C. 2002, \aap, 382, L9

\bibitem[{{Schmitt} {et~al.}(2008){Schmitt}, {Reale}, {Liefke}, {Wolter},
  {Fuhrmeister}, {Reiners}, \& {Peres}}]{schm08}
{Schmitt}, J.~H.~M.~M., {Reale}, F., {Liefke}, C., {Wolter}, U., {Fuhrmeister},
  B., {Reiners}, A., \& {Peres}, G. 2008, \aap, 481, 799

\bibitem[{{Shkolnik} {et~al.}(2008){Shkolnik}, {Liu}, {Reid}, {Hebb},
  {Cameron}, {Torres}, \& {Wilson}}]{shko08}
{Shkolnik}, E., {Liu}, M.~C., {Reid}, I.~N., {Hebb}, L., {Cameron}, A.~C.,
  {Torres}, C.~A., \& {Wilson}, D.~M. 2008, \apj, 682, 1248

\bibitem[{{Skrutskie} {et~al.}(2006){Skrutskie}, {Cutri}, {Stiening},
  {Weinberg}, {Schneider}, {Carpenter}, {Beichman}, {Capps}, {Chester},
  {Elias}, {Huchra}, {Liebert}, {Lonsdale}, {Monet}, {Price}, {Seitzer},
  {Jarrett}, {Kirkpatrick}, {Gizis}, {Howard}, {Evans}, {Fowler}, {Fullmer},
  {Hurt}, {Light}, {Kopan}, {Marsh}, {McCallon}, {Tam}, {Van Dyk}, \&
  {Wheelock}}]{skru06}
{Skrutskie}, M.~F., {et~al.} 2006, \aj, 131, 1163

\bibitem[{{Skumanich}(1972)}]{skum72}
{Skumanich}, A. 1972, \apj, 171, 565

\bibitem[{{Slesnick} {et~al.}(2006){Slesnick}, {Carpenter}, \&
  {Hillenbrand}}]{sles06}
{Slesnick}, C.~L., {Carpenter}, J.~M., \& {Hillenbrand}, L.~A. 2006, \aj, 131,
  3016

\bibitem[{{Song} {et~al.}(2002){Song}, {Weinberger}, {Becklin}, {Zuckerman}, \&
  {Chen}}]{song02}
{Song}, I., {Weinberger}, A.~J., {Becklin}, E.~E., {Zuckerman}, B., \& {Chen},
  C. 2002, \aj, 124, 514

\bibitem[{{Stauffer} {et~al.}(1991){Stauffer}, {Klemola}, {Prosser}, \&
  {Probst}}]{stau91}
{Stauffer}, J., {Klemola}, A., {Prosser}, C., \& {Probst}, R. 1991, \aj, 101,
  980

\bibitem[{{Stauffer} {et~al.}(1997){Stauffer}, {Balachandran}, {Krishnamurthi},
  {Pinsonneault}, {Terndrup}, \& {Stern}}]{stau97}
{Stauffer}, J.~R., {Balachandran}, S.~C., {Krishnamurthi}, A., {Pinsonneault},
  M., {Terndrup}, D.~M., \& {Stern}, R.~A. 1997, \apj, 475, 604

\bibitem[{{Stauffer} {et~al.}(1998){Stauffer}, {Schultz}, \&
  {Kirkpatrick}}]{stau98}
{Stauffer}, J.~R., {Schultz}, G., \& {Kirkpatrick}, J.~D. 1998, \apjl, 499,
  L199+

\bibitem[{{Stern} {et~al.}(1995){Stern}, {Schmitt}, \& {Kahabka}}]{ster95}
{Stern}, R.~A., {Schmitt}, J.~H.~M.~M., \& {Kahabka}, P.~T. 1995, \apj, 448,
  683

\bibitem[{{Stocke} {et~al.}(1991){Stocke}, {Morris}, {Gioia}, {Maccacaro},
  {Schild}, {Wolter}, {Fleming}, \& {Henry}}]{stoc91}
{Stocke}, J.~T., {Morris}, S.~L., {Gioia}, I.~M., {Maccacaro}, T., {Schild},
  R., {Wolter}, A., {Fleming}, T.~A., \& {Henry}, J.~P. 1991, \apjs, 76, 813

\bibitem[{{Torres} {et~al.}(2006){Torres}, {Quast}, {da Silva}, {de La Reza},
  {Melo}, \& {Sterzik}}]{torr06}
{Torres}, C.~A.~O., {Quast}, G.~R., {da Silva}, L., {de La Reza}, R., {Melo},
  C.~H.~F., \& {Sterzik}, M. 2006, \aap, 460, 695

\bibitem[{{Torres} {et~al.}(2008){Torres}, {Quast}, {Melo}, \&
  {Sterzik}}]{torr08}
{Torres}, C.~A.~O., {Quast}, G.~R., {Melo}, C.~H.~F., \& {Sterzik}, M.~F. 2008,
  {Young Nearby Loose Associations} (Handbook of Star Forming Regions, Volume
  II: The Southern Sky ASP Monograph Publications, Vol.~5.~Edited by Bo
  Reipurth, p.757), 757--+

\bibitem[{{Voges} {et~al.}(1999){Voges}, {Aschenbach}, {Boller},
  {Br{\"a}uninger}, {Briel}, {Burkert}, {Dennerl}, {Englhauser}, {Gruber},
  {Haberl}, {Hartner}, {Hasinger}, {K{\"u}rster}, {Pfeffermann}, {Pietsch},
  {Predehl}, {Rosso}, {Schmitt}, {Tr{\"u}mper}, \& {Zimmermann}}]{voge99}
{Voges}, W., {et~al.} 1999, \aap, 349, 389

\bibitem[{{Voges} {et~al.}(2000){Voges}, {Aschenbach}, {Boller}, {Brauninger},
  {Briel}, {Burkert}, {Dennerl}, {Englhauser}, {Gruber}, {Haberl}, {Hartner},
  {Hasinger}, {Pfeffermann}, {Pietsch}, {Predehl}, {Schmitt}, {Trumper}, \&
  {Zimmermann}}]{voge00}
---. 2000, VizieR Online Data Catalog, 9029, 0

\bibitem[{{Vogt} {et~al.}(1994){Vogt}, {Allen}, {Bigelow}, {Bresee}, {Brown},
  {Cantrall}, {Conrad}, {Couture}, {Delaney}, {Epps}, {Hilyard}, {Hilyard},
  {Horn}, {Jern}, {Kanto}, {Keane}, {Kibrick}, {Lewis}, {Osborne},
  {Pardeilhan}, {Pfister}, {Ricketts}, {Robinson}, {Stover}, {Tucker}, {Ward},
  \& {Wei}}]{vogt94}
{Vogt}, S.~S., {et~al.} 1994, in Society of Photo-Optical Instrumentation
  Engineers (SPIE) Conference Series, Vol. 2198, Society of Photo-Optical
  Instrumentation Engineers (SPIE) Conference Series, ed. D.~L. {Crawford} \&
  E.~R. {Craine}, 362--+

\bibitem[{{Webb} {et~al.}(1999){Webb}, {Zuckerman}, {Platais}, {Patience},
  {White}, {Schwartz}, \& {McCarthy}}]{webb99}
{Webb}, R.~A., {Zuckerman}, B., {Platais}, I., {Patience}, J., {White}, R.~J.,
  {Schwartz}, M.~J., \& {McCarthy}, C. 1999, \apjl, 512, L63

\bibitem[{{Wenger} {et~al.}(2007){Wenger}, {Oberto}, {Bonnarel}, {Brouty},
  {Bruneau}, {Brunet}, {Cambresy}, {Dubois}, {Eisele}, {Fernique}, {Genova},
  {Lalo{\"e}}, {Lesteven}, {Loup}, {Ochsenbein}, {Vannier}, {Vollmer},
  {Vonflie}, {Wagner}, {Woelfel}, {Borde}, {Beyneix}, {Chassagnard},
  {Jasniewicz}, \& {Davoust}}]{weng07}
{Wenger}, M., {et~al.} 2007, in Astronomical Society of the Pacific Conference
  Series, Vol. 377, Library and Information Services in Astronomy V, ed.
  S.~{Ricketts}, C.~{Birdie}, \& E.~{Isaksson}, 197--+

\bibitem[{{West} {et~al.}(2008){West}, {Hawley}, {Bochanski}, {Covey}, {Reid},
  {Dhital}, {Hilton}, \& {Masuda}}]{west08}
{West}, A.~A., {Hawley}, S.~L., {Bochanski}, J.~J., {Covey}, K.~R., {Reid},
  I.~N., {Dhital}, S., {Hilton}, E.~J., \& {Masuda}, M. 2008, \aj, 135, 785

\bibitem[{{White} \& {Basri}(2003)}]{whit03}
{White}, R.~J., \& {Basri}, G. 2003, \apj, 582, 1109

\bibitem[{{Wilking} {et~al.}(2005){Wilking}, {Meyer}, {Robinson}, \&
  {Greene}}]{wilk05}
{Wilking}, B.~A., {Meyer}, M.~R., {Robinson}, J.~G., \& {Greene}, T.~P. 2005,
  \aj, 130, 1733

\bibitem[{{Woolf} \& {Wallerstein}(2005)}]{wool05}
{Woolf}, V.~M., \& {Wallerstein}, G. 2005, \mnras, 356, 963

\bibitem[{{Woolf} \& {Wallerstein}(2006)}]{wool06}
---. 2006, \pasp, 118, 218

\bibitem[{{Zickgraf} {et~al.}(2003){Zickgraf}, {Engels}, {Hagen}, {Reimers}, \&
  {Voges}}]{zick03}
{Zickgraf}, F.-J., {Engels}, D., {Hagen}, H.-J., {Reimers}, D., \& {Voges}, W.
  2003, \aap, 406, 535

\bibitem[{{Zickgraf} {et~al.}(2005){Zickgraf}, {Krautter}, {Reffert},
  {Alcal{\'a}}, {Mujica}, {Covino}, \& {Sterzik}}]{zick05}
{Zickgraf}, F.-J., {Krautter}, J., {Reffert}, S., {Alcal{\'a}}, J.~M.,
  {Mujica}, R., {Covino}, E., \& {Sterzik}, M.~F. 2005, \aap, 433, 151

\bibitem[{{Zuckerman} \& {Song}(2004)}]{zuck04}
{Zuckerman}, B., \& {Song}, I. 2004, \araa, 42, 685

\bibitem[{{Zuckerman} {et~al.}(2004){Zuckerman}, {Song}, \&
  {Bessell}}]{zuck04b}
{Zuckerman}, B., {Song}, I., \& {Bessell}, M.~S. 2004, \apjl, 613, L65

\end{thebibliography}
\bibliographystyle{apj}

\clearpage %TABLES for ROSAT PAPER

\begin{deluxetable}{llcccrrclccccc}
\tabletypesize{\scriptsize}
\rotate
\tablecaption{Observed M dwarfs\label{targets}}
\tablewidth{0pt}
\tablehead{
\colhead{Name\tablenotemark{a}} & \colhead{RA \& DEC} & \colhead{Tele-} & \colhead{SpT} & \colhead{Dist.\tablenotemark{c}} &  \colhead{$I$} & \colhead{$J$} & \colhead{log$(F_X/F_J)$\tablenotemark{d}} & \colhead{Binarity\tablenotemark{i}}\\
\colhead{} & \colhead{J2000 (2MASS)} & \colhead{scope\tablenotemark{b}} & \colhead{M--($\pm 0.5$)\tablenotemark{b}} & \colhead{(pc)} & \colhead{$_{USNO}$} & \colhead{$_{2MASS}$} & \colhead{} & \colhead{}
}
\startdata
							
2MASS J00034227--2822410	&	00	03	42.28	-28	22	41.0		&	Keck	&	7.0\tablenotemark{e}	&	29.2$\pm$4.0	&	15.65	&	13.068	&	-1.742	&	--	--		\\
G 217-32                	&	00	07	42.64	60	22	54.3		&	CFHT	&	3.8	&	12.2$\pm$1.7	&	10.32	&	8.911	&	-2.391	&	--	--		\\
LP 348-40	&	00	11	53.03	22	59	04.8		&	CFHT	&	3.5	&	16.0$\pm$2.7	&	10.22	&	8.862	&	-1.824	&			\sout{SB1}	\\
NLTT 614                	&	00	12	57.17	50	59	17.3		&	Keck	&	6.4	&	18.2$\pm$2.7	&	13.65	&	11.406	&	-2.041	&	--	--		\\
1RXS J001557.5--163659	&	00	15	58.08	-16	36	57.9	\tablenotemark{j}	&	CFHT	&	4.1	&	7.3$\pm$1.2	&	10.49	&	8.736	&	-2.430	&			\sout{SB1}	\\
GJ 1006 A                 	&	00	16	14.56	19	51	38.6		&	Keck	&	3.6	&	12.3$\pm$2.7	&	--	&	7.875	&	-2.446	&	P$_{phot}$=4.7901 d	\citep{nort07}\tablenotemark{f}		\\
GJ 3030                 	&	00	21	57.81	49	12	38.0		&	CFHT	&	2.4	&	24.1$\pm$2.7	&	10.12	&	9.139	&	-2.610	&	--	--		\\
GJ 3036                 	&	00	28	53.92	50	22	33.0		&	CFHT	&	3.7	&	17.4$\pm$4.5 	&	10.28	&	8.847	&	-2.222	&	0.426'' VB	\citep{daem07}		\\
NLTT 1875               	&	00	35	4.88	59	53	08.0		&	Keck	&	4.3	&	19.5$\pm$2.7	&	12.58	&	11.039	&	-2.173	&	--	--		\\
G 69-32                   	&	00	54	48.03	27	31	03.6	\tablenotemark{j}	&	Keck	&	4.6	&	17.8$\pm$4.5	&	11.59	&	10.34	&	-2.179	&	--	--		\\
G 132-51B (W)       	&	01	03	42.11	40	51	15.8	\tablenotemark{j}	&	CFHT	&	2.6	&	29.3$\pm$2.7	&	10.32	&	9.372	&	-1.729	&	VB (brighter)	(this work),	\sout{SB1}	\\
G 132-51B (E)               	&	01	03	42.11	40	51	15.8		&	CFHT	&	3.8	&	29.3$\pm$2.7	&	10.32	&	9.372	&	-1.729	&	VB (fainter)	(this work)		\\
G 269-153  (NE)             	&	01	24	27.68	-33	55	08.6	\tablenotemark{j}	&	CFHT	&	4.3	&	12.6$\pm$2.3 	&	10.47	&	9.203	&	-2.096	&	2.065'' VB	\citep{daem07},	\sout{SB1}	\\
G 269-153 (SW)	&	01	24	27.68	-33	55	08.6		&	CFHT	&	4.6	&	12.6$\pm$2.3 	&	10.47	&	9.203	&	-2.096	&	2.065'' VB	\citep{daem07},	\sout{SB1}	\\
G 172-56                	&	01	29	12.57	48	19	35.5		&	Keck	&	5.4	&	44.0$\pm$7.4	&	12.07	&	10.912	&	-1.809	&	--	--		\\
2MASS J01351393--0712517	&	01	35	13.93	-07	12	51.8	\tablenotemark{j}	&	CFHT	&	4.3	&	9.0$\pm$1.5	&	10.46	&	8.964	&	-2.304	&	--	--		\\
G 271-110               	&	01	36	55.17	-06	47	37.9		&	CFHT	&	3.5	&	25.8$\pm$4.2	&	11.02	&	9.707	&	-2.125	&			\sout{SB1}	\\
LHS 6032 (SW)                	&	01	45	18.2	46	32	07.8		&	CFHT	&	1.3	&	26.7$\pm$2.7	&	9.94	&	8.058	&	-2.130	&	VB	(this work)		\\
1RXS J015027.1--185134   	&	01	50	27.12	-18	51	36.0		&	Keck	&	5.4	&	--	&	13.31	&	11.484	&	-1.985	&	--	--		\\
NLTT 6549               	&	01	58	13.61	48	44	19.7		&	CFHT	&	1.5	&	23.4$\pm$2.7	&	10.13	&	9.123	&	-2.554	&	--	--		\\
GJ 1041A (SW)	&	01	59	12.39	03	31	09.2		&	CFHT	&	0.6	&	22.5$\pm$2.7	&	9.8	&	7.906	&	-2.450	&	--	--		\\
GJ 82	&	01	59	23.5	58	31	16.2		&	CFHT	&	4.2	&	9.1$\pm$2.7	&	11.12	&	7.79	&	-2.344	&	--	--		\\
GJ 3136                 	&	02	08	53.6	49	26	56.6		&	CFHT	&	2.9	&	13.7$\pm$1.4	&	9.6	&	8.423	&	-2.167	&	--	--		\\
1RXS J021836.6+121902	&	02	18	36.55	12	18	58.0	\tablenotemark{j}	&	CFHT	&	1.9	&	20.1$\pm$2.7	&	9.9	&	8.797	&	-2.597	&	--	--		\\
GJ 3150                 	&	02	19	2.29	23	52	55.1		&	CFHT	&	3.6	&	18.4$\pm$2.7	&	10.89	&	9.777	&	-2.302	&	--	--		\\
1RXS J022735.8+471021	&	02	27	37.26	47	10	04.5		&	CFHT	&	4.6	&	14.5$\pm$2.7	&	11.47	&	10.306	&	-2.418	&	--	--		\\
1RXS J023138.7+445640   	&	02	31	39.27	44	56	38.8		&	CFHT	&	4.4	&	12.9$\pm$2.4	&	11.1	&	9.97	&	-2.456	&	--	--		\\
G 36-26                   	&	02	36	44.13	22	40	26.5		&	CFHT	&	5.9	&	12.8$\pm$1.1	&	11.39	&	10.081	&	-2.009	&	--	--		\\
GJ 3174                 	&	02	39	17.35	07	28	17.0		&	CFHT	&	3.7	&	19.7$\pm$2.7	&	11.02	&	9.881	&	-2.392	&	--	--		\\
LP 247-13	&	03	15	37.83	37	24	14.3		&	CFHT	&	2.7	&	25.4$\pm$4.0	&	10.49	&	9.317	&	-2.204	&	--	--		\\
G 246-33                 	&	03	19	28.73	61	56	04.6		&	CFHT	&	4.1	&	10.9$\pm$2.7	&	10.92	&	9.511	&	-2.169	&	--	--		\\
1RXS J032230.7+285852	&	03	22	31.66	28	58	29.2		&	Keck	&	4.0	&	18.0$\pm$2.7	&	11.89	&	10.823	&	-2.119	&	--	--		\\
2MASS J03350208+2342356	&	03	35	2.09	23	42	35.6		&	Keck	&	8.5	&	$\approx$59\tablenotemark{k}	&	--	&	12.25	&	--	&	--	--		\\
1RXS J034231.8+121622	&	03	42	31.8	12	16	22.6	\tablenotemark{j}	&	Keck	&	5.2	&	17.9$\pm$2.7	&	11.21	&	10.157	&	-2.144	&			\sout{SB1}	\\
G 80-21                  	&	03	47	23.33	-01	58	19.5	\tablenotemark{j}	&	Keck	&	2.8	&	16.3$\pm$0.8	&	13.02	&	7.804	&	-2.346	&	--	--		\\
II Tau	&	03	49	43.25	24	19	04.7		&	Keck	&	4.6	&	20.4$\pm$3.8	&	10.67	&	9.805	&	-2.338	&	--	--		\\
1RXS J041325.8--013919	&	04	13	25.76	-01	39	41.7	\tablenotemark{j}	&	Keck	&	5.3	&	9.4$\pm$1.5	&	10.94	&	9.375	&	-2.312	&	0.79'' VB	\citep{mcca01}		\\
1RXS J041417.0--090650	&	04	14	17.3	-09	06	54.4	\tablenotemark{j}	&	Keck	&	4.3	&	10.6$\pm$1.7	&	11.12	&	9.63	&	-1.922	&			\sout{SB1}	\\
GJ 3287                  	&	04	27	41.3	59	35	16.7		&	CFHT	&	3.8	&	20.1$\pm$2.7	&	11.32	&	9.975	&	-2.402	&	--	--		\\
GJ 3305                  	&	04	37	37.46	-02	29	28.3	\tablenotemark{j}	&	Keck	&	1.1	&	15.8$\pm$1.6	&	4.94	&	7.299	&	-1.866	&	--	--		\\
GJ 3304                  	&	04	38	12.56	28	13	00.1		&	Keck	&	4.6	&	14.5$\pm$3.8 	&	9.86	&	8.173	&	-2.350	&	0.783'' VB	\citep{beuz04, daem07}		\\
NLTT 13728               	&	04	40	23.25	-05	30	08.3		&	Keck	&	6.0	&	10.3$\pm$0.9	&	13.08	&	10.658	&	-2.139	&	--	--		\\
NLTT 13837               	&	04	44	8.15	14	01	22.9		&	Keck	&	4.3	&	28.8$\pm$4.4	&	11.02	&	9.843	&	-2.360	&	--	--		\\
NLTT 13844               	&	04	45	5.62	43	24	34.2		&	Keck	&	4.6	&	22.5$\pm$2.7	&	13.2	&	10.84	&	-2.320	&	--	--		\\
2MASS J04465175--1116476	&	04	46	51.75	-11	16	47.6		&	Keck	&	4.9	&	13.4$\pm$2.2	&	9.56	&	8.144	&	-2.486	&	VB	(this work)		\\
2MASS J04472312--2750358	&	04	47	23.13	-27	50	35.8		&	Keck	&	0.5	&	10.9$\pm$1.8	&	8.43	&	7.66	&	-2.148	&	--	--		\\
G 81-34                  	&	04	49	29.47	48	28	45.9		&	CFHT	&	4.0	&	13.6$\pm$2.7	&	10.55	&	9.059	&	-2.379	&	--	--		\\
1RXS J045101.0+312734	&	04	51	1.38	31	27	23.9		&	Keck	&	3.7	&	22.2$\pm$2.7	&	10.25	&	9.011	&	-2.209	&	--	--		\\
NLTT 14116               	&	04	52	24.41	-16	49	21.9	\tablenotemark{j}	&	Keck	&	3.3	&	17.4$\pm$1.2	&	10.43	&	7.74	&	-2.389	&	1.48"	G.~Anglada-Escud\'e (priv. comm.)		\\
GJ 3335                  	&	05	09	9.97	15	27	32.5		&	Keck	&	3.5	&	18.3$\pm$2.7	&	10.16	&	8.77	&	-2.324	&	--	--		\\
NLTT 15049               	&	05	25	41.67	-09	09	12.3	\tablenotemark{j}	&	Keck	&	3.8	&	20.2$\pm$4.7 	&	10.13	&	8.454	&	-2.416	&	0.537'' VB	\citep{daem07}		\\
GJ 207.1	&	05	33	44.81	01	56	43.4		&	Keck	&	2.0	&	16.8$\pm$1.2\tablenotemark{g}	&	10.43	&	7.764	&	-2.298	&	--	--		\\
1RXS J055446.0+105559	&	05	54	45.74	10	55	57.1		&	CFHT	&	2.1	&	20.3$\pm$2.7	&	10.23	&	8.832	&	-2.337	&	--	--		\\
2MASS J05575096--1359503	&	05	57	50.97	-13	59	50.3		&	Keck	&	7.0	&	$\approx$60\tablenotemark{k}	&	16.49	&	12.871	&	-1.672	&	--	--		\\
GJ 3372 B                	&	05	59	55.69	58	34	15.6		&	CFHT	&	4.2	&	12.7$\pm$2.7	&	10.5	&	9.028	&	-2.403	&	--	--		\\
G 249-36	&	06	05	29.36	60	49	23.2		&	CFHT	&	4.9	&	10.6$\pm$2.7	&	10.74	&	9.096	&	-2.247	&	--	--		\\
GJ 3395                  	&	06	31	1.16	50	02	48.6		&	CFHT	&	0.8	&	19.6$\pm$2.7	&	9.37	&	7.873	&	-2.448	&	--	--		\\
G 108-36                 	&	06	51	59.02	03	12	55.3		&	CFHT	&	2.5	&	23.4$\pm$2.7	&	10.28	&	9.139	&	-2.306	&	--	--		\\
GJ 3417 (NE)              	&	06	57	57.04	62	19	19.7		&	CFHT	&	5.2	&	9.5$\pm$2.7	&	10.25	&	8.585	&	-2.100	&	VB	(this work)		\\
GJ 2060	&	07	28	51.38	-30	14	49.1		&	Keck	&	1.3	&	15.8$\pm$2.3\tablenotemark{g}	&	8.39	&	6.615	&	-2.207	&	VB - quadruple system	\citep{alle08},	\sout{SB1}	\\
GJ 277B	&	07	31	57.35	36	13	47.8		&	Keck	&	3.3	&	11.5$\pm$0.7\tablenotemark{g}	&	--	&	7.571	&	-2.081	&	--	--		\\
1RXS J073829.3+240014	&	07	38	29.52	24	00	08.8		&	Keck	&	2.7	&	--	&	10.32	&	8.928	&	-2.008	&	--	--		\\
NLTT 18549               	&	07	52	23.9	16	12	15.7		&	Keck	&	$7.0$	&	10.5$\pm$0.5	&	--	&	10.879	&	--	&			\sout{SB1}	\\
2MASS J07572716+1201273	&	07	57	27.16	12	01	27.3		&	Keck	&	2.3	&	23.0$\pm$3.7	&	10.03	&	9.06	&	-2.411	&			\sout{SB1}	\\
2MASS J08031018+2022154	&	08	03	10.18	20	22	15.5	\tablenotemark{j}	&	Keck	&	3.3	&	23.4$\pm$3.8	&	10.34	&	9.242	&	-1.899	&			\sout{SB1}	\\
GJ 316.1	&	08	40	29.75	18	24	09.2		&	Keck	&	6.0	&	14.1$\pm$0.2	&	--	&	11.053	&	-1.033	&			\sout{SB1}	\\
NLTT 20303               	&	08	48	36.45	-13	53	08.4		&	Keck	&	2.6	&	16.4$\pm$1.7	&	9.22	&	8.748	&	-1.701	&			\sout{SB1}	\\
1RXS J091744.5+461229 	&	09	17	44.73	46	12	24.7		&	Keck	&	1.7	&	18.9$\pm$3.0	&	--	&	8.126	&	-2.009	&			\sout{SB1}	\\
G 43-2                     	&	09	48	50.2	15	38	44.9		&	Keck	&	$2.0$\tablenotemark{h}	&	24.3$\pm$2.7	&	--	&	9.303	&	-2.419	&	--	--		\\
NLTT 22741               	&	09	51	4.6	35	58	09.8		&	Keck	&	4.5	&	23.0$\pm$2.7	&	11.89	&	10.577	&	-2.349	&	M4.5/L6 wide VB	\citep{reid06}		\\
GJ 3577 A (W)               	&	09	59	18.8	43	50	25.6		&	Keck	&	3.5\tablenotemark{e}	&	26.2$\pm$2.7	&	10.72	&	9.682	&	-1.922	&	VB	(this work)		\\
G 196-3A                 	&	10	04	21.49	50	23	13.6		&	Keck	&	3.0\tablenotemark{e}	&	14.9$\pm$2.7	&	10.74	&	8.081	&	-2.063	&	--	--		\\
GJ 2079	&	10	14	19.19	21	04	29.8		&	Keck	&	0.7	&	20.4$\pm$0.7\tablenotemark{g}	&	8.64	&	7.074	&	-2.640	&	--	--		\\
1RXS J101432.0+060649	&	10	14	31.95	06	06	41.0		&	Keck	&	4.1	&	17.7$\pm$2.7	&	9.94	&	8.879	&	-2.169	&	VB ($<$ 0.7")	(this work)		\\
GJ 388	&	10	19	36.35	19	52	12.2		&	Keck	&	3.5\tablenotemark{e}	&	4.9$\pm$0.1\tablenotemark{g}	&	7.78	&	5.449	&	-2.389	&	--	--		\\
G 44-9	&	10	20	44.07	08	14	23.4		&	Keck	&	5.9	&	20.2$\pm$2.7	&	12.13	&	10.354	&	-2.365	&	--	--		\\
2MASS J10364483+1521394    	&	10	36	44.84	15	21	39.5		&	Keck	&	4.0\tablenotemark{e}	&	19.6$\pm$4.6 	&	10.03	&	8.748	&	-2.186	&	1.061'' VB	\citep{daem07}		\\
GJ 3629                    	&	10	51	20.6	36	07	25.6		&	Keck	&	3.0\tablenotemark{e}	&	16.7$\pm$2.7	&	10.63	&	9.422	&	-2.187	&	--	--		\\
GJ 3639                  	&	11	03	10	36	39	08.5		&	Keck	&	3.5\tablenotemark{e}	&	18.6$\pm$2.7	&	10.44	&	9.464	&	-2.327	&	--	--		\\
NLTT 26114               	&	11	03	21.25	13	37	57.1		&	Keck	&	$3.0$\tablenotemark{h}	&	14.1$\pm$1.5	&	--	&	8.759	&	-2.287	&	--	--		\\
G 119-62	&	11	11	51.76	33	32	11.2		&	Keck	&	3.5\tablenotemark{e}	&	13.6$\pm$1.3	&	9.71	&	8.297	&	-2.234	&	--	--		\\
1RXS J111300.1+102518	&	11	13	0.6	10	25	05.9		&	Keck	&	$3.0$\tablenotemark{h}	&	20.2$\pm$2.7	&	11.39	&	10.032	&	-2.201	&	--	--		\\
GJ 3653                    	&	11	15	54.04	55	19	50.6		&	Keck	&	0.5\tablenotemark{e}	&	22.3$\pm$2.7	&	9.56	&	8.09	&	-2.381	&	--	--		\\
2MASS J11240434+3808108	&	11	24	4.35	38	08	10.9		&	Keck	&	$4.5$\tablenotemark{h}	&	19.2$\pm$3.1	&	10.64	&	9.928	&	-2.241	&	--	--		\\
G 10-52                    	&	11	48	35.49	07	41	40.4		&	Keck	&	$3.5$\tablenotemark{h}	&	16.0$\pm$2.7	&	10.62	&	9.476	&	-2.423	&	--	--		\\
2MASS J12065663+7007514 (E)	&	12	06	56.63	70	07	51.4	\tablenotemark{j}	&	CFHT	&	4.4	&	16.8$\pm$1.9	&	9.1	&	9.251	&	-1.758	&	VB	(this work)		\\
G 122-74                   	&	12	12	11.36	48	49	03.2		&	Keck	&	$3.5$\tablenotemark{h}	&	23.3$\pm$2.7	&	10.04	&	9.258	&	-2.536	&	VB	(this work)		\\
GJ 3729                  	&	12	29	2.9	41	43	49.7		&	Keck	&	3.5\tablenotemark{e}	&	15.2$\pm$2.7	&	9.87	&	8.786	&	-2.496	&	--	--		\\
GJ 3730                  	&	12	29	27.13	22	59	46.7		&	Keck	&	4.0\tablenotemark{e}	&	16.7$\pm$2.7	&	11.04	&	9.823	&	-2.105	&	--	--		\\
1RXS J124147.5+564506	&	12	41	47.37	56	45	13.8	\tablenotemark{j}	&	Keck	&	2.5\tablenotemark{e}	&	26.2$\pm$4.2	&	10.4	&	9.483	&	-2.091	&	--	--		\\
GJ 490 B	&	12	57	39.35	35	13	19.5		&	Keck	&	4.0\tablenotemark{e}	&	18.0$\pm$1.1\tablenotemark{g}	&	14.78	&	8.872	&	-1.648	&	--	--		\\
GJ 490 A	&	12	57	40.3	35	13	30.6		&	Keck	&	0.5\tablenotemark{e}	&	18.0$\pm$1.1\tablenotemark{g}	&	8.84	&	7.401	&	-2.236	&	P$_{phot}$=3.17 d	\citep{pett80, nort07}\tablenotemark{f}		\\
NLTT 32659 (E)                	&	13	02	5.87	12	22	21.6		&	Keck	&	3.7	&	28.4$\pm$2.7	&	9.96	&	9.089	&	-2.337	&	VB	(this work)		\\
NLTT 32659 (W)                	&	13	02	5.87	12	22	21.6		&	Keck	&	1.6	&	28.4$\pm$2.7	&	9.96	&	9.089	&	-2.337	&	VB	(this work),	\sout{SB1}	\\
GJ 1167 A                 	&	13	09	34.95	28	59	06.6	\tablenotemark{j}	&	Keck	&	4.8	&	11.5$\pm$2.4\tablenotemark{g}	&	10.67	&	9.476	&	-2.308	&			\sout{SB1}	\\
GJ 3786                  	&	13	27	19.67	-31	10	39.4	\tablenotemark{j}	&	Keck	&	3.5\tablenotemark{e}	&	19.6$\pm$4.6 	&	10.89	&	9.329	&	-2.327	&	0.544'' VB	\citep{daem07},	\sout{SB1}	\\
2MASS J13292408--1422122	&	13	29	24.08	-14	22	12.3	\tablenotemark{j}	&	CFHT	&	2.8	&	22.6$\pm$3.6	&	10.42	&	9.061	&	-2.531	&	--	--		\\
2MASS J14215503--3125537	&	14	21	55.04	-31	25	53.7		&	Keck	&	3.9	&	--	&	10.98	&	9.66	&	-2.151	&			\sout{SB1}	\\
2MASS J1442809--0424078	&	14	44	28.1	-04	24	07.8	\tablenotemark{j}	&	Keck	&	3.0	&	27.5$\pm$4.4	&	11.01	&	9.728	&	-2.411	&			\sout{SB1}	\\
1RXS J150907.2+590422	&	15	09	6.96	59	04	28.2	\tablenotemark{j}	&	CFHT	&	2.2	&	25.8$\pm$4.2	&	9.66	&	9.249	&	-2.356	&	--	--		\\
GJ 9520	&	15	21	52.92	20	58	39.5		&	Keck	&	1.0	&	11.4$\pm$0.3	&	8.41	&	6.61	&	-2.485	&			\sout{SB1}	\\
NLTT 40561               	&	15	33	50.62	25	10	10.6		&	Keck	&	3.5	&	23.2$\pm$2.7	&	11.31	&	10.433	&	-2.282	&			\sout{SB1}	\\
G 167-54                 	&	15	43	48.48	25	52	37.7		&	Keck	&	4.1	&	21.1$\pm$2.7	&	11.3	&	10.022	&	-2.305	&			\sout{SB1}	\\
2MASS J15534211--2049282 (S)	&	15	53	42.12	-20	49	28.2		&	Keck	&	3.4	&	$\approx$145\tablenotemark{k}	&	13.44	&	11.256	&	--	&	VB	(this work),	\sout{SB1}	\\
GJ 3928                  	&	15	55	31.78	35	12	02.9	\tablenotemark{j}	&	CFHT	&	5.3	&	12.6$\pm$3.6 	&	10.51	&	8.928	&	-2.274	&	1.571'' VB	\citep{daem07}		\\
NLTT 43695  (E)           	&	16	51	9.95	35	55	07.1	\tablenotemark{j}	&	Keck	&	4.6	&	34.5$\pm$3.9	&	11.67	&	10.334	&	-2.094	&	VB	(this work),	\sout{SB1}	\\
LP 331-57	&	17	03	52.83	32	11	45.6	\tablenotemark{j}	&	CFHT	&	2.4	&	18.0$\pm$1.2 	&	10.94	&	7.886	&	-2.659	&	P$_{phot}$=15.4221 d, 1.13'' VB	\citep{nort07, daem07}\tablenotemark{f}		\\
GJ 616.2	&	17	19	52.98	26	30	02.6		&	Keck	&	5.6	&	10.8$\pm$0.2\tablenotemark{g}	&	8.78	&	8.229	&	-2.036	&			\sout{SB1}	\\
GJ 669A	&	17	19	54.22	26	30	03.0		&	Keck	&	3.4	&	12.0$\pm$0.9\tablenotemark{g}	&	9.6	&	7.273	&	-2.418	&			\sout{SB1}	\\
1RXS J173130.9+272134	&	17	31	29.75	27	21	23.3		&	Keck	&	2.6	&	16.1$\pm$2.7	&	14.94	&	12.094	&	-1.507	&			\sout{SB1}	\\
G 227-22                 	&	18	02	16.6	64	15	44.6	\tablenotemark{j}	&	Keck	&	6.1	&	7.1$\pm$0.6	&	9.92	&	8.541	&	-2.536	&			\sout{SB1}	\\
GJ 4044                  	&	18	13	6.57	26	01	51.9		&	Keck	&	4.5	&	12.1$\pm$2.7	&	10.39	&	8.899	&	-2.018	&	VB of LP 390-16, P$_{phot}$=2.2838 d	(this work)	\sout{SB1}	\\
LP 390-16	&	18	13	6.57	26	01	51.9		&	CFHT	&	3.8	&	13.2$\pm$1.5	&	10.39	&	8.899	&	-2.018	&	VB of GJ 4044	(this work)		\\
1RXS J183203.0+203050 (N)	&	18	32	2.91	20	30	58.1		&	Keck	&	4.9	&	21.6$\pm$2.7	&	11.45	&	10.653	&	-1.565	&	VB	(this work),	\sout{SB1}	\\
1RXS J183203.0+203050 (S)	&	18	32	2.91	20	30	58.1		&	Keck	&	5.1	&	21.6$\pm$2.7	&	11.45	&	10.653	&	-1.565	&	VB	(this work),	\sout{SB1}	\\
1RXS J184410.0+712909 (E)	&	18	44	10.2	71	29	17.6		&	CFHT	&	3.9	&	22.2$\pm$2.7	&	10.73	&	10.138	&	-2.502	&	VB	(this work)		\\
1RXS J184410.0+712909 (W)	&	18	44	10.2	71	29	17.6		&	CFHT	&	4.1	&	22.2$\pm$2.7	&	10.73	&	10.138	&	-2.502	&	VB	(this work)		\\
GJ 9652 A                	&	19	14	39.26	19	19	02.6		&	Keck	&	4.5	&	19.1$\pm$1.1\tablenotemark{g}	&	10.46	&	7.579	&	-2.335	&			\sout{SB1}	\\
2MASS J19303829--1335083	&	19	30	38.3	-13	35	08.4		&	Keck	&	6.0	&	25.0$\pm$4.0	&	13.61	&	11.53	&	-2.269	&			\sout{SB1}	\\
1RXS J193124.2--213422    	&	19	31	24.34	-21	34	22.6	\tablenotemark{j}	&	Keck	&	2.4	&	19.0$\pm$3.1	&	10.23	&	8.694	&	-2.469	&			\sout{SB1}	\\
1RXS J193528.9+374605	&	19	35	29.23	37	46	08.2		&	Keck	&	3.0	&	9.4$\pm$2.7	&	9.91	&	7.562	&	-2.514	&			\sout{SB1}	\\
1RXS J194213.0--204547	&	19	42	12.82	-20	45	47.8	\tablenotemark{j}	&	Keck	&	5.1	&	10.8$\pm$1.7	&	11.07	&	9.598	&	-2.249	&			\sout{SB1}	\\
G 125-36                 	&	19	50	15.93	31	46	59.9		&	Keck	&	2.1	&	24.2$\pm$2.7	&	10.34	&	9.178	&	-2.306	&			\sout{SB1}	\\
2MASS J20003177+5921289	&	20	00	31.77	59	21	29.0		&	Keck	&	4.1	&	20.9$\pm$3.4	&	9.99	&	9.636	&	-2.243	&			\sout{SB1}	\\
1RXS J204340.6--243410 (NE)	&	20	43	41.15	-24	33	53.5	\tablenotemark{j}	&	Keck	&	3.7	&	24.7$\pm$2.8	&	10.08	&	8.597	&	-2.243	&	VB	(this work),	\sout{SB1}	\\
1RXS J204340.6--243410 (SW)	&	20	43	41.15	-24	33	53.5		&	Keck	&	4.1	&	24.7$\pm$2.8	&	10.08	&	8.597	&	-2.243	&	VB	(this work),	\sout{SB1}	\\
NLTT 49856               	&	20	46	43.61	-11	48	13.2	\tablenotemark{j}	&	Keck	&	4.5	&	14.6$\pm$2.2	&	10.48	&	9.349	&	-2.222	&			\sout{SB1}	\\
2MASS J20530910--0133039	&	20	53	9.1	-01	33	04.0	\tablenotemark{j}	&	CFHT	&	5.6	&	16.5$\pm$1.5	&	11.39	&	10.659	&	-2.362	&			\sout{SB1}	\\
NLTT 50066               	&	20	53	14.65	-02	21	21.9	\tablenotemark{j}	&	CFHT	&	2.9	&	23.7$\pm$3.8	&	10.68	&	9.329	&	-2.601	&	--	--		\\
GJ 4185B	&	21	16	3.79	29	51	46.0		&	Keck	&	3.3	&	16.1$\pm$2.7	&	10.45	&	9.295	&	-1.816	&			\sout{SB1}	\\
GJ 4185 A                	&	21	16	5.77	29	51	51.1		&	Keck	&	3.3	&	8.4$\pm$2.7	&	9.84	&	8.448	&	-2.155	&			\sout{SB1}	\\
GJ 4231                    	&	21	52	10.4	05	37	35.7		&	Keck	&	2.4	&	14.6$\pm$2.7	&	13.62	&	8.248	&	-2.359	&			\sout{SB1}	\\
1RXS J221419.3+253411	&	22	14	17.66	25	34	06.6		&	Keck	&	4.3	&	23.9$\pm$2.7	&	11.33	&	10.177	&	-2.355	&			\sout{SB1}	\\
GJ 4282 (E)         	&	22	33	22.65	-09	36	53.8	\tablenotemark{j}	&	CFHT	&	2.5	&	26.7$\pm$4.1 	&	9.91	&	8.534	&	-2.173	&	1.571'' VB	\citep{daem07}		\\
GJ 4282 (W)         	&	22	33	22.65	-09	36	53.8		&	CFHT	&	2.6	&	26.7$\pm$4.1 	&	9.91	&	8.534	&	-2.173	&	1.571'' VB	\citep{daem07}		\\
2MASS J22344161+4041387 	&	22	34	41.62	40	41	38.8		&	Keck	&	6.0	&	$\approx$325\tablenotemark{l}	&	14.24	&	12.573	&	-2.470	&	0.16" VB	\citep{alle09}	\sout{SB1}	\\
LP 984-91	&	22	44	57.94	-33	15	01.6	\tablenotemark{j}	&	Keck	&	4.5	&	23.7$\pm$2.0	&	13.51	&	7.786	&	-1.988	&			\sout{SB1}	\\
GJ 873	&	22	46	49.81	44	20	03.1		&	CFHT	&	3.2	&	5.1$\pm$0.0	&	8.88	&	6.106	&	-1.928	&			\sout{SB1}	\\
NLTT 54873               	&	22	47	37.64	40	41	25.4	\tablenotemark{j}	&	CFHT	&	3.8	&	21.7$\pm$2.7	&	11.44	&	10.35	&	-1.980	&			\sout{SB1}	\\
GJ 875.1	&	22	51	53.49	31	45	15.3	\tablenotemark{j}	&	CFHT	&	2.7	&	14.3$\pm$0.6	&	10.5	&	7.697	&	-2.351	&			\sout{SB1}	\\
2MASS J22581643--1104170	&	22	58	16.43	-11	04	17.1		&	CFHT	&	2.7	&	23.0$\pm$3.7	&	10.27	&	9.071	&	-2.469	&	--	--		\\
GJ 9809                  	&	23	06	4.83	63	55	34.0		&	CFHT	&	0.3	&	24.9$\pm$1.0\tablenotemark{g}	&	9.6	&	7.815	&	-2.281	&			\sout{SB1}	\\
NLTT 56194               	&	23	13	47.28	21	17	29.4		&	Keck	&	7.5	&	20.2$\pm$2.7	&	13.03	&	11.421	&	-2.097	&			\sout{SB1}	\\
NLTT 56566               	&	23	20	57.66	-01	47	37.3	\tablenotemark{j}	&	CFHT	&	3.8	&	22.6$\pm$6.7 	&	10.76	&	9.355	&	-2.362	&	0.099'' VB	\citep{daem07}		\\
GJ 4338 B                	&	23	29	22.58	41	27	52.2		&	CFHT	&	4.2	&	14.8$\pm$0.4\tablenotemark{g}	&	8.48	&	8.017	&	-2.041	&			\sout{SB1}	\\
GJ 4337 A                	&	23	29	23.46	41	28	06.9		&	CFHT	&	2.9	&	14.8$\pm$0.4\tablenotemark{g}	&	10	&	7.925	&	-2.077	&			\sout{SB1}	\\
GJ 1290                  	&	23	44	20.84	21	36	05.0		&	CFHT	&	3.4	&	22.0$\pm$2.2\tablenotemark{g}	&	10.41	&	9.07	&	-2.406	&			\sout{SB1}	\\
1RXS J235005.6+265942	&	23	50	6.39	26	59	51.9	\tablenotemark{j}	&	CFHT	&	4.0	&	26.5$\pm$4.3	&	11.22	&	10.142	&	-2.192	&			\sout{SB1}	\\
G 68-46	&	23	51	22.28	23	44	20.8	\tablenotemark{j}	&	CFHT	&	4.0	&	18.1$\pm$3.3	&	10.79	&	9.683	&	-2.469	&			\sout{SB1}	\\
1RXS J235133.3+312720    	&	23	51	33.67	31	27	23.0		&	CFHT	&	2.0	&	35.0$\pm$5.6	&	10.92	&	9.821	&	-2.228	&			\sout{SB1}	\\
1RXS J235452.2+383129	&	23	54	51.47	38	31	36.3		&	CFHT	&	3.1	&	13.2$\pm$2.1	&	10.22	&	8.937	&	-2.189	&			\sout{SB1}	\\
GJ 4381                  	&	23	57	49.9	38	37	46.9		&	CFHT	&	2.8	&	8.9$\pm$2.7	&	9.91	&	8.691	&	-2.206	&			\sout{SB1}	\\
G 273-191 (N)	&	23	58	13.66	-17	24	33.8	\tablenotemark{j}	&	CFHT	&	1.9	&	26.1$\pm$2.0	&	9.72	&	8.311	&	-2.332	&	1.904'' VB	\citep{daem07},	\sout{SB1}	\\
G 273-191 (S)              	&	23	58	13.66	-17	24	33.8		&	CFHT	&	1.9	&	26.1$\pm$2.0 	&	9.72	&	8.311	&	-2.332	&	1.904'' VB	\citep{daem07},	\sout{SB1}	\\
G 130-31                 	&	23	59	19.86	32	41	24.5		&	Keck	&	5.6	&	23.9$\pm$2.7	&	11.8	&	10.451	&	-2.241	&			\sout{SB1}	\\
\enddata
\tablenotetext{a}{Those target names with directions in parentheses were resolved as visual binaries (VB) and when possible, both components were observed. If a target was resolved as a VB by another group (see last column) but not resolved at the telescope, the resulting spectrum is a composite of the both components. Likewise, the published photometry is of the combined system.}
\tablenotetext{b}{Each of the two spectrographs used, Keck+HIRES and CFHT+ESPaDOnS, produces a different measurement error for the TiO index and thus, the errors of the of the calculated SpTs are $\delta$(SpT) = 0.33, 0.11, respectively. However, the calibration is based on a sample binned to 0.5 subclasses, limiting the true uncertainty to this value. 
}
\tablenotetext{c}{Photometric distances from \cite{reid02b, reid07a} unless otherwise noted. \cite{daem07} revised the distances to those VBs resolved by them. (See last column). Distances of those VBs resolved by us during acquisition have been corrected for their binarity assuming equal flux for each component.}
\tablenotetext{d}{Fractional X-ray flux from the ROSAT All-Sky Survey.  Three targets were not detected by ROSAT but have been included in the sample due to previously published indications of youth.  (See column 7 of Table~\ref{ages}.)}
\tablenotetext{e}{Due to either poor S/N or bad seeing, we were unable to measure reliable TiO indices for these targets and defer to the published SpT ($\pm 0.5$ subclasses) based on low-resolution data (\citealt{reid02b, reid07a}).}
%\tablenotetext{e}{These targets were resolved as visual binaries using Gemini-North's Altair AO system by \citealt{daem07} who published these revised photometric distances, with the exception of 2MASS 2234 which has been observed to be a binary by Allers et al. (in prep).}
\tablenotetext{f}{\cite{nort07} attribute the photometric period to an eclipsing binary of BY Dra-type, however we see no evidence that this target is a spectroscopic binary.}	
\tablenotetext{g}{Distances from the Hipparcos \& Tycho Catalogues (\citealt{perr97}) using trigonometric parallaxes.}
\tablenotetext{h}{These targets were plagued with poor seeing and/or low S/N {\it and} no previously published SpT.  Therefore, we did a by-eye comparison of the TiO bands with spectra of known SpT. The error for these is $\pm$1 subclass.}
\tablenotetext{i}{Visual binaries are marked as ``VB''. Stars observed during two epochs with no RV variation are marked as ``\sout{SB1}''.}
\tablenotetext{j}{\cite{riaz06} obtained low-resolution spectra of these targets.}
\tablenotetext{k}{The photometric distances have been corrected here for overluminosity due the very young stellar ages listed in Table 2 using evolutionary models of \cite{bara98}.  The binarity of 2M1553 as also taken into account.}
\tablenotetext{l}{Distance from \cite{alle09}.}

\end{deluxetable}

\begin{deluxetable}{lccccclcr}
\tabletypesize{\scriptsize}
\rotate
\tablecaption{Measured Quantities and Stellar Ages\label{ages}}
\tablewidth{0pt}
\tablehead{
\colhead{Name} & \colhead{SpT} & \colhead{CaH\tablenotemark{a}} & \colhead{CaH\tablenotemark{a}} & \colhead{K I EW\tablenotemark{a}} & \colhead{\ha\/ EW\tablenotemark{a}} & \colhead{Youth\tablenotemark{b}} & \colhead{Youth Indicator} & \colhead{Age\tablenotemark{c}} \\
\colhead{} & \colhead{M-- ($\pm 0.5$)} & \colhead{wide} & \colhead{narr} & \colhead{(\AA)} & \colhead{(\AA)} & \colhead{Index} & \colhead{from literature} & \colhead{(Myr)}
}
\startdata
							
%Name for Table	&	coordinates	&	SpT	&	Distance (pc)	&	IMag & J (2MASS)	& log(fx/fj) &	CaHwide	&	CaHnarrow	&  K I 7700 EW (A)	& Halpha+/-0.5 A ?	& youth code	\\	
2MASS J00034227--2822410	&	7.0	&	--	&	--	&	2.35	&	-2.19	&	?0000	&	--	--	&	100-300	\\
G 217-32                	&	3.8	&	1.45	&	1.65	&	0.91	&	-3.42	&	01000	&	--	--	&	35-300	\\
LP 348-40	&	3.5	&	1.26	&	1.40	&	1.01	&	0.13	&	11000	&	--	--	&	$\gtrsim$2000	\\
NLTT 614                	&	6.4	&	1.36	&	1.53	&	1.94	&	-9.42	&	10100	&	--	--	&	90-300	\\
1RXS J001557.5--163659	&	4.1	&	1.43	&	1.61	&	1.91	&	-4.36	&	00100	&	--	--	&	35-300	\\
GJ 1006 A                 	&	3.6	&	1.22	&	1.43	&	0.83	&	-2.70	&	11000	&	--	--	&	35-300	\\
GJ 3030                 	&	2.4	&	1.25	&	1.36	&	1.84	&	-2.00	&	10000	&	--	--	&	20-150	\\
GJ 3036                 	&	3.7	&	1.39	&	1.57	&	1.05	&	-4.56	&	01000	&	--	--	&	35-300	\\
NLTT 1875               	&	4.3	&	1.31	&	1.47	&	1.42	&	-2.63	&	10000	&	--	--	&	40-300	\\
G 69-32                   	&	4.6	&	1.34	&	1.54	&	1.06	&	-5.48	&	11000	&	--	--	&	60-300	\\
G 132-51B (W)       	&	2.6	&	1.30	&	1.41	&	0.99	&	-2.46	&	11000	&	--	--	&	20-150	\\
G 132-51B (E)               	&	3.8	&	1.39	&	1.56	&	1.04	&	-4.78	&	01000	&	--	--	&	35-300	\\
G 269-153  (NE)             	&	4.3	&	1.42	&	1.60	&	1.76	&	-6.62	&	01000	&	--	--	&	40-300	\\
G 269-153 (SW)	&	4.6	&	1.44	&	1.61	&	1.82	&	-6.77	&	00000	&	--	--	&	60-300	\\
G 172-56                	&	5.4	&	1.46	&	1.68	&	1.09	&	-9.72	&	00000	&	--	--	&	60-300	\\
2MASS J01351393--0712517	&	4.3	&	1.34	&	1.50	&	1.30	&	-5.36	&	10000	&	--	--	&	40-300	\\
G 271-110               	&	3.5	&	1.40	&	1.58	&	1.79	&	-5.03	&	00000	&	--	--	&	25-300	\\
LHS 6032 (SW)                	&	1.3	&	1.23	&	1.32	&	1.35	&	-2.35	&	00000	&	--	--	&	15-150	\\
1RXS J015027.1--185134   	&	5.4	&	1.42	&	1.64	&	2.49	&	-8.35	&	00000	&	--	--	&	60-300	\\
NLTT 6549               	&	1.5	&	1.21	&	1.31	&	0.89	&	-2.16	&	11000	&	--	--	&	15-150	\\
GJ 1041A (SW)	&	0.6	&	1.11	&	1.16	&	0.68	&	0.59	&	11000	&	--	--	&	15-150\tablenotemark{d}	\\
GJ 82	&	4.2	&	1.35	&	1.50	&	1.57	&	-4.37	&	10000	&	--	--	&	35-300	\\
GJ 3136                 	&	2.9	&	1.35	&	1.49	&	1.30	&	-5.63	&	00000	&	flare star	\citep{gers99}	&	20-300	\\
1RXS J021836.6+121902	&	1.9	&	1.22	&	1.33	&	0.84	&	-1.45	&	11000	&	--	--	&	20-150	\\
GJ 3150                 	&	3.6	&	1.42	&	1.59	&	1.84	&	-5.48	&	00000	&	--	--	&	35-300	\\
1RXS J022735.8+471021	&	4.6	&	1.50	&	1.65	&	2.03	&	-8.85	&	00000	&	--	--	&	60-300	\\
1RXS J023138.7+445640   	&	4.4	&	1.33	&	1.47	&	1.08	&	-5.11	&	11000	&	--	--	&	40-300	\\
G 36-26                   	&	5.9	&	1.43	&	1.62	&	2.21	&	-3.45	&	10000	&	--	--	&	90-300	\\
GJ 3174                 	&	3.7	&	1.35	&	1.51	&	1.85	&	-2.40	&	10000	&	--	--	&	35-300	\\
LP 247-13	&	2.7	&	1.29	&	1.41	&	0.89	&	-3.99	&	11000	&	weak-line T Tau? 	\citep{li98}	&	20-100	\\
G 246-33                 	&	4.1	&	1.44	&	1.60	&	2.12	&	-4.65	&	00000	&	--	--	&	35-300	\\
1RXS J032230.7+285852	&	4.0	&	1.31	&	1.47	&	1.81	&	-8.14	&	10000	&	--	--	&	35-300	\\
2MASS J03350208+2342356	&	8.5	&	1.29	&	1.43	&	3.15	&	-10.72	&	11111	&	Li detection	\citep{reid02b}	&	10	\\
1RXS J034231.8+121622	&	5.2	&	1.42	&	1.68	&	1.48	&	-5.66	&	00000	&	--	--	&	60-300	\\
G 80-21                  	&	2.8	&	1.33	&	1.54	&	0.81	&	-2.76	&	01000	&	AB Dor member	\citep{zuck04b}	&	30--50	\\
II Tau	&	4.6	&	1.42	&	1.69	&	1.87	&	-7.22	&	00000	&	Pleiades member	\citep{stau91}	&	120	\\
1RXS J041325.8--013919	&	5.3	&	1.44	&	1.69	&	1.79	&	-9.00	&	00000	&	--	--	&	60-300	\\
1RXS J041417.0--090650	&	4.3	&	1.50	&	1.79	&	1.67	&	-14.15	&	00110	&	--	--	&	$<$40	\\
GJ 3287                  	&	3.8	&	1.38	&	1.55	&	1.47	&	-2.04	&	00000	&	flare star	\citep{gers99}	&	35-300	\\
GJ 3305                  	&	1.1	&	1.20	&	1.36	&	0.79	&	-2.10	&	01010	&	$\beta$ Pic member 	\citep{torr06}	&	12	\\
GJ 3304                  	&	4.6	&	1.50	&	1.79	&	1.97	&	-4.40	&	00000	&	--	--	&	60-300	\\
NLTT 13728               	&	6.0	&	1.62	&	1.90	&	2.75	&	-19.11	&	00111	&	--	--	&	$<$90	\\
NLTT 13837               	&	4.3	&	1.52	&	1.80	&	1.79	&	-4.90	&	00000	&	--	--	&	40-300	\\
NLTT 13844               	&	4.6	&	1.42	&	1.74	&	2.06	&	-3.63	&	00000	&	--	--	&	60-300	\\
2MASS J04465175--1116476	&	4.9	&	1.63	&	1.98	&	1.09	&	-4.11	&	01000	&	--	--	&	60-300	\\
2MASS J04472312--2750358	&	0.5	&	1.14	&	1.28	&	0.75	&	0.47	&	01000	&	--	--	&	$\gtrsim$400	\\
G 81-34                  	&	4.0	&	1.37	&	1.52	&	1.86	&	-3.30	&	10000	&	--	--	&	35-300	\\
1RXS J045101.0+312734	&	3.7	&	1.43	&	1.70	&	1.39	&	-2.31	&	00000	&	--	--	&	35-300	\\
NLTT 14116               	&	3.3	&	1.36	&	1.58	&	1.32	&	-6.96	&	00000	&	AB Dor member	\citep{torr08}	&	30--50	\\
GJ 3335                  	&	3.5	&	1.31	&	1.49	&	1.07	&	-3.08	&	11000	&	--	--	&	25-300	\\
NLTT 15049               	&	3.8	&	1.43	&	1.68	&	1.56	&	-2.91	&	00000	&	--	--	&	35-300	\\
GJ 207.1	&	2.0	&	1.35	&	1.64	&	1.53	&	-4.96	&	00000	&	flare star	\citep{gers99}	&	20-150	\\
1RXS J055446.0+105559	&	2.1	&	1.28	&	1.39	&	0.53	&	-3.31	&	11000	&	--	--	&	20-150	\\
2MASS J05575096--1359503	&	7.0	&	1.15	&	1.49	&	0.13	&	-25.56	&	11011	&	mis-classified as giant	\citep{cruz07}	&	10	\\
GJ 3372 B                	&	4.2	&	1.42	&	1.58	&	1.60	&	-3.38	&	00000	&	--	--	&	40-300	\\
G 249-36	&	4.9	&	1.32	&	1.52	&	1.26	&	-6.11	&	10100	&	--	--	&	60-300	\\
GJ 3395                  	&	0.8	&	1.17	&	1.26	&	0.59	&	-1.46	&	01000	&	--	--	&	20-150	\\
G 108-36                 	&	2.5	&	1.27	&	1.40	&	0.78	&	-0.18	&	11000	&	--	--	&	20-150\tablenotemark{d}	\\
GJ 3417 (NE)              	&	5.2	&	1.36	&	1.52	&	0.26	&	-3.06	&	11000	&	--	--	&	60-300	\\
GJ 2060	&	1.3	&	1.24	&	1.41	&	0.88	&	-3.10	&	01000	&	AB Dor member	\citep{zuck04b}	&	30--50	\\
GJ 277B	&	3.3	&	1.26	&	1.39	&	1.57	&	-2.20	&	10000	&	--	--	&	25-300	\\
1RXS J073829.3+240014	&	2.7	&	1.35	&	1.51	&	1.88	&	-3.78	&	00000	&	--	--	&	20-300	\\
NLTT 18549               	&	7.0	&	1.46	&	1.63	&	2.81	&	-22.26	&	10101	&	strong \ha\/ emission	\citep{cruz03}	&	100	\\
2MASS J07572716+1201273	&	2.3	&	1.28	&	1.46	&	0.93	&	-0.32	&	01000	&	--	--	&	$\gtrsim$1200	\\
2MASS J08031018+2022154	&	3.3	&	1.41	&	1.64	&	1.57	&	-6.10	&	00000	&	--	--	&	25-300	\\
GJ 316.1	&	6.0	&	1.59	&	1.96	&	3.46	&	-22.68	&	00101	&	--	--	&	100	\\
NLTT 20303               	&	2.6	&	1.31	&	1.57	&	1.15	&	-2.93	&	00000	&	--	--	&	20-300	\\
1RXS J091744.5+461229 	&	1.7	&	1.25	&	1.40	&	1.04	&	-3.89	&	01000	&	--	--	&	20-150	\\
G 43-2                     	&	2.0	&	--	&	--	&	1.94	&	-2.13	&	?0000	&	--	--	&	20-150	\\
NLTT 22741               	&	4.5	&	--	&	--	&	2.02	&	-4.98	&	?0100	&	100--200 Myr	\citep{reid06}	&	40-300	\\
GJ 3577 A (W)               	&	3.5	&	--	&	--	&	1.98	&	-6.07	&	?0000	&	--	--	&	25-300	\\
G 196-3A                 	&	3.0	&	--	&	--	&	1.07	&	-3.77	&	?0000	&	--	--	&	25-300	\\
GJ 2079	&	0.7	&	--	&	--	&	1.11	&	-0.99	&	?1000	&	--	--	&	$\gtrsim$400	\\
1RXS J101432.0+060649	&	4.1	&	--	&	--	&	1.38	&	-4.09	&	?0000	&	--	--	&	35-300	\\
GJ 388	&	3.5	&	--	&	--	&	1.39	&	-2.99	&	?0000	&	--	--	&	25-300	\\
G 44-9	&	5.9	&	--	&	--	&	1.13	&	-5.30	&	?0000	&	--	--	&	90-300	\\
2MASS J10364483+1521394    	&	4.0	&	--	&	--	&	2.07	&	-4.99	&	?0000	&	--	--	&	35-300	\\
GJ 3629                    	&	3.0	&	--	&	--	&	1.25	&	-3.07	&	?0000	&	--	--	&	25-300	\\
GJ 3639                  	&	3.5	&	--	&	--	&	2.05	&	-4.80	&	?0000	&	--	--	&	25-300	\\
NLTT 26114               	&	3.0	&	--	&	--	&	1.85	&	-2.56	&	?0000	&	--	--	&	25-300	\\
G 119-62	&	3.5	&	--	&	--	&	1.80	&	-3.79	&	?0000	&	--	--	&	25-300	\\
1RXS J111300.1+102518	&	3.0	&	--	&	--	&	1.61	&	-3.11	&	?0000	&	--	--	&	25-300	\\
GJ 3653                    	&	0.5	&	--	&	--	&	0.90	&	-2.06	&	?1000	&	--	--	&	15-150	\\
2MASS J11240434+3808108	&	4.5	&	--	&	--	&	1.74	&	-3.19	&	?0000	&	--	--	&	40-300	\\
G 10-52                    	&	3.5	&	--	&	--	&	1.54	&	-4.36	&	?0000	&	--	--	&	25-300	\\
2MASS J12065663+7007514 (E)	&	4.4	&	1.35	&	1.48	&	1.71	&	-0.17	&	10000	&	--	--	&	$\gtrsim$4500	\\
G 122-74                   	&	3.5	&	--	&	--	&	0.94	&	0.37	&	?0000	&	--	--	&	$\gtrsim$2000	\\
GJ 3729                  	&	3.5	&	--	&	--	&	1.88	&	-2.23	&	?0000	&	--	--	&	25-300	\\
GJ 3730                  	&	4.0	&	--	&	--	&	1.27	&	-1.42	&	?0000	&	--	--	&	35-300	\\
1RXS J124147.5+564506	&	2.5	&	--	&	--	&	1.67	&	-3.67	&	?0000	&	--	--	&	20-150	\\
GJ 490 B	&	4.0	&	--	&	--	&	2.20	&	-4.06	&	?0000	&	--	--	&	35-300	\\
GJ 490 A	&	0.5	&	--	&	--	&	1.05	&	-1.29	&	?1000	&	--	--	&	15-150	\\
NLTT 32659 (E)                	&	3.7	&	1.29	&	1.48	&	1.66	&	-3.69	&	10000	&	--	--	&	35-300	\\
NLTT 32659 (W)                	&	1.6	&	1.18	&	1.32	&	0.74	&	-2.26	&	11000	&	--	--	&	20-150	\\
GJ 1167 A                 	&	4.8	&	1.40	&	1.62	&	2.66	&	-4.82	&	00000	&	flare star	\citep{gers99}	&	60-300	\\
GJ 3786                  	&	3.5	&	--	&	--	&	2.11	&	-3.43	&	?0000	&	--	--	&	25-300	\\
2MASS J13292408--1422122	&	2.8	&	1.33	&	1.48	&	1.44	&	-3.08	&	00000	&	--	--	&	20-150	\\
2MASS J14215503--3125537	&	3.9	&	1.29	&	1.48	&	1.55	&	-6.36	&	10100	&	--	--	&	35-300	\\
2MASS J1442809--0424078	&	3.0	&	1.31	&	1.50	&	1.29	&	-4.63	&	00000	&	--	--	&	20-150	\\
1RXS J150907.2+590422	&	2.2	&	1.29	&	1.41	&	1.09	&	-2.55	&	01000	&	--	--	&	20-150	\\
GJ 9520	&	1.0	&	1.14	&	1.27	&	0.76	&	-2.14	&	01000	&	--	--	&	15-150	\\
NLTT 40561               	&	3.5	&	1.30	&	1.47	&	1.57	&	-7.27	&	10100	&	--	--	&	25-300	\\
G 167-54                 	&	4.1	&	1.33	&	1.58	&	2.70	&	-0.70	&	10000	&	--	--	&	$\gtrsim$4500	\\
2MASS J15534211--2049282 (S)	&	3.4	&	1.23	&	1.41	&	0.55	&	-78.35	&	11111	&	low g 	(I.N. Reid, priv. comm.)	&	3	\\
GJ 3928                  	&	5.3	&	1.37	&	1.51	&	2.11	&	-5.94	&	10100	&	--	--	&	60-300	\\
NLTT 43695  (E)           	&	4.6	&	1.27	&	1.51	&	1.88	&	-3.19	&	10000	&	--	--	&	60-300	\\
LP 331-57	&	2.4	&	1.30	&	1.43	&	1.45	&	-1.62	&	00000	&	--	--	&	20-150	\\
GJ 616.2	&	5.6	&	1.33	&	1.58	&	2.31	&	-7.43	&	10000	&	--	--	&	90-300	\\
GJ 669A	&	3.4	&	1.23	&	1.43	&	1.56	&	-0.90	&	10000	&	--	--	&	$\gtrsim$2000	\\
1RXS J173130.9+272134	&	2.6	&	--	&	--	&	4.89	&	-7.31	&	?0000	&	--	--	&	20-300	\\
G 227-22                 	&	6.1	&	1.52	&	1.76	&	2.51	&	-7.70	&	00000	&	--	--	&	90-300	\\
GJ 4044                  	&	4.5	&	1.42	&	1.64	&	1.98	&	-6.99	&	00000	&	flare star	\citep{gers99}	&	40-300	\\
LP 390-16	&	3.8	&	1.38	&	1.55	&	1.79	&	-4.34	&	00000	&	--	--	&	35-300	\\
1RXS J183203.0+203050 (N)	&	4.9	&	1.40	&	1.63	&	2.61	&	-3.14	&	00000	&	--	--	&	60-300	\\
1RXS J183203.0+203050 (S)	&	5.1	&	1.47	&	1.71	&	2.29	&	-10.92	&	00000	&	--	--	&	60-300	\\
1RXS J184410.0+712909 (E)	&	3.9	&	1.37	&	1.53	&	1.59	&	-5.24	&	00100	&	--	--	&	35-300	\\
1RXS J184410.0+712909 (W)	&	4.1	&	1.43	&	1.70	&	2.15	&	-3.17	&	00000	&	--	--	&	35-300	\\
GJ 9652 A                	&	4.5	&	1.34	&	1.55	&	1.60	&	-5.71	&	10000	&	--	--	&	60-300	\\
2MASS J19303829--1335083	&	6.0	&	1.38	&	1.57	&	3.06	&	-4.90	&	10000	&	--	--	&	90-300	\\
1RXS J193124.2--213422    	&	2.4	&	1.21	&	1.36	&	1.24	&	-5.36	&	10000	&	--	--	&	20-150	\\
1RXS J193528.9+374605	&	3.0	&	1.25	&	1.45	&	1.60	&	-2.58	&	10000	&	--	--	&	20-300	\\
1RXS J194213.0--204547	&	5.1	&	1.38	&	1.58	&	2.34	&	-6.49	&	10000	&	--	--	&	60-300	\\
G 125-36                 	&	2.1	&	1.22	&	1.39	&	1.13	&	-0.24	&	00000	&	--	--	&	$\gtrsim$1200	\\
2MASS J20003177+5921289	&	4.1	&	1.43	&	1.64	&	1.87	&	-5.67	&	00000	&	--	--	&	35-300	\\
1RXS J204340.6--243410 (NE)	&	3.7	&	1.23	&	1.37	&	0.73	&	-5.24	&	11000	&	--	--	&	35-300	\\
1RXS J204340.6--243410 (SW)	&	4.1	&	1.31	&	1.49	&	0.79	&	-8.03	&	11000	&	--	--	&	35-300	\\
NLTT 49856               	&	4.5	&	1.37	&	1.63	&	1.76	&	-6.50	&	00100	&	--	--	&	40-300	\\
2MASS J20530910--0133039	&	5.6	&	1.53	&	1.85	&	1.92	&	-3.15	&	00000	&	--	--	&	90-300	\\
NLTT 50066               	&	2.9	&	1.32	&	1.50	&	1.38	&	-2.57	&	00000	&	--	--	&	20-300	\\
GJ 4185B	&	3.3	&	1.31	&	1.49	&	1.49	&	-3.21	&	00000	&	--	--	&	25-300	\\
GJ 4185 A                	&	3.3	&	1.32	&	1.51	&	1.77	&	-5.50	&	00100	&	--	--	&	25-300	\\
GJ 4231                    	&	2.4	&	1.21	&	1.37	&	1.22	&	-4.19	&	10000	&	--	--	&	20-150	\\
1RXS J221419.3+253411	&	4.3	&	1.38	&	1.59	&	1.70	&	-5.53	&	00000	&	--	--	&	40-300	\\
GJ 4282 (E)         	&	2.5	&	1.33	&	1.47	&	1.02	&	-4.41	&	01000	&	--	--	&	20-150	\\
GJ 4282 (W)         	&	2.6	&	1.32	&	1.45	&	1.09	&	-4.94	&	01000	&	--	--	&	20-150	\\
2MASS J22344161+4041387 	&	6.0	&	1.20	&	1.27	&	0.14	&	-46.00	&	11011	&	low g, strong \ha\/	(\citealt{cruz03}, Allers et al. 2009)	&	$\sim$1	\\
LP 984-91	&	4.5	&	1.33	&	1.50	&	0.62	&	-7.80	&	11000	&	$\beta$ Pic member	\citep{torr06}	&	12	\\
GJ 873	&	3.2	&	1.36	&	1.52	&	1.59	&	-3.63	&	00000	&	flare star	\citep{gers99}	&	25-300	\\
NLTT 54873               	&	3.8	&	1.39	&	1.55	&	1.79	&	-3.27	&	00100	&	--	--	&	35-300	\\
GJ 875.1	&	2.7	&	1.29	&	1.41	&	1.30	&	-3.38	&	10000	&	--	--	&	20-300	\\
2MASS J22581643--1104170	&	2.7	&	1.29	&	1.42	&	1.37	&	-1.22	&	10000	&	--	--	&	20-300	\\
GJ 9809                  	&	0.3	&	1.12	&	1.18	&	1.00	&	-1.27	&	01010	&	AB Dor member, Li detection	\citep{zuck04b}	&	30--50	\\
NLTT 56194               	&	7.5	&	1.35	&	1.54	&	2.72	&	-7.92	&	10000	&	--	--	&	100-300	\\
NLTT 56566               	&	3.8	&	1.34	&	1.47	&	1.65	&	-4.18	&	10000	&	--	--	&	35-300	\\
GJ 4338 B                	&	4.2	&	1.41	&	1.55	&	2.00	&	-5.05	&	00000	&	flare star	\citep{gers99}	&	40-300	\\
GJ 4337 A                	&	2.9	&	1.30	&	1.42	&	1.47	&	-3.48	&	10000	&	--	--	&	20-300	\\
GJ 1290                  	&	3.4	&	1.32	&	1.46	&	1.59	&	-1.69	&	10000	&	--	--	&	25-300	\\
1RXS J235005.6+265942	&	4.0	&	1.45	&	1.62	&	1.76	&	-6.07	&	00000	&	--	--	&	35-300	\\
G 68-46	&	4.0	&	1.38	&	1.54	&	1.53	&	-5.69	&	00000	&	--	--	&	35-300	\\
1RXS J235133.3+312720    	&	2.0	&	1.25	&	1.36	&	1.35	&	-2.82	&	10000	&	--	--	&	20-150	\\
1RXS J235452.2+383129	&	3.1	&	1.35	&	1.51	&	1.53	&	-3.39	&	00000	&	--	--	&	25-300	\\
GJ 4381                  	&	2.8	&	1.31	&	1.45	&	1.46	&	-3.73	&	00000	&	--	--	&	20-300	\\
G 273-191 (N)	&	1.9	&	1.24	&	1.35	&	1.10	&	-3.26	&	11000	&	--	--	&	20-150	\\
G 273-191 (S)              	&	1.9	&	1.24	&	1.36	&	1.04	&	-2.98	&	01000	&	--	--	&	20-150	\\
G 130-31                 	&	5.6	&	1.27	&	1.44	&	1.83	&	-6.61	&	10000	&	--	--	&	90-300	\\

\enddata

\tablenotetext{a}{Each of the two spectrographs used, Keck+HIRES and CFHT+ESPaDOnS, produces different error bars for measured values. As discussed in Section~\ref{spt}, the smaller errors for the CFHT data are due to the more stable fiber-fed instrumental set-up of ESPaDOnS. (See Table~\ref{targets}.)  The measurement errors for Keck and CFHT respectively are: $\delta$(CaH-wide) = 0.035, 0.008, $\delta$(CaH-narr) = 0.051, 0.013, $\delta$(K~I EW) = 0.24, 0.16, $\delta$(\ha\/ EW) = 0.45, 0.21. Negative \ha\/ EWs indicate the line is in emission.}
\tablenotetext{b}{Youth index in binary format in order of least to most restrictive age indicator: low-g from CaH, low-g from K I, He I emission, Li detection, strong \ha\/ emission (EW$_{\rm H\alpha}<-10$ \AA). Note that determining low-g targets of SpT $\gtrsim$ M6 from the CaH and KI indicators is difficult unless extremely young. See Section~\ref{grav} for more details. Also, due to either poor S/N or bad seeing, we were unable to measure reliable CaH indices from those spectra (denoted by a `?').}
\tablenotetext{c}{Age limits are based on youth diagnostics discussed in Section~\ref{age_dating}, unless a target has been previously published to be a member of known young moving group, in which case the age of the group is adopted. For stars with no additional indications of youth, the lower limit is based on the lack of lithium in the spectra and the upper limit is the {\it statistical} age determined from their SpT and X-ray emission, i.e.~A star of SpT earlier than M2.5 has an upper age limit of 150 Myr and one of later SpT 300 Myr. For the 11 stars with no significant \ha\/ emission (i.e.~$>-1$ \AA), we list their activity lifetimes ($\geq$400 Myr) as determined by \cite{west08}. These ages have large error bars on the order of 500 Myr.}
%\tablenotetext{d}{\cite{torr06} detects a Li EW of 0.12 \AA\/ in their spectrum of this target but we cannot definitively see it.}
\tablenotetext{d}{LP 348-40, GJ 1041A (SW) and G 108-36 appear to have low surface gravity using both the CaH and K I diagnostics but do not have any \ha\/ emission.  According to \cite{west08}, the lack of \ha\/ implies an age $>$ 2000, 400 and 1200 ($\pm$400) Myr, respectively.}
%\tablenotetext{e}{These stars are kinematically linked to the AB Dor moving group with a published age of 30--50 Myr \citep{lope06}.}

\end{deluxetable}

\begin{deluxetable}{lccccccrcrclccccc}
\tabletypesize{\scriptsize}
%\rotate
\tablecaption{RV Standards \label{standards}}
\tablewidth{0pt}
\tablehead{
\colhead{Name} & \colhead{Tele-} & \colhead{SpT} &  \colhead{$J$} & \colhead{log($F_X/F_J$)} & \colhead{CaH\tablenotemark{a}} & \colhead{CaH\tablenotemark{a}} & \colhead{K I\tablenotemark{a}} &\colhead{\ha\/\tablenotemark{a}}  &\colhead{[Fe/H]\tablenotemark{b}} &\colhead{[Fe/H]} &\colhead{RV}  \\
\colhead{} & \colhead{scope\tablenotemark{a}} & \colhead{M-- $\pm0.5$} & \colhead{$_{2MASS}$} & \colhead{} & \colhead{wide} & \colhead{narr} & \colhead{\AA} & \colhead{\AA} & \colhead{publ.} & \colhead{J\&A\tablenotemark{c}} & \colhead{Source\tablenotemark{b}}}
\startdata
							
%Name for Table	&	SpT	&	J (2MASS)	& log(fx/fj) &	CaHwide	&	CaHnarrow	&  K I 7700 EW (A)	& Halpha+/-0.5 A ?	\\
GJ 406	&	Keck	&	6.0	&	7.085	&	-2.928	&	1.50	&	1.69	&	3.45	&	-9.03	&	--		&	high	&	[4]	\\
LHS 2065	&	Keck	&	9.0	&	11.212	&	-3.336	&	1.09	&	1.21	&	1.94	&	-30.49	&	--		&	--	&	[5]	\\
GJ 436	&	Keck	&	2.7	&	6.900	&	--	&	1.25	&	1.39	&	1.02	&	0.46	&	+0.02$\pm$0.20	[1]	&	high	&	[4]	\\
GJ 179	&	CFHT	&	3.7	&	7.814	&	--	&	1.27	&	1.38	&	1.36	&	0.09	&	+0.02$\pm$0.30	[2]	&	high	&	[6]	\\
GJ 908	&	CFHT	&	1.3	&	5.827	&	-4.110	&	1.15	&	1.25	&	0.91	&	0.40	&	--0.52$\pm$0.20	[1]	&	low	&	[4]	\\
VB 10	&	Keck	&	8.0	&	9.908	&	--	&	1.39	&	1.61	&	3.57	&	-4.29	&	--0.05$\pm$0.20\tablenotemark{c}	[1]	&	--	&	[7]	\\
GJ 205	&	CFHT	&	1.2	&	4.999	&	-3.972	&	1.16	&	1.25	&	0.81	&	0.46	&	+0.21$\pm$0.13 	[3]	&	high	&	[6]	\\
GJ 687	&	CFHT	&	3.2	&	5.335	&	-4.347	&	1.24	&	1.36	&	1.15	&	0.16	&	+0.15$\pm$0.09 	[3]	&	high	&	[4]	\\
GJ 821	&	CFHT	&	1.0	&	7.688	&	--	&	1.16	&	1.27	&	0.75	&	0.54	&	--0.65$\pm$0.30	[2]	&	low	&	[4]	\\
GJ 273	&	CFHT	&	3.8	&	5.714	&	--	&	1.28	&	1.41	&	1.25	&	0.00	&	--0.16$\pm$0.20	[1]	&	solar	&	[4]	\\

\enddata
\tablenotetext{a}{Each of the two spectrographs used, Keck+HIRES and CFHT+ESPaDOnS, produces different error bars for the CaH indices and K~I and \ha\/ EWs. (See Table~\ref{targets}.)  The errors for Keck and CFHT respectively are: $\delta$(CaH-wide) = 0.035, 0.008, $\delta$(CaH-narr) = 0.051, 0.013, $\delta$(K~I EW) = 0.24, 0.16, $\delta$(\ha\/ EW) = 0.45, 0.21).}
\tablenotetext{b}{References: [1] \cite{bonf05}, [2] \cite{casa08}, [3] \cite{wool05}, [4] \cite{nide02}, [5] \cite{reid02b}, [6] \cite{marc89}.}
\tablenotetext{c}{As discussed in the text, \cite{john09} have shown that [Fe/H] determinations using the \cite{bonf05}, and thus the \cite{casa08}, calibration are strongly underestimated.  Here we have indicated whether the RV standard has higher or lower than solar metallicity using the new Johnson \& Apps photometric calibration (Figure~\ref{rvstd_colors}). Those that are left blank have colors beyond the limits of the new calibration, which is valid for 3.9 $<$ (V--K) $<$ 6.6.}
\tablenotetext{d}{Based on metallicity measurement for VB 10's companion M dwarf, GJ 752A.}

\end{deluxetable}

\begin{deluxetable}{llccccccccccc}
\tabletypesize{\scriptsize}
\rotate
\tablecaption{Observed $\beta$ Pic Members\label{betapic}}
\tablewidth{0pt}
\tablehead{
\colhead{Name} & \colhead{RA \& DEC} & \colhead{M subclass} &  \colhead{$J$} & \colhead{log($F_X/F_J$)} & \colhead{CaH} & \colhead{CaH} & \colhead{K I EW} &\colhead{\ha\/ EW}  & \colhead{Li EW\tablenotemark{a}} & \colhead{Youth\tablenotemark{b}} \\
\colhead{} & \colhead{} & \colhead{$\pm$0.5} &  \colhead{$_{2MASS}$} & \colhead{} & \colhead{wide, $\pm 0.008$} & \colhead{narr, $\pm 0.013$} & \colhead{$\pm$0.16 \AA} & \colhead{$\pm$0.21 \AA} & \colhead{$\pm$0.025 \AA} & \colhead{Index}
}
\startdata
							
%Name for Table	&	coordinates	&	SpT	&	J (2MASS)	& log(fx/fj) &	CaHwide	&	CaHnarrow	&  K I 7700 EW (A)	& Halpha+/-0.5 A ?	& Li EW & youth code	\\

AU Mic	&	20 45 9.5 --31 20 27     	&	0.9	&	5.436	&	-2.125	&	1.13	&	1.20	&	0.96	&	-1.41	&	0.074	&	11010	\\
AT Mic (N)	&	20 41 51.1 --32 26 7     	&	4.5	&	5.807	&	-2.195	&	1.37	&	1.52	&	0.87	&	-11.04	&	--	&	11101	\\
AT Mic (S)	&	20 41 51.1 --32 26 10    	&	4.9	&	5.807	&	-2.195	&	1.38	&	1.55	&	1.08	&	-8.13	&	--	&	11101	\\
HIP 11437 B	&	2 27 28.1 +30 58 41      	&	2.2	&	8.817	&	-1.821	&	1.24	&	1.36	&	0.89	&	-3.95	&	0.133	&	11010	\\
BD-13 6424	&	23 32 30.9 --12 15 52    	&	0.8	&	7.45	&	-2.311	&	1.14	&	1.21	&	0.75	&	-1.83	&	0.171	&	11010	\\
GJ 182	&	4 59 34.8 +1 47 2	&	0.5	&	7.117	&	-2.450	&	1.12	&	1.18	&	0.74	&	-1.23	&	0.296	&	11010	\\
HIP 11437 A	&	2 27 28.1 +30 58 41      	&	-0.2	&	7.87	&	-1.821	&	1.04	&	1.07	&	0.53	&	-0.16	&	0.271	&	11010	\\
HIP 12545	&	2 41 25.8 +5 59 19	&	-0.1	&	7.904	&	-2.203	&	1.06	&	1.09	&	0.59	&	-0.69	&	0.447	&	11010	\\

\enddata

\tablenotetext{a}{The lithium abundances have not been corrected for possible contamination with the Fe I line at 6707.44 \AA. Uncertainties in the setting of continuum levels prior to measurement induce EW errors of about 10--20 m\AA\/  \citep{zuck04} with a strong dependence on the S/N in the region. We therefore conservatively consider our detection limit to be 0.05 \AA.}
\tablenotetext{b}{Youth index in binary format in the following order: low-g from CaH, low-g from K I, He I emission, Li detection, strong \ha\/ emission (EW$_{\rm H\alpha}<-10$ \AA).}

\end{deluxetable}

\begin{deluxetable}{lcccccccccccc}
\tabletypesize{\scriptsize}
%\rotate
\tablecaption{Targets with detected lithium\label{lithium_stars}}
\tablewidth{0pt}
\tablehead{
\colhead{Name} & \colhead{SpT} &  \colhead{Li EW\tablenotemark{a}} & \colhead{He I EW} & \colhead{\ha\/ EW} & \colhead{\ha\/ 10\%-width} & \colhead{Accretor?\tablenotemark{b}}    \\
\colhead{} & \colhead{M-- ($\pm 0.5$)} & \colhead{$\pm$0.025 \AA} & \colhead{$\pm$0.05 \AA}  & \colhead{$\pm$0.45 \AA}  & \colhead{$\pm$4 km/s} & \colhead{} }
\startdata
							
%Name for Table	&	SpT	&	Li EW	& He I & Halpha EW & Halpha 10% width & Accretor?	\\
2MASS J03350208+2342356	&	8.5	&	0.615	&	--2.89	&	--10.7	&	273	&	yes	\\
2MASS J05575096--1359503	&	7.0	&	0.285	&	--	&	--25.6	&	208	&	possible	\\
2MASS J15534211--2049282 (S)	&	3.4	&	0.630	&	--0.81	&	--70.9, --85.8	&	421, 470	&	yes	\\
2MASS J22344161+4041387 	&	6.0	&	0.825	&	--	&	--52.7,--39.2	&	314, 298	&	yes	\\
\\
1RXS J041417.0--090650	&	4.3	&	0.078	&	--0.54	&	--5.1, --23.2	&	86, 242	&	no	\\
GJ 3305                  	&	1.1	&	0.086	&	--	&	--2.1	&	113	&	no	\\
GJ 9809                  	&	0.3	&	0.087	&	--	&	--1.3	&	105	&	no	\\
NLTT 13728               	&	6.0	&	0.126	&	--0.28	&	--19.1	&	89	&	no	\\
\enddata

\tablenotetext{a}{The lithium abundances have not been corrected for possible contamination with the Fe I line at 6707.44 \AA. Uncertainties in the setting of continuum levels prior to measurement induce EW errors of about 10--20 m\AA\/ \citep{zuck04} with a strong dependence on the S/N in the region. We therefore consider our detection limit to be 0.05 \AA.}
\tablenotetext{b}{Accreting or not, based on criteria discussed in text and devised by \cite{moha05} and \cite{whit03}.}
%\tablenotetext{c}{No ROSAT detection perhaps due to the newly-discovered binarity putting the target outside of ROSAT's sensitivity.  However, we kept it in our sample due the previous indicators of youth.}

\end{deluxetable}
 \clearpage

%%%%%FIGURES%%%%%%

\begin{figure}
\epsscale{.80}
\plotone{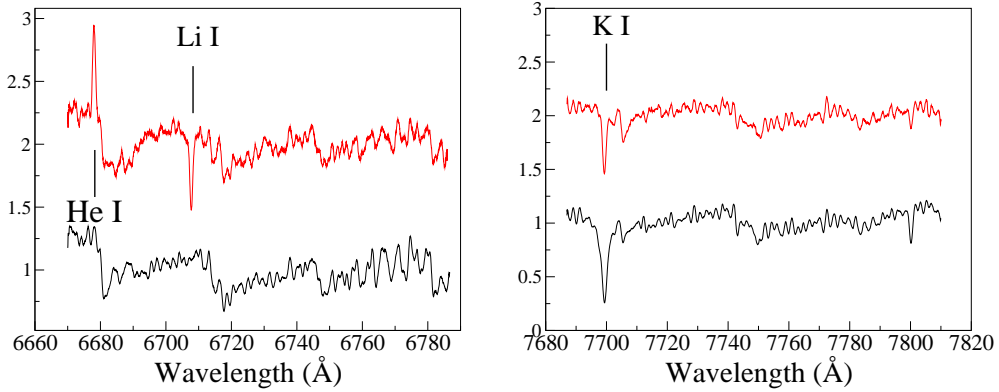}
\caption{Two orders of Keck/HIRES spectra for two young M4
  targets from our sample, 2MASS~J1553 (S) (red) and 1RXS~J142155.3 (black).  {\it Top
    spectra:} The strong Li~I ($\lambda$6708) absorption, He I
  ($\lambda$6678) emission, and the lower K I ($\lambda$7700) EW all
  indicate a very young age for 2MASS~J1553 (S), $\lesssim$40~Myr based on
  \cite{chab96} models for pre-main sequence lithium
  depletion.  {\it Bottom spectra:} 1RXS~J142155.3 shows no clear
  spectroscopic signatures of low gravity, indicating it is on or near
  the main-sequence.  However, its X-ray flux is significantly
  enhanced compared to Hyades stars and nearby (many Gyr old) field
  stars, indicating an age of $\lesssim$300~Myr.
\label{Li_KI}}
\end{figure}

\begin{figure}
\epsscale{.60}
\plotone{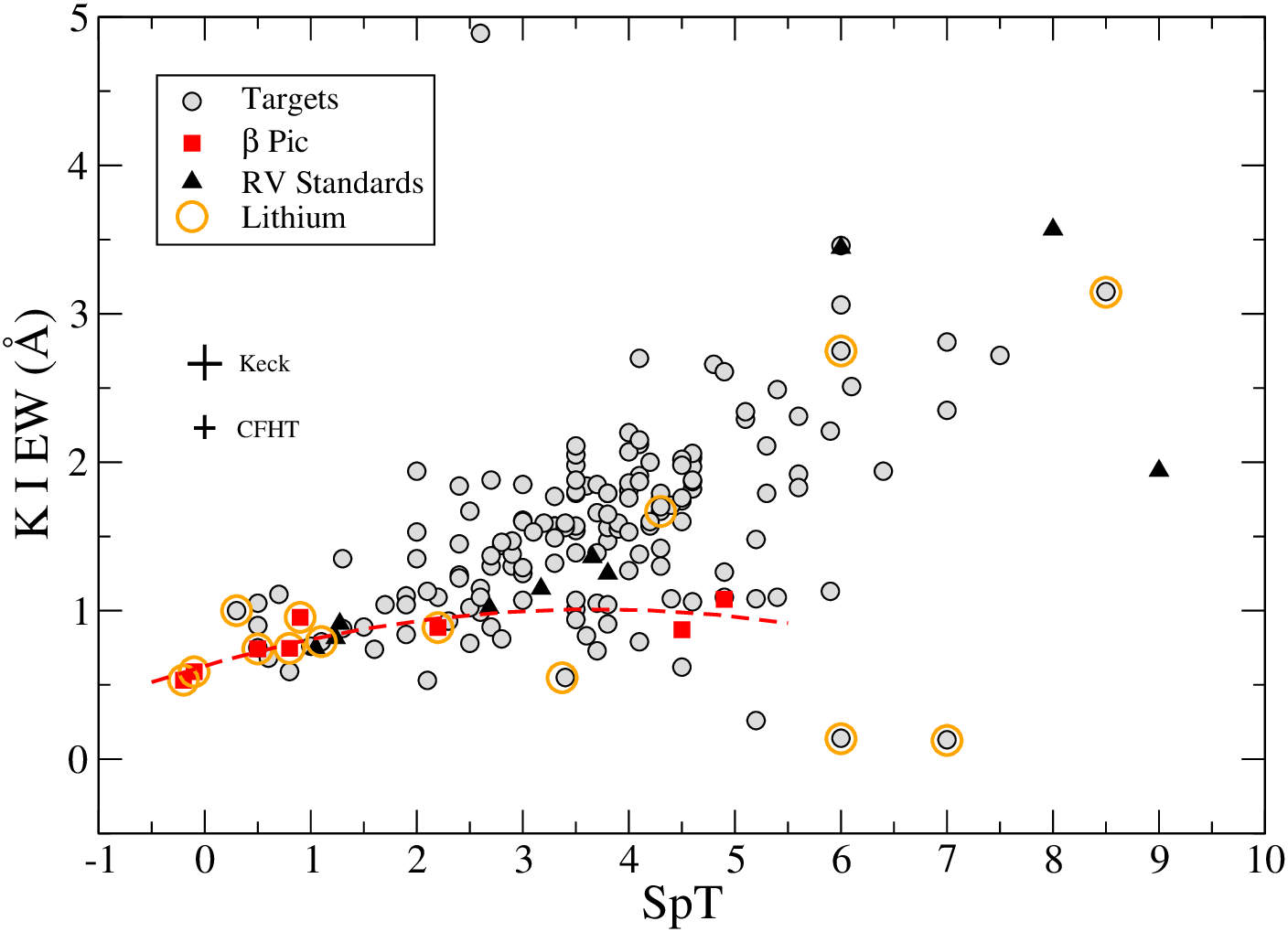}
\caption{K I equivalent width as a function of M-type subclass. The red dashed curve is a polynomial fit to the $\beta$ Pic observations: EW$_{\rm KI}$ = --0.02821 SpT$^2$ + 0.20731 SpT + 0.62911.
\label{SpT_KI}}
\end{figure}

\begin{figure}
\epsscale{.60}
\plotone{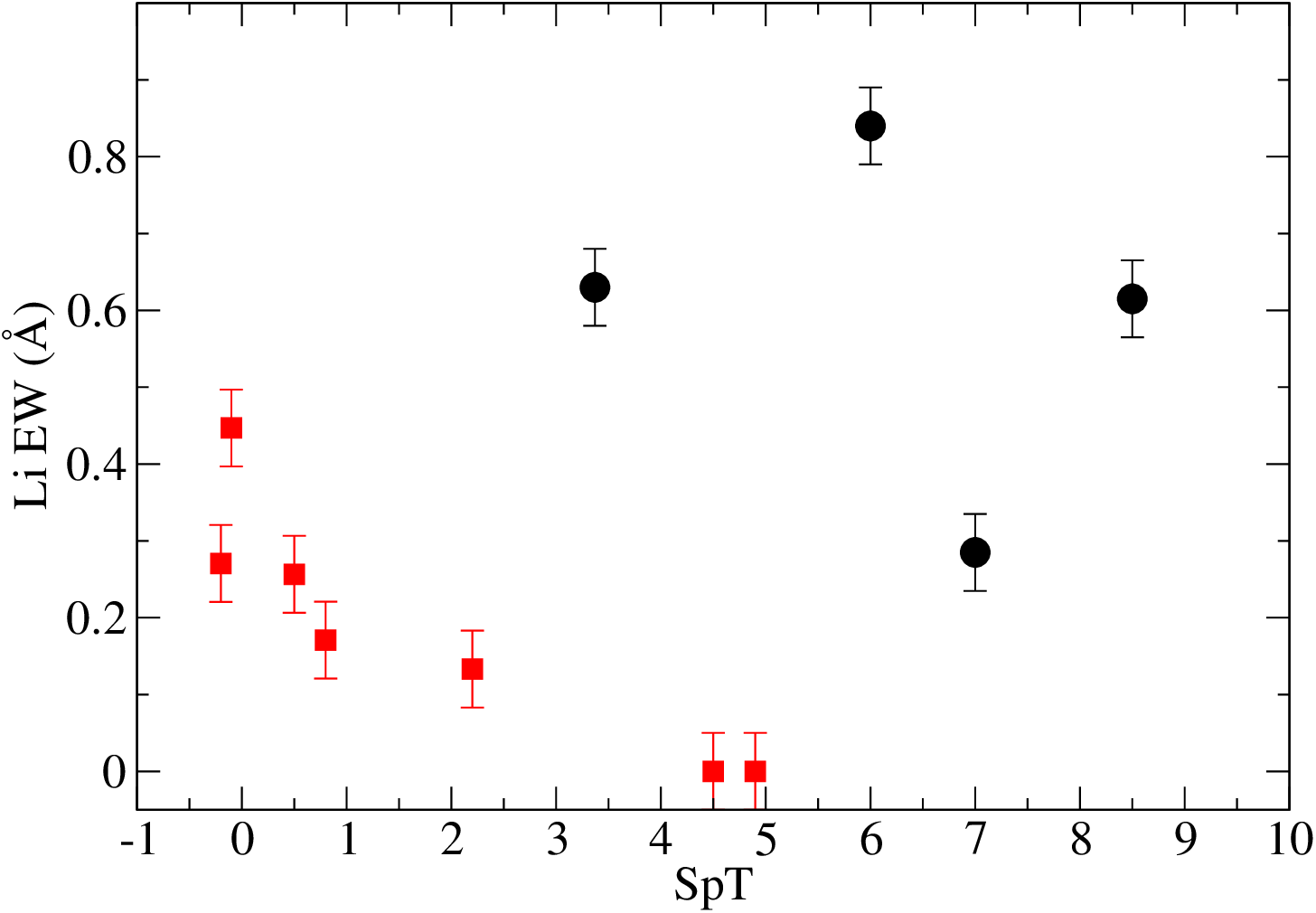}
\caption{Measured lithium EWs for the 8 targets with clear Li absorption (black circles) and for the known $\beta$ Pic members (red squares) that we observed.
\label{lithium_EW}}
\end{figure}

\begin{figure}
\epsscale{.60}
\plotone{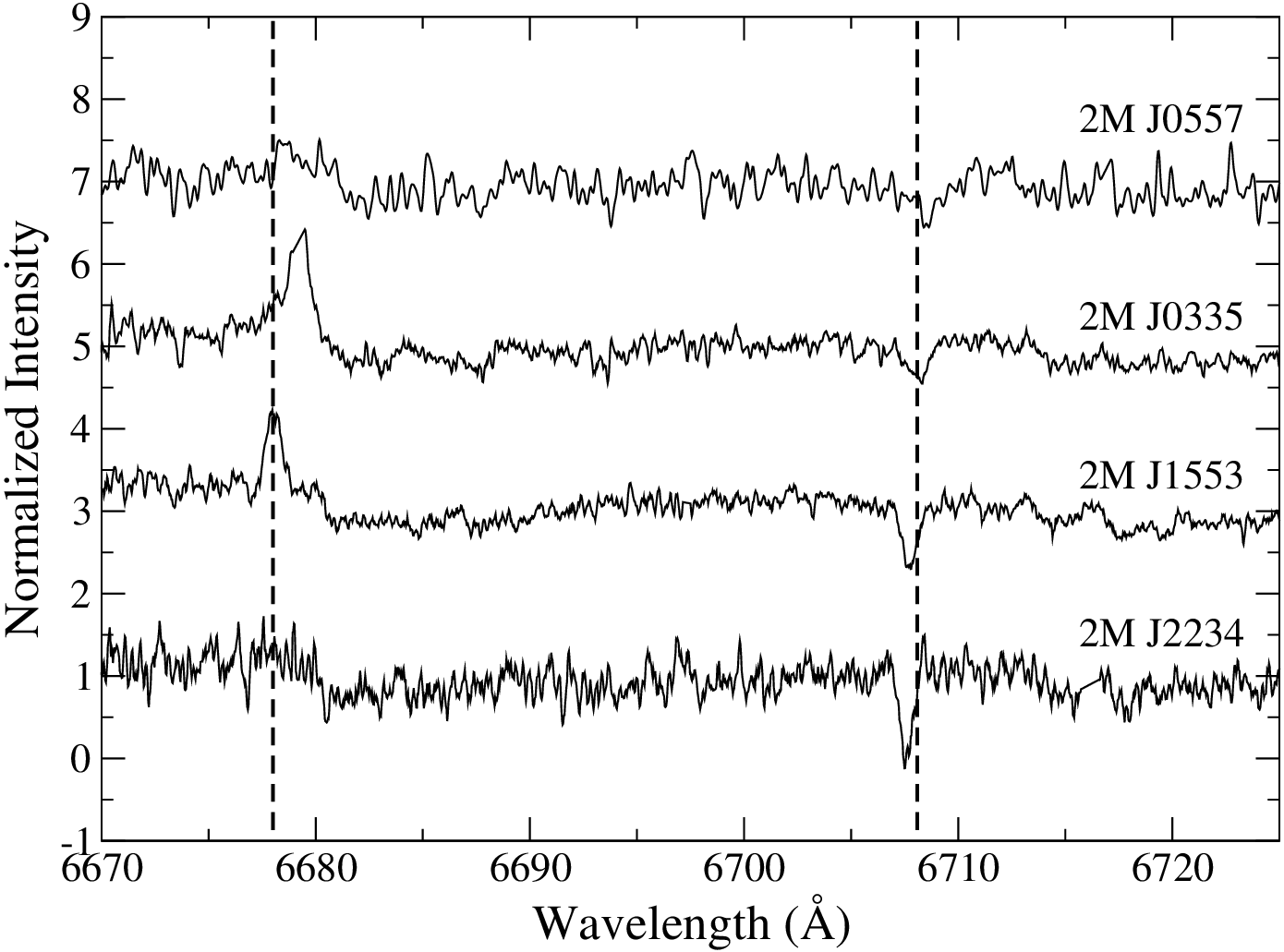}
\caption{Keck/HIRES spectra of the four accreting targets listed in Table~\ref{lithium_stars}. The vertical dashed lines mark the rest wavelengths of He I ($\lambda$6678) and Li I ($\lambda$6708). The shifts of the Li line centers are due to the stars' radial velocities.
\label{lithium_spectra}}
\end{figure}

\begin{figure}
\epsscale{.60}
\plotone{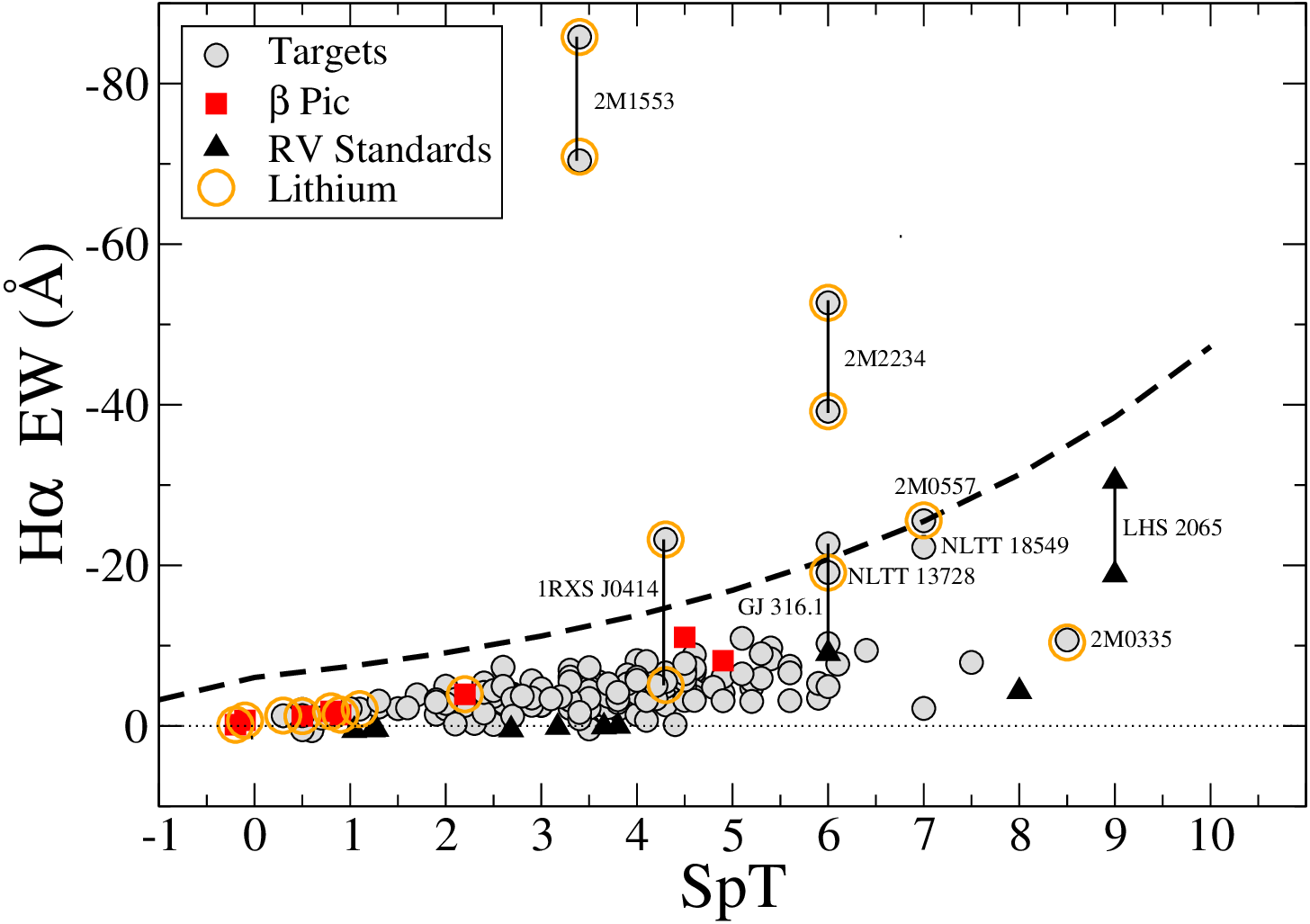}
\caption{Measured EW$_{\rm H\alpha}$. Of the 65 M dwarfs observed twice, 4 had H$\alpha$ variability in excess of 10 \AA, plus the standard star LHS 2065. Each is plotted twice and connected with a solid line. The dashed curve represents the empirical accretion boundary determined by \cite{barr03}. See text for further discussion of the labeled targets.
\label{SpT_Ha}}
\end{figure}

\begin{figure}
\epsscale{.80}
\plotone{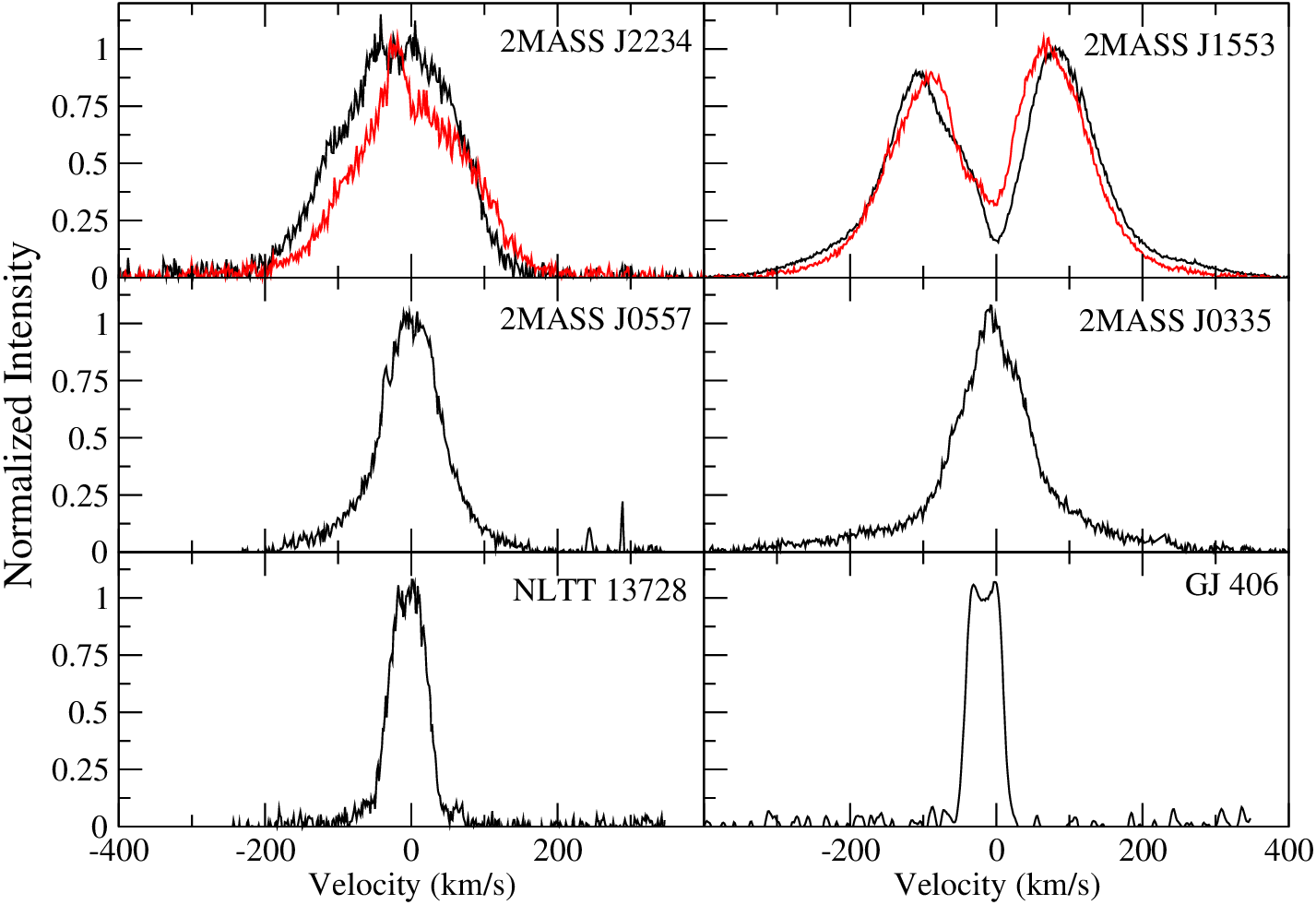}
\caption{The top four \ha\/ velocity profiles are those of the targets with strong 
lithium absorption, while NLTT 13728 has weaker EW$_{\rm Li}$ and GJ 406, a RV standard, has none. Both observations are shown for 2MASS J2234 and 2MASS J1553. The profiles have been continuum-subtracted and normalized to range from 0.0 
(continuum level) to 1.0 (emission peak level). Their 10\%-widths (the full velocity span at 0.1 of the peak flux) are 306, 446, 207, 273, 89 and 72 km/s, top left to bottom right. 
According to criteria devised by \cite{whit03} and \cite{moha05}, 2MASS~J2234, 2MASS~J1553 and 2MASS~J0335 are 
accreting T Tauri stars while 2MASS J0557 probably is as well.   
The corresponding age limit for each of these objects (again top left to bottom right) is $\sim$1 (Allers et al., in press), 3 (\citealt{whit03}), 10, 10 (\citealt{barr03}), and 300 Myr, with the standard star GJ 406 likely $>$ 1 Gyr. 
\label{Halpha_vel}}
\end{figure}

\end{document}